\documentclass{LMCS}

\usepackage{amssymb,amsmath}
\usepackage{latexsym} 

\usepackage{pifont} 
\usepackage{url}

\usepackage{enumerate,hyperref}

\newenvironment{proofof}[1]{\par\noindent\textbf{Proof of {#1}.}}{\mbox{}\qed\par\medskip}
\renewenvironment{proof}{\par\noindent\textbf{Proof.}}{\mbox{}\qed\par\medskip}

\theoremstyle{plain}
\newtheorem{theorem}[thm]{Theorem}
\newtheorem{lemma}[thm]{Lemma}

\newtheorem{claim}[thm]{Claim}
\newtheorem{corollary}[thm]{Corollary}

\usepackage{graphics}

\usepackage[dvipsnames,usenames]{xcolor}

\newcommand{\cB}{\ensuremath{\mathcal{B}}}%
\newcommand{\cBc}{\ensuremath{\mathcal{B}c}}%
\newcommand{\cBcc}{\ensuremath{\mathcal{B}cc}}%

\newcommand{\cBC}{\ensuremath{\mathcal{C}}}
\newcommand{\cBCc}{\ensuremath{\cBC{}c}} 
\newcommand{\cBCcc}{\ensuremath{\cBC{}cc}} 

\newcommand{\SF}{\ensuremath{\mathcal{S}4}}%
\newcommand{\SFU}{\ensuremath{\mathcal{S}4_u}}%
\newcommand{\conT}{\ensuremath{\SFU{}c}}%
\newcommand{\conTc}{\ensuremath{\SFU{}cc}}%

\newcommand{\cBCm}{\ensuremath{\mathcal{C}^m}}
\newcommand{\cBCmc}{\ensuremath{\cBCm{}c}}%
\newcommand{\cBCmcc}{\ensuremath{\cBCm{}cc}}%

\newcommand{\BRCCE}{\ensuremath{\mathcal{BRCC}\text{-}8}}%

\newcommand{\RCCE}{\ensuremath{\mathcal{RCC}\text{-}8}}%
\newcommand{\RCCEcc}{\ensuremath{\RCCE{}cc}}%
\newcommand{\RCCEc}{\ensuremath{\RCCE{}c}}%

\newcommand{\fA}{\mathfrak{A}}%
\newcommand{\fB}{\mathfrak{B}}%
\newcommand{\fM}{\mathfrak{M}}%
\newcommand{\cK}{\mathcal{K}}%
\newcommand{\R}{\mathbb{R}}%
\newcommand{\cR}{\mathcal{R}}%
\newcommand{\cL}{\mathcal{L}}%
\newcommand{\cLc}{\mathcal{L}c}%
\newcommand{\cLcc}{\mathcal{L}cc}%

\newcommand{\ti}[2][]{#2^{\circ_{#1}}}
\newcommand{\tc}[2][]{#2^{-_{#1}}}
\newcommand{\compl}[2][]{\overline{#2}}

\newcommand{\Sat}{\textit{Sat}}

\newcommand{\All}{\textsc{All}}
\newcommand{\AlekF}{\textsc{Alek}} 
\newcommand{\Con}{\textsc{Con}}
\newcommand{\Regc}{\textsc{RegC}}
\newcommand{\ConR}{\textsc{ConRegC}}

\newcommand{\NLogSpace}{\textsc{NLogSpace}}
\newcommand{\NP}{\textsc{NP}}
\newcommand{\PSpace}{\textsc{PSpace}}
\newcommand{\ExpTime}{\textsc{ExpTime}}
\newcommand{\NExpTime}{\textsc{NExpTime}}

\newcommand{\zero}{\mathbf{0}}%
\newcommand{\one}{\mathbf{1}}%

\newcommand{\Res}{{S}}%
\newcommand{\RegC}{\textsc{RC}}

\newcommand{\RVar}{\mathcal{R}}

\newcommand{\tp}{\mathfrak{t}}

\newcommand{\since}{\mathbin{\mathcal{S}}}
\newcommand{\until}{\mathbin{\mathcal{U}}}

\newcommand{\coor}[1]{\textit{coor}(#1)}%
\newcommand{\neighb}[1]{\textit{4-nb}(#1)}%

\newcommand{\Exp}{\textsc{ExpTime}}
\newcommand{\NExp}{\textsc{NExpTime}}

\newcommand{\NPPSpace}{\!\!\small$\leq$\PSpace{},$\geq$\NP{}\!\!\!}
\newcommand{\CPLX}[3][Thm.]{\rule[-8pt]{0pt}{20pt}\!\!\raisebox{-3pt}[0pt][7pt]{\begin{tabular}{c}#2\\[-6pt]{\scriptsize #1~\ref{#3}}\end{tabular}}\!\!}

\newcommand{\fzc}{\thicklines\circle{3.5}}
\newcommand{\foc}{\thicklines\circle*{3.5}}

\title[Spatial logics with connectedness predicates]{Spatial logics with connectedness predicates\rsuper*}

\author[R.~Kontchakov]{Roman Kontchakov\rsuper a}	
\address{{\lsuper a}Department of Computer Science and Information Systems, Birkbeck College London}	
\email{roman@dcs.bbk.ac.uk}  

\author[I.~Pratt-Hartmann]{Ian Pratt-Hartmann\rsuper b}	
\address{{\lsuper b}Department of Computer Science, Manchester University}	
\email{ipratt@cs.man.ac.uk}  

\author[F.~Wolter]{Frank Wolter\rsuper c}	
\address{{\lsuper c}Department of Computer Science, University of Liverpool}	
\email{frank@csc.liv.ac.uk}  

\author[M.~Zakharyaschev]{Michael Zakharyaschev\rsuper d}	
\address{{\lsuper d}Department of Computer Science and Information Systems, Birkbeck College London}	
\email{michael@dcs.bbk.ac.uk}  

\keywords{Spatial logic, topological space, connectedness, computational complexity}
\subjclass{F.4.1, I.2.4}
\titlecomment{{\lsuper*}Some results of the paper were presented at LPAR 2008 and ECAI 2000.}

\def\doi{6 (3:7) 2010}
\lmcsheading%
{\doi}
{1--43}
{}
{}
{Jun.~17, 2009}
{Aug.~18, 2010}
{}

\begin{document}

\maketitle

\begin{abstract}
We consider quantifier-free spatial logics, designed for qualitative
spatial representation and reasoning in AI, and extend them with the
means to represent topological connectedness of regions and restrict
the number of their connected components. We investigate the
computational complexity of these logics and show that the
connectedness constraints can increase complexity from \NP{} to
\PSpace{}, \ExpTime{} and, if component counting is allowed, to
\NExpTime{}.
\end{abstract}


\section{Introduction}
\label{sec:intro}

The field of Artificial Intelligence known as \emph{qualitative spatial
reasoning} is concerned with the problem of representing and
manipulating spatial information about everyday, middle-sized
entities. In recent decades, much activity in this field has centred
on {\em spatial logics}---formal languages whose variables range over
geometrical objects (not necessarily points), and whose non-logical
primitives represent geometrical relations and operations involving
those objects. (For a recent survey, see~\cite{Cohn&Renz08}.)  The
hope is that, by using a formalism couched entirely at the level of
these geometrical objects, we can avoid the expressive---hence
computationally expensive---logical machinery required to reconstruct
them in terms of sets of points.

\begin{figure}[ht]
\begin{center}
\begin{picture}(0,0)%
\includegraphics{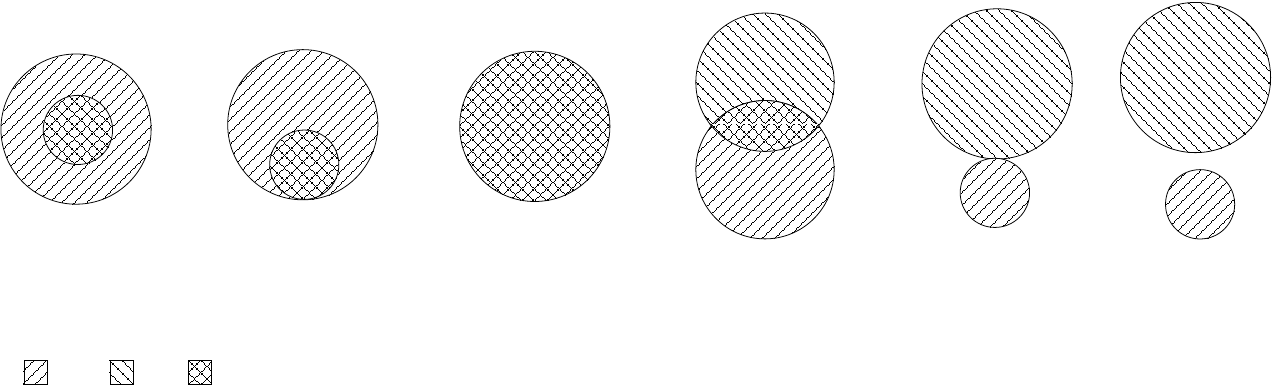}%
\end{picture}%
\setlength{\unitlength}{1973sp}%
\begin{picture}(12205,3685)(522,-3148)\footnotesize
\put(526,-2311){${\mathsf{NTTP}}(X,Y)$ }%
\put(2626,-2311){${\mathsf{TTP}}(X,Y)$ }%
\put(4801,-2311){${\mathsf{EQ}}(X,Y)$ }%
\put(7276,-2311){${\mathsf{PO}}(X,Y)$ }%
\put(11326,-2311){${\mathsf{DC}}(X,Y)$ }%
\put(9451,-2311){${\mathsf{EC}}(X,Y)$ }%
\put(1051,-3111){$X$}%
\put(1876,-3111){$Y$}%
\put(2626,-3111){$X$ and $Y$}%
\end{picture}%
\end{center}
\caption{The $\RCCE$-relations illustrated for disc-homeomorphs in $\R^2$}
\label{fig:ccircles}
\end{figure}
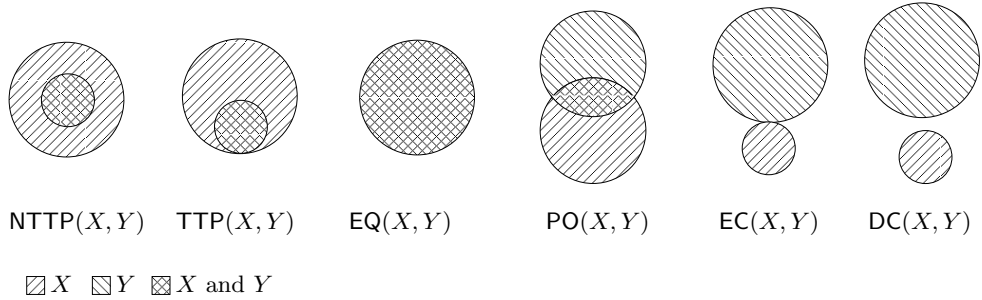

What might such qualitative spatial relations typically be?  Probably
the most intensively studied collection is the set of six topological
relations illustrated---for the case of closed disc-homeomorphs in
$\R^2$---in Fig.~\ref{fig:ccircles}.  These relations---$\mathsf{DC}$
(disconnection), $\mathsf{EC}$ (external connection), $\mathsf{PO}$
(partial overlap), $\mathsf{EQ}$ (equality), $\mathsf{TPP}$
(tangential proper part) and $\mathsf{NTPP}$ (non-tangential proper
part)---were popularized in the seminal treatments of spatial logics
by Egenhofer and
Franzosa~\cite{Egenhofer&Franzosa91} and Randell~{\em et al.}~\cite{Randelletal92}.
Counting the converses of the two asymmetric relations $\mathsf{TPP}$
and $\mathsf{NTPP}$, the resulting eight relations are frequently
referred to under the moniker $\RCCE$ (for {\em region connection
  calculus}). To see how this collection of relations gives rise to a
spatial logic, let $r_1$, $r_2$ and $r_3$ be disc-homeomorphs in
$\R^2$, and suppose that $r_1$, $r_2$ stand in the relation
$\mathsf{TPP}$, while $r_1$, $r_3$ stand in the relation
$\mathsf{NTPP}$. A little experimenting with diagrams suffices to show
that $r_2$, $r_3$ must stand in one of the three relations
$\mathsf{PO}$, $\mathsf{TPP}$ or $\mathsf{NTPP}$. As we might put it,
the $\RCCE$-formula
\begin{equation*}
\big ( \mathsf{TPP}(r_1,r_2) \wedge \mathsf{NTPP}(r_1,r_3) \big ) \to
 \big ( \mathsf{PO}(r_2,r_3) \vee \mathsf{TPP}(r_2,r_3) \vee
  \mathsf{NTPP}(r_2,r_3) \big )
\end{equation*}
is valid over the spatial domain of disc-homeomorphs in the plane: all
assignments of such regions to the variables $r_1$, $r_2$ and $r_3$
make it true. Similar experimentation shows that, by contrast, the
formula
\begin{equation*}
\mathsf{TPP}(r_1,r_2) \wedge \mathsf{NTPP}(r_1,r_3) \land \mathsf{EC}(r_2,r_3),
\end{equation*}
is {\em unsatisfiable}: no assignments of disc-homeomorphs to $r_1$,
$r_2$ and $r_3$ make this formula true.

More generally, let $\cL$ be a formal language featuring some
collection of predicates and function symbols having (fixed)
interpretations as geometrical relations and operations.  The formulas
of $\cL$ may then be interpreted over any collection of subsets of
some space $T$ for which the relevant geometrical notions make sense:
we refer to the elements of such a domain of interpretation as \emph{regions}.
Let $\cK$ be a class of domains of interpretation for
$\cL$. The notion of the \emph{satisfaction} of an $\cL$-formula by a
tuple of regions, and, derivatively, the notions of \emph{satisfiability}
and \emph{validity} of an $\cL$-formula with respect
to $\cK$, can then be understood in the usual way. We call the pair
$(\cL, \cK)$ a \emph{spatial logic}. If all the primitives of $\cL$ are
topological in character---as in the case of $\RCCE$---we speak of a
\emph{topological logic}. For languages featuring negation, the notions
of satisfiability and validity are dual in the usual sense.  The
primary question arising in connection with any spatial logic is: how
do we recognize the satisfiable (dually, the valid) formulas? From an
algorithmic point of view, we are particularly concerned with the
decidability and complexity of these problems.

A second example will make this abstract characterization more
concrete.  Constraints featuring $\RCCE$ predicates give us no means
to \emph{combine} regions into new ones; and it is natural to ask what
happens when this facility is provided. Let $T$ be a topological
space.  A subset of $T$ is \emph{regular closed} if it is the
topological closure of an open set in $T$. The collection of regular
closed sets forms a Boolean algebra with binary operations $+$,
$\cdot$ and a unary operation $-$.  Intuitively, we are to think of
$r_1 +r_2$ as the \emph{agglomeration} of $r_1$ and $r_2$, $r_1 \cdot
r_2$ as the \emph{common part} of $r_1$ and $r_2$, and $-r$ as the
\emph{complement} of $r$.  Further, the $\RCCE$-relations illustrated
above can be generalized, in a natural way, so that they apply to
regular closed subsets of any topological space $T$. (Details are
given below.)  By augmenting the language $\RCCE$ with the function
symbols $+$, $\cdot$ and $-$, we obtain the more expressive formalism
originally introduced in~\cite{Wolter&Z00ecai} under the name $\BRCCE$
({\em Boolean} $\RCCE$), which we may interpret over any algebra of
regular closed subsets of some topological space. Again, we can ask
which of these formulas are satisfiable or valid over the class of
domains in question.  For instance, the formula
\begin{equation*}
\mathsf{EC}(r_1+r_2, r_3) \to
\big (\mathsf{EC}(r_1, r_3) \vee \mathsf{EC}(r_2, r_3) \big ),
\end{equation*}
asserting that, if the sum of two regions $r_1$ and $r_2$ stands in the
relation of external connection to a region $r_3$, then one of
the summands does as well, turns out to be valid; by contrast, the formula
\begin{equation*}
\mathsf{EC}(r_1, r_2) \wedge \mathsf{EC}(r_1, -r_2),
\end{equation*}
asserting that $r_1$ is externally connected to both $r_2$ and its
complement, is unsatisfiable.

Arguably, the topological primitive with the longest pedigree in the
spatial logic literature is the relation now generally referred to as
$C$ (for \emph{contact}).  Intuitively, two regions are to be thought
of as being in contact just in case they either overlap or have
touching boundaries. This relation was originally introduced by
Whitehead (\cite{Whitehead29}, pp.~294, ff.) under the name {\em
  extensive connection}, and formed the starting point for his
region-based reconstruction of space. More recently, it has been
studied within the framework of \emph{Boolean contact algebras}~\cite{DV1,DW05}.
It turns out that, in the presence of the
operations $+$, $\cdot$ and $-$, all of the $\RCCE$-relations can be
expressed in terms of the relations of equality and contact, and \emph{vice versa}.
Accordingly, and in order to unify these two lines of
research, we shall denote the language $\BRCCE$ by $\cBC$ in this
paper.

\begin{figure}
\begin{center}
\begin{picture}(0,0)%
\includegraphics{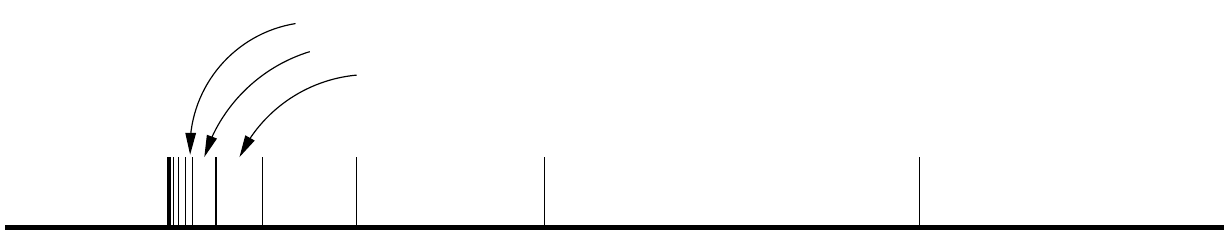}%
\end{picture}%
\setlength{\unitlength}{2960sp}%
\begin{picture}(7866,1479)(1918,-1594)\small
\put(2401,-1411){$r_1$}%
\put(6151,-1411){$r_3$}%
\put(8626,-1411){$r_4$}%
\put(4576,-1411){$r_2$}%
\put(3676,-1411){$r_4$}%
\put(3961,-496){$r_2$}%
\put(4231,-661){$r_3$}%
\put(3826,-271){$r_4$}%
\end{picture}%
\end{center}
\caption{A configuration of regular closed regions in $\R$ satisfying
  the $\RCCEc$-formula $c(r_1) \wedge \bigwedge_{1 \leq i < j \leq 4}
  \mathsf{EC}(r_i,r_j)$: the region $r_1$ is connected, while $r_2$, $r_3$
  and $r_4$ have infinitely many components, with a common accumulation point
  on the boundary of the $r_1$-region.}
\label{fig:icky}
\end{figure}
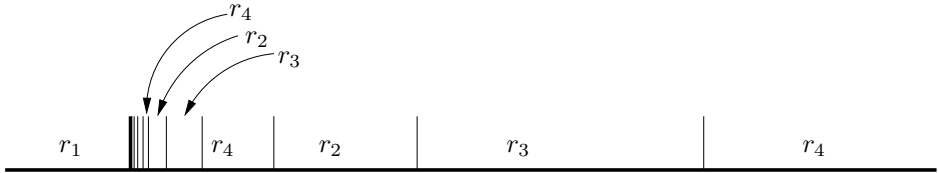

One familiar topological property that has been notable by its absence
from the spatial logic literature, however, is \emph{connectedness}
(or, as it is occasionally called, `self-connect\-edness'~\cite{Borgo96}).
This lacuna is particularly surprising
given the recognized significance of this concept in qualitative
spatial reasoning~\cite{Cohn&Renz08}. The availability of
connectedness as a primitive relation greatly expands the expressive
power of topological logics, and in particular increases their
sensitivity to the underlying domain of quantification. For example,
let the connectedness predicate $c$ be added to the language $\RCCE$,
yielding the language $\RCCEc$; and consider the $\RCCEc$-formula
\begin{equation*}
\bigwedge_{1 \leq i \leq 3} c(r_i) \wedge
\bigwedge_{1 \leq i < j \leq 3} \mathsf{EC}(r_i,r_j).
\end{equation*}
This formula states that regions $r_1$, $r_2$ and $r_3$ are connected,
and that any two of them touch at their boundaries without
overlapping. It is easily seen to be satisfiable over the domain of
regular closed sets in $\R^2$; however, it is not satisfiable over the
domain of regular closed sets in $\R$. For a non-empty, regular closed
subset of $\R$ is connected if and only if it is a non-punctual,
closed interval (possibly unbounded); and it is obvious that no three
such intervals can touch in pairs without overlapping. More tellingly,
consider the $\RCCEc$-formula
\begin{equation*}
c(r_1) \wedge \bigwedge_{1 \leq i < j \leq 4} \mathsf{EC}(r_i,r_j),
\end{equation*}
stating that $r_1$ is connected, and that any two of $r_1, \ldots,
r_4$ touch at their boundaries without overlapping. This formula \emph{is}
satisfiable over the regular closed subsets of $\R$, as shown in
Fig.~\ref{fig:icky}.
However, such an arrangement is only possible provided at least two of
the regions $r_2$, $r_3$ and $r_4$ have infinitely many
components. If---for whatever reason---our spatial ontology does not
countenance regions with infinitely many components, the formula
becomes unsatisfiable. Thus, simple logics featuring the connectedness
predicate are sensitive to the underlying topological space, and
indeed to the choice of subsets of that space we count as regions. By
contrast, we shall see below that topological logics lacking the
connectedness predicate---even ones much more expressive than
$\RCCE$---are remarkably insensitive to the spatial domains over which
they are interpreted. More generally, the above examples give a hint
of the interesting mathematical challenges which the property of
connectedness presents us with in the context of almost any
topological logic.

It is surprising that only sporadic attempts have been made to
investigate the expressive power and computational complexity of
topological logics able to talk about the connectedness of
regions~\cite{cc94,Shehtman99,Wolter&Z00ecai,PrattHartmann02}.  The
present paper rectifies this omission by introducing the unary
predicates $c$ and $c^{\leq k}$ (for $k \geq 1$).
We read $c(r)$ as
`region $r$ is connected' and $c^{\le k}(r)$ as `region $r$ has at
most $k$ connected components.' Our aim is to provide a systematic
study of the impact of these predicates on the computational
complexity of the satisfiability problem for topological logics.  We
restrict attention in this paper to {\em quantifier-free}
languages---i.e.~those in which formulas are Boolean combinations of
atomic formulas---in line with the constraint satisfaction approach
of~\cite{Renz&Nebel07}---since first-order spatial logics are
generally undecidable
\cite{KK:Grzegorczyk51,KK:Dornheim98kr,Davis06,Lutz&WolterLMCS}.\footnote{One
  of the notable exceptions in this regard is Tarski's theory of {\em
    elementary geometry}, which can be regarded as a first-order
  spatial logic whose domain of interpretation is the set of points in
  the Euclidean plane. The precise computational complexity of this
  logic---essentially the first-order fragment of the system set out
  by Hilbert~\cite{hilbert09}---is still unknown, with the current
  lower bound being \NExpTime{} \cite{Fischer&Rabin74} and the upper
  bound \textsc{ExpSpace} \cite{expspace}.}  For an overview of
first-order topological logics, see~\cite{PH:HSL}.

Specifically, we consider three principal \emph{base languages},
characterized by various collections of topological primitives, and
investigate the effect of augmenting each of these base languages with
the predicates $c$ and $c^{\le k}$.  The weakest of these base
languages, denoted $\cB$, features only the region-combining operators
$+$, $\cdot$ and $-$, together with the equality predicate. Thus,
$\cB$ is essentially just the language of Boolean algebra equations:
as such, this language can express no really characteristic
topological properties; further, its satisfiability problem, when
interpreted over the class of regular closed algebras of topological
spaces, is easily seen to be \NP-complete.  If, however, we add the
connectedness predicate $c$, we obtain the language $\cBc$---a
fully-fledged topological logic able to simulate (in a sense explained
below) the contact relation $C$, and hence all the
$\RCCE$-relations. More ambitiously, we can add to $\cB$ all of the
predicates $c^{\le k}$ (for $k \geq 1$), to obtain the language
$\cBcc$. An indication of the resulting increase in expressiveness is
that the satisfiability problem for the same class of interpretations
jumps from \NP{} to \ExpTime{} (in the case of $\cBc$) and \NExpTime{}
(in the case of $\cBcc$).

Our next base language is $\cBC$ (alias $\BRCCE$), which we
encountered above.  When interpreted over the class of all regular
closed algebras of topological spaces, the satisfiability problem for
this language is still NP-complete~\cite{Wolter&Z00ecai}.  By
extending $\cBC$ with the predicates $c$ and $c^{\le k}$ (for $k
\geq1$), however, we obtain the more expressive languages $\cBCc$ and
$\cBCcc$, whose satisfiability problems for the same class of
interpretations again jump from \NP{} to \ExpTime{} and \NExpTime{},
respectively.

Our final base language has its roots in the seminal paper by McKinsey
and Tarski~\cite{McKinsey&Tarski44}.  Following the modal logic
tradition, we call it $\SFU$, that is, Lewis' system $\SF$ extended
with the universal modality. (For more information on the relationship
between spatial and modal logic see
\cite{Benthem&Guram07hb,Gabelaiaetal05} and references therein.)  The
variables of this language may be taken to range over any collection
of subsets of a topological space (not just regular closed sets), and
its primitives include the operations of union, intersection,
complement and topological interior and closure. Since the property of
being regular closed is expressible in $\SFU$, this language may be
regarded as being more expressive than $\cBC$.  When interpreted over
the class of power sets of topological spaces, the satisfiability
problem for $\SFU$ is~\PSpace{}-complete.  By extending $\SFU$ with
the predicates $c$ and $c^{\le k}$ (for $k \geq 1$), however, we
obtain the languages $\conT$ and $\conTc$, whose satisfiability
problems, for the same class of interpretations, once again jump to
\ExpTime{} and \NExpTime{}, respectively.

Thus, the addition of connectedness predicates to topological logics
leads to greater expressive power and higher computational
complexity. However, this increase in complexity is `stable': over the
most general classes of interpretations, the extensions of such
different formalisms as $\cB$ and $\SFU$ with connectedness predicates
are of the same complexity. Another interesting result is that, by
restricting these languages to formulas with just one connectedness
constraint of the form $c(r)$, we obtain logics that are still
in \PSpace, while two such constraints lead to \ExpTime-hardness. In
fact, if the connectedness predicate is applied only to regions that
are known to be pairwise disjoint, then it does not matter how many
times this predicate occurs in the formula: satisfiability is still
in \PSpace.

The rest of this paper is organized as
follows. Section~\ref{sec:without} presents the syntax and semantics
of our base languages (together with some of their variants), and
Section~\ref{sec:with} extends these languages with connectedness
predicates. Section~\ref{sec:aleksandrov} introduces the first main
ingredient of our proofs---a representation theorem allowing us to
work with Aleksandrov topological spaces rather than arbitrary
ones. Such spaces can be represented by Kripke frames with
quasi-ordered accessibility relations, and topological connectedness
in these frames corresponds to graph-theoretic connectedness in the
(non-directed) graphs induced by the accessibility relation.  Based on
this observation, we can prove the upper bounds in a more-or-less
standard way using known techniques from modal and description logic;
by contrast, the lower bounds are more involved and
unexpected. Section~\ref{sec:complexity} presents the proofs of these
complexity results. Section~\ref{sec:Euclidean} considers the
computational behaviour of our topological logics when interpreted
over various \emph{Euclidean} spaces $\R^n$, and lists some open
problems.

\section{Background: topological logics without connectedness}
\label{sec:without}

A \emph{topological space} is a pair $(T,\mathcal{O})$, where $T$ is a
set and $\mathcal{O}$ a collection of subsets of $T$ containing
$\emptyset$ and $T$, and closed under arbitrary unions and finite
intersections. The elements of $\mathcal{O}$ are referred to as {\em
open sets}; their complements are {\em closed} sets.  If $\mathcal{O}$
is clear from context, we refer to the topological space
$(T,\mathcal{O})$ simply as $T$. Given any $X \subseteq T$, the
\emph{interior} of $X$, denoted $\ti{X}$, is the largest open set
included in $X$, and the \emph{closure} of $X$, denoted $\tc{X}$, is
the smallest closed set including $X$. These sets always exist. It is
convenient, where the space $T$ is clear from context, to denote $T$
by $\one$, the empty set by $\zero$, and, for any $X \subseteq T$, the
complement $T \setminus X$ by $\compl{X}$. Evidently, $\tc{X} =
\compl{(\ti{(\compl{X})})}$.  If $X \subseteq T$, the {\em subspace
topology} on $X$ is the collection of sets $\mathcal{O}_X = \{O \cap X
\mid O \in \mathcal{O}\}$. It is readily checked that
$(X,\mathcal{O}_X)$ is a topological space.

\begin{figure}[t]
\setlength{\unitlength}{.05cm}
\begin{picture}(215,40)
\put(0,10){\includegraphics{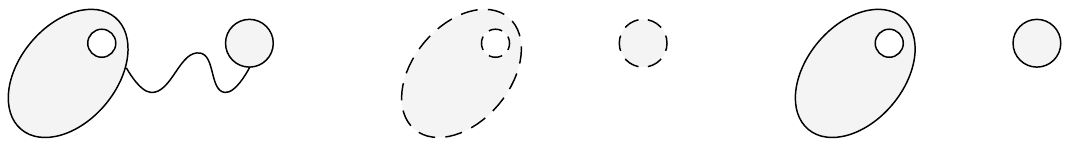}}
\put(30,0){\small (\textit{a})}%
\put(110,0){\small (\textit{b})}%
\put(190,0){\small (\textit{c})}%
\put(-5,37){$X$}%
\put(75,37){$\ti{X}$}%
\put(155,37){$\tc{{\ti{X}}}$}%
\end{picture}
\caption[]{(\textit{a}) A closed set  in the plane, (\textit{b}) its interior and (\textit{c}) the closure of its interior.\\ Note that $\ti{X} \subseteq \tc{{\ti{X}}} \subseteq X$.}
\label{fig:1:intuitionrc}
\end{figure}
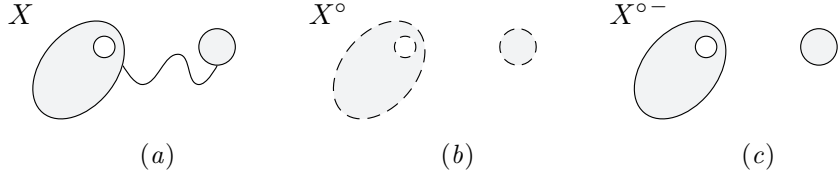

Let $T$ be a topological space. A subset of $T$ is called
\emph{regular closed} if it is the closure of an open set.  We denote
the set of regular closed subsets of $T$ by $\RegC(T)$. It is a
standard result (for example,~\cite{Koppelberg89}, pp.~25--27) that,
for any topological space $T$, the collection of sets $\RegC(T)$ forms
a Boolean algebra, with the top and bottom elements $\one = T$ and
$\zero = \emptyset$, respectively, Boolean operations given by
\begin{equation}\label{eq:booleanFunctions}
X+Y = X \cup Y, \qquad\qquad
X \cdot Y = \tc{{\ti{(X \cap Y)}}}\hspace*{-1em},  \qquad\qquad
-X = \tc{(\compl{X})}\hspace*{-0.5em},
\end{equation}
and Boolean order $\leq$ coinciding with the subset relation. In the
context of the Euclidean plane $\mathbb R^2$, the regular closed sets
are---roughly speaking---those closed sets with no `filaments' or
`isolated points' (Fig.~\ref{fig:1:intuitionrc}).  When dealing
with the Boolean algebra $\RegC(T)$, for some topological space $T$,
we generally write $X+Y$ in preference to $X \cup Y$ (though these
are, formally, equivalent); similarly, we generally write $X \leq Y$
in preference to $X \subseteq Y$.
%

We establish a general framework for defining the topological
languages studied in this paper.  Fix a countably infinite set
$\RVar$. We refer to the elements of $\RVar$ as {\em region variables}
(or, more simply: {\em variables}) and denote them by $r$, $s$, etc.\ possibly with sub- or superscripts.  Let $F$ be any set of
function symbols (of fixed arities) and $P$ any set of predicate symbols (of
fixed arities). In practice, the symbols in $F$ and $P$ may be assumed
to have fixed topological interpretations, along the lines indicated
in Section~\ref{sec:intro}. For example, $F$ might contain
function symbols denoting the operations $+$, $\cdot$ and $-$ on
regular closed sets defined in~\eqref{eq:booleanFunctions};
likewise, $P$ might contain predicates denoting the $\RCCE$ relations.
The $\cL(F,P)$-\emph{terms}, $\tau$, are given by the rule:
\begin{equation*}
\tau \quad ::= \quad r \quad \mid \quad \ f^n (\tau_1, \dots, \tau_n),
\end{equation*}
where $r$ is a variable in $\RVar$, $f^n$ a function symbol of arity
$n$ in $F$, and the $\tau_i$ $\cL(F,P)$-terms. The $\cL(F,P)$-\emph{formulas},
$\varphi$, are given by the rule:
\begin{equation*}
\varphi \quad ::= \quad p^n(\tau_1, \dots, \tau_n) \quad \mid \quad \varphi_1 \land \varphi_2 \quad \mid \quad \varphi_1 \lor \varphi_2
\quad \mid \quad \varphi_1 \to \varphi_2 \quad \mid \quad \neg\varphi ,
\end{equation*}
where $p^n$ is a predicate symbol of arity $n$ in $P$, and
$\varphi_1$, $\varphi_2$ are $\cL(F,P)$-formulas.  The {\em
  topological language} $\cL(F,P)$ is the set of
$\cL(F,P)$-formulas. We shall write $\cL$ in place of $\cL(F,P)$ if
$F$ and $P$ are understood. As usual, formulas of the form
$p^n(\tau_1, \dots, \tau_n)$ and $\neg p^n(\tau_1, \dots, \tau_n)$ are
called \emph{literals}. Notice that, for the purposes of this paper,
topological languages involve no quantifiers.

We now turn to the semantics of these languages.  A \emph{topological
  frame} is a pair of the form $(T, \Res)$, where $T$ is a topological
space, and $\Res \subseteq 2^T$.  We refer to the elements of $\Res$
as \emph{regions}; there is no requirement for $\Res$ to be closed
under any operations.  A \emph{topological model} on $(T,\Res)$ is a
triple $\mathfrak M =(T,\Res,\cdot^{\mathfrak M})$, where
$\cdot^{\mathfrak M}$ is a map from $\RVar$ to $\Res$, referred to as
a \emph{valuation}. Assuming the function symbols in $F$ and
predicates in $P$ to have standard interpretations, any topological
model $\fM$ determines the truth-value of an $\cL$-formula in the
obvious way. We write $\fM \models \varphi$ if the formula $\varphi$
is true in $\mathfrak M$.

Let $\cK$ be a class of topological frames and $\varphi$ a formula of
a topological language $\cL$. We say that $\varphi$ is \emph{satisfiable}
over $\cK$ if $\fM \models \varphi$ for some
topological frame $(T,\Res)$ in $\cK$ and some topological model $\fM$
on $(T,\Res)$; dually, $\varphi$ is \emph{valid} over $\cK$ if $\fM
\models \varphi$ for every topological frame
$(T,\Res)$ in $\cK$ and every topological model $\fM$ on
$(T,\Res)$. As usual, $\varphi$ is valid if and only if $\neg \varphi$
is not satisfiable.  The {\em satisfiability problem} for
$\cL$-formulas over topological frames in $\cK$ is the decision
problem for the set
\begin{equation*}
\Sat (\cL,\cK) ~=~ \bigl\{ \varphi \in \mathcal{L} \mid \text{$\varphi$ is satisfiable over $\cK$}\,\bigr\},
\end{equation*}
that is: given an $\cL$-formula $\varphi$, decide whether it is
satisfiable in a topological model based on a topological frame from
$\cK$.  A \emph{topological logic} is a pair $(\cL, \cK)$, where $\cL$
is a topological language (whose primitives are taken to have fixed
topological interpretations) and $\cK$ a class of topological frames.

In the sequel, except where indicated to the contrary, we generally
speak of \emph{frames}, \emph{models}, \emph{logics} etc., taking the qualifier
`topological' to be implicit.

The primary motivation for introducing the notion of a frame
$(T,\Res)$ is to provide a mechanism for confining attention to those
subsets $\Res$ of the space $T$ which we regard as {\em bona fide}
regions.  For example, it is frequently observed (see,
e.g.,~\cite{Galton00}) that no clear sense can be given to the question
of whether a given physical object occupies a topologically closed,
semi-closed or open region of space.  Consequently, spatial logics in
AI conventionally identify regions differing only with respect to
boundary points. A convenient way to finesse the issue of boundary
points in a topological space $T$ is to restrict attention
to the regular closed sets $\RegC(T)$. For, given any closed subset
$X$ of $T$, there exists a unique $Y \in \RegC(T)$ such that $\ti{X}
\subseteq Y \subseteq X$ (see Fig.~\ref{fig:1:intuitionrc}).
Moreover, these regular closed sets, as
noted above, form a Boolean algebra with $+$, $\cdot$ and $-$
providing reasonable reconstructions of the intuitive operations of
agglomeration, intersection, and complementation, respectively. In this
paper, we shall be principally concerned with the classes of frames
$\All$ and $\Regc$ given by
\begin{align*}
\All  &=  \{(T, 2^T) \mid \mbox{$T$ a topological space} \},\\
\Regc &= \{(T, \RegC(T)) \mid \mbox{$T$ a topological space} \}.
\end{align*}
One word of caution: the Boolean algebras $\RegC(\R^2)$ and
$\RegC(\R^3)$ include many sets which are not at all obviously suited
to model regions occupied by physical objects. For this reason, we
may decide to interpret our languages over topological frames
$(T,\Res)$ where $\Res$ is a {\em sub-algebra} of $\RegC(T)$, a
restriction which turns out to have interesting mathematical
consequences (see, e.g.,~\cite{PH:HSL}).

With these semantic preliminaries behind us, we now survey some of the
most familiar topological logics occurring in the AI literature.  This
survey will occupy the remainder of this section.  Remember: our aim
in the sequel is to investigate the effect of increasing the
expressive resources available to these logics by adding predicates
expressing connectedness and related notions.

\subsection*{The logic $\RCCE$:}
We begin with a formal account of the topological language $\RCCE$,
which we encountered in Section~\ref{sec:intro}.  As a preliminary, we
define the following six binary relations on $\RegC(T)$, where $T$ is
any topological space:
\begin{equation}
\begin{array}{rcl}
\mathsf{DC}(X, Y)  &\mbox{ iff }  &X \cap Y = \emptyset,\\
\mathsf{EC}(X,Y)  &\mbox{ iff }  &X \cap Y \neq \emptyset \text{\ \  but }
        \ti{X} \cap \ti{Y} = \emptyset,\\
\mathsf{PO}(X,Y) &\mbox{ iff } &\ti{X} \cap \ti{Y}, \ \ti{X} \setminus Y \mbox{ and } \ti{Y} \setminus X \text{ are all non-empty,}\\
\mathsf{EQ}(X,Y)  &\mbox{ iff }  &X = Y,\\
\mathsf{TPP}(X,Y)  &\mbox{ iff }  &X \subseteq Y  \text{ but }
       X \not \subseteq \ti{Y} \text{ and } Y \not\subseteq X,\\
\mathsf{NTPP}(X,Y)  &\mbox{ iff }  &X \subseteq \ti{Y}  \text{ but } Y \not\subseteq X.
\end{array}
\label{eq:rcc8}
\end{equation}
All of these relations except $\mathsf{TPP}$ and $\mathsf{NTPP}$ are
symmetric.  Counting the converses of $\mathsf{TPP}$ and
$\mathsf{NTPP}$, we thus obtain eight binary relations altogether:
these eight relations are easily seen to be jointly exhaustive and
mutually exclusive over non-empty elements of $\RegC(T)$.  In
Fig.~\ref{fig:ccircles}, we illustrated them in the special
case where the relata are closed disc-homeomorphs in the plane. We
remark in passing that, when restricted to closed disc-homeomorphs in
the plane, these relations are actually the atoms of a finite relation
algebra~\cite{D+O00,Li-Ying03b}.

We now define the language $\RCCE$ by
\begin{equation*}
\RCCE ~=~ \cL(\emptyset,
\{\mathsf{DC}, \mathsf{EC}, \mathsf{PO}, \mathsf{EQ}, \mathsf{TPP}, \mathsf{NTPP}\}),
\end{equation*}
with the symbols $\mathsf{DC}$, $\mathsf{EC}$, $\mathsf{PO}$,
$\mathsf{EQ}$, $\mathsf{TPP}$, $\mathsf{NTPP}$ taken to be binary
predicates. There
are no function symbols; so $\RCCE$-terms are simply variables.

We always interpret this language over (some sub-class of)
$\Regc$---that is: variables are always taken to range over (certain)
regular closed sets of (certain) topological spaces.  The semantics
for $\RCCE$ may then be given by specifying the interpretations of the
predicates in obvious way, thus:
\begin{equation*}
\begin{array}{lcl}
\fM \models \mathsf{DC}(r_1,r_2)  &\mbox{ iff }
    &\mathsf{DC}(r_1^\fM, r_2^\fM), \\
\fM \models \mathsf{EC}(r_1,r_2)  &\mbox{ iff }
    &\mathsf{EC}(r_1^\fM, r_2^\fM),\\
{\it etc}. &&
\end{array}
\end{equation*}
Note the overloading of the symbols $\mathsf{DC}$,
$\mathsf{EC}$, etc.~here: on the left-hand sides of these
equations, they are predicates of $\RCCE$; on the right-hand side,
they denote the relations on $\RegC(T)$ defined in~\eqref{eq:rcc8}.
Since these predicates will always be used with their standard
meanings, no confusion need arise. In the literature, the language $\RCCE$ is
sometimes subject to the additional restriction that variables range
only over \emph{non-empty} regular closed subsets of the space in
question. We do not impose this requirement, remarking, however, that
non-emptiness is anyway expressible in $\RCCE$ (on our interpretation)
by adding conjuncts of the form $\neg \mathsf{DC}(r,r)$.

%

It is known that
$\Sat(\RCCE,\Regc)$ is \NP-complete~\cite{Renz98}. (Proofs of
all the complexity results mentioned in this section are discussed in
Section~\ref{sec:aleksandrov}.)


\subsection*{The logic $\BRCCE$:}
Since, as we have observed, the regular closed subsets of a
topological space form a Boolean algebra, it makes sense to augment
$\RCCE$ with function symbols denoting the obvious operations and
constants of this Boolean algebra. The language $\BRCCE$ (for {\em
Boolean} $\RCCE$,~\cite{Wolter&Z00ecai}) is defined by:
\begin{equation*}
\BRCCE ~=~ \cL(\{+, \cdot, -, \zero, \one \},
\{\mathsf{DC}, \mathsf{EC}, \mathsf{PO}, \mathsf{EQ}, \mathsf{TPP}, \mathsf{NTPP}\}),
\end{equation*}
where the function symbols $+$ and $\cdot$ are binary, $-$ is unary,
and $\zero$ and $\one$ are nullary (i.e.~individual constants).  Again, we
confine attention to the class of frames $\Regc$, with the
function symbols interpreted as the obvious operations on regular
closed sets. Formally:
\begin{equation*}
\begin{array}{cllll}
(\tau_1 + \tau_2)^\fM &= &\tau_1^\fM + \tau_2^\fM, \hspace{1cm} &
(-\tau)^\fM ~=~ -(\tau^\fM), \hspace{1.5cm} &
\zero^\fM ~=~ \zero,\\
(\tau_1 \cdot \tau_2)^\fM &= &\tau_1^\fM \cdot \tau_2^\fM, &
& \one^\fM ~=~ \one.
\end{array}
\end{equation*}
Note the overloading of the symbols: on the left-hand side of these
equations, they are function symbols of $\BRCCE$; on the right-hand
side, they denote the corresponding Boolean algebra operations in
$\RegC(T)$ as defined in~\eqref{eq:booleanFunctions}. The predicates
are interpreted in the same way as for $\RCCE$.


Despite its increased expressive power, $\BRCCE$ is in the same
complexity class as $\RCCE$---at least when interpreted over arbitrary
topological spaces. That is: the problem
$\Sat(\BRCCE,\Regc)$ is
\NP-complete~\cite{Wolter&Z00ecai}. (However, as we shall see below,
this situation changes even under very mild restrictions on the class
of frames.)


\subsection*{The logic $\cBC$:}
We can re-formulate $\BRCCE$ more elegantly using the binary
predicates $=$ (equality) and $C$ (\emph{contact}).  The language
$\cBC$ is defined by:
\begin{equation*}
\cBC = \cL(\{+, \cdot, -, \zero, \one \}, \{C, =\}).
\end{equation*}
As with $\RCCE$ and $\BRCCE$, we confine our attention to the class of
frames $\Regc$. The equality predicate $=$ denotes identity (as
usual), and the contact predicate $C$ is interpreted as follows:
\begin{equation*}
\fM \models C(\tau_1, \tau_2) \quad \mbox{ iff } \quad
\tau_1^\fM \cap \tau_2^\fM \neq \emptyset.
\end{equation*}
That is: two regions are taken to be in contact just in case they
intersect.  Notice that $C(\tau_1,\tau_2)$ is not equivalent to the
condition $\tau_1 \cdot \tau_2 \neq \zero$ (a shorthand for
$\neg(\tau_1 \cdot \tau_2 = \zero$)), which states that the
\emph{interiors} of $\tau_1$ and $\tau_2$ intersect.  In the context
of any logic involving the function symbols $+$, $\cdot$, $-$,
$\zero$, $\one$ and the equality predicate, we standardly write
$\tau_1 \leq \tau_2$ as an abbreviation for $\tau_1 \cdot
(-\tau_2) = \zero$;  $\neg(\tau_1 \leq \tau_2)$ is abbreviated by $\tau_1 \nleq \tau_2$.

Evidently, $C(\tau_1,\tau_2)$ is equivalent to $\neg
\mathsf{DC}(\tau_1,\tau_2)$, and $=$ is just another symbol for $\mathsf{EQ}$; hence, $\BRCCE$ is at least as
expressive as $\cBC$. Conversely, it is easy to verify that the four
remaining $\RCCE$-relations can easily be equivalently expressed in~$\cBC$, as follows:
\begin{equation*}
\begin{array}{rll}
{\sf EC}(\tau_1,\tau_2) &\leftrightarrow & (\tau_1 \cdot \tau_2 = \zero) \ \wedge \ C(\tau_1,\tau_2),\\
{\sf PO}(\tau_1,\tau_2) &\leftrightarrow & (\tau_1 \cdot \tau_2 \neq \zero) \ \wedge \ (\tau_1 \nleq \tau_2) \ \land \ (\tau_2 \nleq \tau_1),\\
{\sf TPP}(\tau_1,\tau_2) &\leftrightarrow &
(\tau_1 \leq \tau_2) \ \wedge \ C(\tau_1, -\tau_2) \  \land \ (\tau_2 \nleq \tau_1),\\
{\sf NTPP}(\tau_1,\tau_2) &\leftrightarrow & \neg C(\tau_1,-\tau_2) \ \land \ (\tau_2 \nleq \tau_1).
\end{array}%
\end{equation*}
Hence, we may regard the languages $\cBC$ and $\BRCCE$ as equivalent.

The predicate $C$ has an interesting history. Originally introduced by
Whitehead~\cite{Whitehead29} under the name `extensive connection,' it
provided the inspiration for many of the early approaches to
topological logics in AI.  To avoid confusion with the familiar
topological property of \emph{connectedness}, Whitehead's relation is
now generally referred to as \emph{contact}. Investigation of the
contact-structure of the regular closed algebras of topological spaces
gave rise to the study of so-called \emph{Boolean connection algebras}
(BCAs). The relationship between BCAs and the topological spaces that
generate them is now well-understood~\cite{Roeper97,DW05,DV1,DV2}. Our
logic $\cBC$ (that is: $\BRCCE$) is, in essence, the quantifier-free
fragment of the first-order theory of BCAs. For an up-to-date account
of this work, see~\cite{Balbianietal07}.


\subsection*{The logic $\cBC^m$: }
As formulas in $\cBC$ are built from $\cBC$-terms using the binary
predicates $\tau_1=\tau_2$ and $C(\tau_1,\tau_2)$, they are not
capable of expressing, for example, the predicate $\mathsf{EC}_3(\tau_1,\tau_2,\tau_3)$ stating that \emph{three} regions $\tau_1,\tau_2,\tau_3$ are externally connected and have some common border. One
way to extend the expressive power of $\cBC$ is to generalize the
contact predicate and consider its extension $\cBC^m$ with arbitrary
$k$-ary contact relations $C^k(\tau_1,\dots,\tau_k)$, for $k \geq 2$.
The language $\cBC^m$ is defined by:
\begin{equation*}
\cBC^m ~=~ \cL(\{+, \cdot, -, \zero, \one \}, \{C^k \mid k \geq 2 \} \cup \{=\}).
\end{equation*}
Again, we confine attention to the class of frames $\Regc$. The
predicates $C^k$ are interpreted as follows:
\begin{equation*}
\fM \models C^k(\tau_1, \dots, \tau_k) \quad \mbox{ iff } \quad
\tau_1^\fM \cap \cdots \cap \tau_k^\fM \neq \emptyset.
\end{equation*}
The ternary predicate $\mathsf{EC}_3(\tau_1,\tau_2,\tau_3)$ above can now be expressed in a straightforward way:
$$
\mathsf{EC}_3(\tau_1,\tau_2,\tau_3) ~=~ C^3(\tau_1,\tau_2,\tau_3) \land \mathsf{EC}(\tau_1,\tau_2) \land \mathsf{EC}(\tau_1,\tau_3) \land \mathsf{EC}(\tau_2,\tau_3),
$$
which is not expressible in $\mathcal{C}$.
Obviously, the predicates $C$ and $C^2$ have identical semantics;
thus, $\cBC$ is a sub-language of $\cBCm$. Again, the increased
expressive power makes no difference to the complexity class:
$\Sat(\cBC^m,\Regc)$ is still \NP-complete \cite{Gabelaiaetal05}.


\subsection*{The logic $\cB$:}
We mention at this point a sub-language of $\cBC$ so inexpressive that
no distinctively topological facts can be expressed in it, but which
will nevertheless prove significant in the sequel.  The language
$\cB$, again interpreted over sub-classes of $\Regc$, is defined by:
\begin{equation*}
\cB ~=~ \cL(\{+, \cdot, -, \zero, \one \}, \{=\}).
\end{equation*}
Thus, $\cB$ is the language of the variety of Boolean algebras.  In the present context,
it can be seen as capturing the essential content of \emph{mereology}---the
logic of `part-whole' relations.  (For a discussion
of the relationship between mereology and Boolean algebra, see
\cite{Tarski35,Grzegorczyk55}.) Trivially, $\Sat(\cB,\Regc)$ is
\NP-complete.


\subsection*{The logic $\SFU$: }
Returning to matters topological, we come to the most expressive
topological logic to have been considered in the literature.
The language $\SFU$ is defined by:
\begin{equation*}
\SFU ~=~
   \cL(\{\cup, \cap, \overline{\, \cdot\, }, \ti{\cdot}, \tc{\cdot}, \zero, \one \},\{=\}),
\end{equation*}
and we write $\tau_1 \subseteq \tau_2$ as an abbreviation for $\tau_1
\cap \compl{\tau}_2 = \zero$.  We interpret the terms of this language
as follows:
\begin{equation}
\begin{array}{lrll}
(\tau_1 \cap \tau_2)^\fM = \tau_1^\fM \cap \tau_2^\fM, \hspace{1cm} &
(\compl{\tau})^\fM &= \compl{\tau^\fM} = T \setminus \tau^\fM,
  \hspace{1cm} & \zero^\fM = \zero = \emptyset,\\
(\tau_1 \cup \tau_2)^\fM = \tau_1^\fM \cup \tau_2^\fM,  &
(\ti{\tau})^\fM &= \ti{(\tau^\fM)}, & \one^\fM = \one = T,\\
& (\tc{\tau})^\fM &= \tc{(\tau^\fM)},
\end{array}
\label{eq:semRC}
\end{equation}
where $\fM$ is a model over some frame $(T, \Res)$. As before, we
have deliberately equivocated between
function symbols in our formal language and the operations they
denote.  Since these operations do not in general preserve the
property of being regular closed, it is unnatural to confine attention
to the frame class $\Regc$. Accordingly, we standardly interpret
$\SFU$ over the class $\All$ of all frames.

The richer term-language of $\SFU$ means that, even though $=$ is the
only predicate, we are still able to formulate distinctively
topological (not just mereological) statements. Consider, for example,
the formula
\begin{equation*}
\bigl(\ti{r}_1 \cap \tc{r}_2 \neq \zero\bigr) \ \rightarrow \ \bigl(\ti{r}_1 \cap r_2 \neq \zero\bigr).
\end{equation*}
This formula states that, if an open set $\ti{r}_1$ intersects the
closure of a set $r_2$, then it also intersects $r_2$. Thus, it is
valid over the class of frames $\All$.

The language $\SFU$ may be regarded as the richest of all the
languages considered here, in the following sense.  Given a
$\cBC^m$-term $\tau$, we define inductively the $\SFU$-term
$\tau^\dag$ as follows:
\begin{gather*}
\one^\dag = \one,\qquad \zero^\dag = \zero,\qquad r^\dag ~=~ \tc{{\ti{r}}} \ \ (r \text{ a variable}),\\
(-\tau_1)^\dag ~=~ \tc{\bigl(\overline{\tau_1^\dag}\bigr)},\qquad (\tau_1\cdot\tau_2)^\dag ~=~ \tc{{\ti{{(\tau_1^\dag\cap\tau_2^\dag)}}}}, \qquad (\tau_1+\tau_2)^\dag ~=~ \tau_1^\dag\cup\tau_2^\dag.
\end{gather*}
If $\varphi$ is a $\cBCm$-formula, let $\varphi^\dag$ be the
$\SFU$-formula obtained by replacing every occurrence of $\tau_1 =
\tau_2$ in $\varphi$ with $\tau_1^\dag = \tau_2^\dag$ and every
occurrence of $C^k(\tau_1,\dots,\tau_k)$ in $\varphi$ with $\tau_1^\dag
\cap \dots \cap \tau_k^\dag \ne \zero$. For any topological space $T$,
the regular closed subsets of $T$ are exactly the sets of the form
$\tc{{\ti{X}}}$, where $X \subseteq T$.  Hence, as the variable $r$
ranges over $2^T$, the $\SFU$-term $r^\dag = \tc{(\ti{r})}$ ranges
over exactly the regular closed subsets of $T$.  Using this
observation, it is readily checked that $\varphi$ is satisfiable
over a frame $(T,\RegC(T))$ if and only if $\varphi^\dag$ is satisfiable over
the frame $(T,2^T)$. Thus, we may informally regard any logic $(\cL,
\Regc)$, where $\cL$ is a fragment of $\cBC^m$, as contained within
the logic $(\SFU, \All)$.\footnote{That \RCCE{} is a simple fragment of $\SFU$ was first
observed by Bennett~\cite{Bennett94}; see also \cite{RenzN97,Nutt99}
(in fact, \RCCE{} and \BRCCE{} can be embedded into the modal logic $\mathcal{S}5$~\cite{Wolter&Z02millenium}).}

Furthermore, the logic $(\SFU, \All)$ has essentially the same
expressive power as the modal logic \textbf{S4} (under
McKinsey and Tarski's~\cite{McKinsey&Tarski44} topological
interpretation) extended with the universal and existential modalities
$\forall$ and $\exists$ of~\cite{Goranko&Passy92}. More precisely,
define ${\bf S4}_u$ to be the set of terms formed using the variables
in $\cR$ together with the function symbols
\begin{equation*}
   \cup, \ \ \cap, \ \ \overline{\, \cdot\, }, \ \ \ti{\cdot}, \ \ \tc{\cdot}, \ \ \forall, \ \ \exists, \ \ \zero, \ \ \one.
\end{equation*}
Here, $\exists$ and $\forall$ are unary, with the remaining symbols
having their usual arities.  Given any interpretation $\fM$, we define
$\tau^\fM$ for any ${\bf S4}_u$-term $\tau$ using~\eqref{eq:semRC} together with:
$$
(\exists \tau)^\mathfrak M ~=~
\begin{cases}
T & \text{if $\tau^\mathfrak M \ne \emptyset$},\\
\emptyset & \text{if $\tau^\mathfrak M = \emptyset$};
\end{cases} \qquad
(\forall \tau)^\mathfrak M ~=~
\begin{cases}
T & \text{if $\tau^\mathfrak M = T$},\\
\emptyset & \text{if $\tau^\mathfrak M \ne T$}.
\end{cases}
$$
Thus, $\exists$ is interpreted as the discriminator function, and $\forall$
as its dual. We say that an ${\bf S4}_u$-term $\tau$ is \emph{valid}
if $\tau^\mathfrak M = T$ for any model $\mathfrak M$ over any
topological frame $(T, \Res)$. By replacing each equality $\tau_1 =
\tau_2$ with the term $\forall \big( (\tau_1 \cap \tau_2) \cup
(\overline{\tau_1} \cap \overline{\tau_2}) \big)$ and the Boolean
connectives with the corresponding function symbols, we obtain a
validity-preserving embedding of $\SFU$ into ${\bf S4}_u$-terms. On
the other hand, it is well known (see,
e.g.,~\cite{Aiello&Benthem02,Hughes&Cresswell96}) that every ${\bf
  S4}_u$-term can be equivalently transformed to a term without
occurrences of $\forall$ and $\exists$ in the scope of $\ti{}$,
$\tc{}$, $\forall$ and $\exists$. Any such term can easily be
rewritten as an equivalent $\SFU$-formula by replacing $\forall\tau$
and $\exists \tau$ with $\tau = \one$ and $\tau \ne \zero$,
respectively, and by replacing Boolean function symbols with the
corresponding Boolean connectives. (Note, however, that this
transformation in general results in an exponential increase in size.)
As the validity and satisfiability problems for both ${\bf S4}$ and
${\bf S4}_u$ are known to
be \PSpace{}-complete~\cite{Ladner77,Nutt99,Arecesetal00}, it follows
that the problem $\Sat(\SFU,\All)$ is
\PSpace{}-complete as well.

In this section, we introduced the languages $\RCCE$, $\cB$,
$\BRCCE$ (=$\cBC$), $\cBCm$ and $\SFU$.  The variables
of these languages range over certain distinguished subsets of some
topological space, and their non-logical symbols denote various fixed
primitive topological relations and operations. The topological space
in question and its collection of distinguished subsets together form
a (topological) frame; and the pair of a language and a class of
frames is a (topological) logic. In particular, we interpreted
$\RCCE$, $\cB$, $\BRCCE$ (=$\cBC$) and $\cBCm$ over the frame-class
$\Regc$, and the language $\SFU$ over the frame-class $\All$. We
explained how all of these logics can be seen as fragments of
$(\SFU,\All)$, which has essentially the expressive power of the modal
logic ${\bf S4}_u$.  We observed that the complexity of the
satisfiability problems for all of these logics is known.  However,
none of the above languages can express the property of being a
connected region. Our question is: what happens to the
complexity of satisfiability when that facility is provided?


\section{Topological logics with connectedness}
\label{sec:with}

A topological space $T$ is \emph{connected} just in case it is not the
union of two non-empty, disjoint, open sets. Note that this definition can be expressed by the following $\SFU$-formula (see~\cite{Shehtman99}):
\begin{equation*}
\big(\ti{r} \cup \ti{(\overline{r})} = \one\big) \to \big ( (r=\one ) \lor (r = \zero ) \big).
\end{equation*}
A subset $X \subseteq T$
is \emph{connected in $T$} if the topological space $X$ (with the
subspace topology) is connected. If $X \subseteq T$, a maximal
connected subset of $X$ is called a (\emph{connected})
\emph{component} of $X$. Every set is the disjoint union of its
components (of which there is always at least one); a set is connected
just in case it has exactly one component.  If $T$ and $T'$ are
topological spaces, a function $f\colon T \rightarrow T'$ is \emph{continuous}
if the inverse image under $f$ of every open subset of
$T'$ is open in $T$ (equivalently:
if
the inverse image of every closed subset of $T'$ is closed in
$T$).  The image of a connected set under a continuous function $f$ is
always connected;  in fact, if $X \subseteq T$ has $k\geq 1$ components then $f(X)$ has at most $k$ components.

The simplest way to introduce connectedness into topological logics
is to restrict attention to frames over connected topological
spaces.  For example, consider the classes of frames given by
\begin{align*}
\Con  &=  \{(T, 2^T) \mid \mbox{$T$ a connected topological space} \}, \\
\ConR  &=  \{(T, \RegC(T)) \mid \mbox{$T$ a connected topological space} \}.
\end{align*}
Thus, it makes sense to consider the problems $\Sat(\cL,\ConR)$ for
$\cL$ any of $\RCCE$, $\cB$, $\cBC$ or $\cBC^m$, as well as the problem
$\Sat(\SFU,\Con)$.  An alternative, and more flexible, approach,
however, is to expand the languages in question. Let $c$ be a unary
predicate.  If $\cL$ is one of the topological languages introduced in
Section~\ref{sec:without}, denote by $\cLc$ ($\cL$ {\em with
connectedness}) the result of augmenting the topological primitives of
$\cL$ by $c$.  Formally, if $\cL = \cL(F,P)$,
\begin{equation*}
\cLc ~=~ \cL(F, P \cup \{c\}).
\end{equation*}
The predicate $c$ is given the expected fixed interpretation as follows:
\begin{equation*}
\fM \models c(\tau) \quad \mbox{ iff } \quad \tau^\fM \mbox{ is connected}.
\end{equation*}
Thus, from the languages $\RCCE$, $\cB$, $\cBC$,  $\cBC^m$ and $\SFU$, we obtain
$\RCCEc$, $\cBc$, $\cBCc$,  $\cBC^m\!c$, $\conT$.

Consider, for example, the language $\cBc$, which includes the formula
\begin{equation}\label{eq:eg5}
\big ( c(r_1) \ \wedge \ c(r_2) \ \wedge \ (r_1 \cdot r_2 \ne \zero) \big ) \ \rightarrow \ c(r_1+r_2).
\end{equation}
It is a well-known fact that, if $T$ is a topological space, any
two
connected subsets of $T$ with non-empty intersection have a connected
union; further, if $X$ is connected, and $X \subseteq Y \subseteq
\tc{X}$, then $Y$ is also connected. But if $r$ and $s$ are regular closed
subsets of $T$, then $\ti{r} \cup \ti{s} \subseteq r+s
\subseteq \tc{(\ti{r} \cup \ti{s})}$.  It follows that~\eqref{eq:eg5}
is valid over $\Regc$.  At the other end of the expressive spectrum,
consider the language $\conT$, which includes the formula
\begin{equation}\label{eq:eg0}
\big ( c(r_1) \ \wedge \ ( r_1 \subseteq r_2 ) \ \wedge \ ( r_2 \subseteq \tc{r_1} ) \big ) \ \rightarrow \ c(r_2).
\end{equation}
Using one of the facts we have just alluded to, it follows that~\eqref{eq:eg0} is
valid over $\All$.

The predicate $c$ can be generalized in the following way. Let
$c^{\leq k}$ be a unary predicate, where $k \geq 1$ is represented in
binary. If $\cL$ is one of the languages introduced in
Section~\ref{sec:without}, we denote by $\cLcc$ ($\cL$ \emph{with
  component counting}) the result of augmenting the topological
primitives of $\cL$ by all of the predicates $c^{\leq k}$ ($k \geq
1$).  Formally, if $\cL = \cL(F,P)$,
\begin{equation*}
\cLcc ~=~ \cL(F, P \cup \{c^{\leq k} \mid k \geq 1 \}).
\end{equation*}
The predicates $c^{\leq k}$ are given fixed interpretations as
follows, where $\fM$ is a model over some frame $(T, \Res)$:
\begin{eqnarray*}
   \fM \models c^{\leq k}(\tau)\quad \mbox{ iff \quad  $\tau^\fM$ has at most $k$
   components in $T$}.
\end{eqnarray*}
Thus, from the languages $\RCCE$, $\cB$, $\cBC$, $\cBC^m$ and $\SFU$, we obtain
$\RCCEcc$, $\cBcc$, $\cBCcc$, $\cBCmcc$, $\conTc$. We write $\neg c^{\leq k}(\tau)$
as $c^{\geq k+1}(\tau)$ and abbreviate $c^{\leq 1}(\tau)$ by
$c(\tau)$. Thus, we may regard $\cLc$ as a sub-language of $\cLcc$.
To illustrate, consider the $\cBcc$-formula
\begin{equation}\label{eq:eg6}
\big ( c^{\leq k}(r_1) \ \wedge \ c^{\leq l}(r_2) \ \wedge \ (r_1 \cdot r_2 \ne \zero) \big ) \ \rightarrow \
      c^{\leq l+k-1}(r_1+r_2).
\end{equation}
Using the same argument as for~\eqref{eq:eg5}, this formula is
easily shown to be valid over $\Regc$.

For rich topological languages, such as $\conT$, the predicates
$c^{\leq k}$ give us---in some sense---no expressive power that $c$
does not already give us.  Let $\tau$ be an $\conT$-term and
$r_1,\dots,r_k$ variables not occurring in $\tau$. Consider the
$\conT$-formulas
\begin{gather}\label{eq:comp:leq}
\bigl(\tau = \bigcup_{1 \leq i \leq k} r_i\bigr)  \ \  \wedge \ \
\bigwedge_{1 \leq i \leq k}c(r_i),\\
\label{eq:comp:geq} \bigl(\tau = \hspace*{-0.3em}\bigcup_{1 \leq i
\leq k+1} \hspace*{-0.3em} r_i\bigr) \ \  \wedge  \ \ \bigwedge_{1
\leq i \leq k+1} \hspace*{-0.5em} \bigl(r_i \ne \zero\bigr)
\ \  \wedge  \bigwedge_{1 \leq i < j \leq k+1}\hspace*{-0.6em} \bigl(\tau \cap \tc{r_i} \cap \tc{r_j} = \zero\bigr) 
\end{gather}
together with some model $\fM$ over a topological frame $(T,
\Res)$.  Let us assume that $(T, \Res)$ has the property that, if $r
\in \Res$, and $s$ is a component of $r$, then $s \in \Res$---a very
natural requirement for topological frames.  If~\eqref{eq:comp:leq} is
true in $\fM$, then $\tau^\fM$ is
seen to have at most $k$
components.  Conversely, if $\tau^\fM$ has at most $k$ components,
then, by modifying the regions assigned to $r_1, \ldots, r_k$ if
necessary, we easily obtain a model $\fM'$
satisfying~\eqref{eq:comp:leq}. It follows that, if $\varphi$ is an
$\conTc$-formula, then any instance of $c^{\leq k}(\tau)$ having
positive polarity may be equisatisfiably replaced
by~\eqref{eq:comp:leq} (with fresh variables $r_1, \ldots, r_k$).
Similarly, any instance of $c^{\geq k}(\tau)$ having positive polarity
may be equisatisfiably likewise replaced by~\eqref{eq:comp:geq}.
However, while the number of symbols in the predicate $c^{\leq k}$ is
proportional to $\log k$, the number of symbols in~\eqref{eq:comp:leq}
is proportional to $k$. That is: $\conT$-formulas are in general
exponentially longer than the $\conTc$-formulas they replace.  So,
although the component-counting predicates $c^{\leq k}$ can usually be
eliminated in this way, doing so may affect the complexity of the
satisfiability problem.

The main contribution of this paper is to determine the computational
complexity of the satisfiability problems for topological logics based
on the languages $\cLc$ and $\cLcc$, where $\cL$ is any of the
languages introduced in Section~\ref{sec:without}. To date, there are
only two known complexity results for such logics. On the one hand,
according to \cite{PrattHartmann02}, satisfiability of
$\conTc$-formulas over $\All$ is \NExpTime-complete, which gives the
\NExpTime{} upper bound for all of the other logics considered in this
section.  On the other, according to \cite{Wolter&Z00ecai},
satisfiability of $\cBC$-formulas is
\PSpace-complete over $\ConR$.



\section{Aleksandrov spaces.}
\label{sec:aleksandrov}

The close connection between spatial logics and the modal logic $\textbf{S4}_u$ mentioned above suggests that instead of topological semantics one may try to employ Kripke semantics (which gives rise to topological spaces with a very transparent structure) and the corresponding modal logic machinery.

Recall that \emph{Kripke frames} for $\textbf{S4}_u$ are pairs of the
form $(W,R)$, where $W$ is a set, and $R$ is a reflexive
and transitive relation on $W$; such frames are also called \emph{quasi-orders}.
Every quasi-order $(W,R)$ can be
regarded as a topological space by declaring $X \subseteq W$ to be
\emph{open} if and only if $X$ is upward closed with respect to $R$, that is,
if $x \in X$ and $xRy$ then $y \in X$. In other words, for every $X
\subseteq W$,
\begin{equation*}
\ti{X} ~=~ \bigl\{ x \in X \mid \forall y \in W\ (xRy \to y\in X) \bigr\}.
\end{equation*}
The most important property distinguishing topological spaces $T$ induced by quasi-orders is that arbitrary (not only finite) intersections of open sets in $T$ are open. Topological spaces with this property are called \emph{Aleksandrov spaces}. It is also known (see, e.g.,~\cite{Bourbaki66}) that every Aleksandrov space is induced by a
quasi-order.

Another important feature of such topological spaces is that the topological notion of connectedness in $T$ coincides with the graph-theoretic notion of connectedness in the undirected graph induced by $(W,R)$. More precisely, one can easily check that a set $X\subseteq W$ is connected in $T$ if and only if, for any points $x,y\in X$, there is a path $x= x_1, \dots, x_n = y$ such that, for all $i$, $1 \leq i < n$, we have $x_i \in X$ and either $x_iRx_{i+1}$ or $x_{i+1}Rx_i$.

Henceforth, we shall identify an Aleksandrov space with the
quasi-order generating it, alternating freely between topological and
graph-theoretic perspectives. Denote by $\AlekF$ the class of \emph{finite}
Aleksandrov frames.  A topological model
based on an Aleksandrov space will be called an \emph{Aleksandrov
model}.

The next lemma, originating in
\cite{McKinsey&Tarski44} and \cite{Kripke63}, shows that, for many
topological logics, it suffices to work with finite Aleksandrov
spaces. It can be proved by the standard filtration argument (see, e.g.,
\cite{Chagrov&Z97}).
\begin{lemma}\label{prop:KripkeSemantics}
For every finite set $\Theta$ of $\SFU$-terms closed under subterms
and every topological model $\mathfrak{M} = (T,\Res,\cdot^{\mathfrak M})$,
there exist an Aleksandrov model $\mathfrak{A} = (T_A, 2^{T_A},
\cdot^{\mathfrak A})$ and a continuous function $f\colon T \to T_A$ such that $|T_A| \leq 2^{O(|\Theta|)}$ and $\tau^\mathfrak{A} = f(\tau^\mathfrak{M})$, for every $\tau\in \Theta$.
\end{lemma}

This lemma has a number of important consequences. First, it follows
immediately that $\Sat(\SFU,\All) = \Sat(\SFU,\AlekF)$. Using the
translation $\cdot^\dag$ of $\cB$-terms and $\cBCm$-formulas into
$\SFU$ defined in Section~\ref{sec:without}, we obtain
$\Sat(\cBCm,\Regc) = \Sat(\cBCm,\AlekF \cap \Regc)$,
etc. The \PSpace{} upper bound for $\Sat(\SFU,\All)$ follows from
Lemma~\ref{prop:KripkeSemantics} and the fact (well-known in modal
logic) that the model $\mathfrak{A}$ in
Lemma~\ref{prop:KripkeSemantics} can be `unravelled' into a forest of
trees of clusters\hbox to 0pt{,}\footnote{A cluster in a quasi-order
  $(W,R)$ is any set of the form $\{x\in W \mid xRy \text{ and }
  yRx\}$, for some $y\in W$.}\, with the length of branches not
exceeding the maximal size of terms in $\Theta$---the so-called
\emph{tree-model property} of $\SFU$. Finally, the fact that $f$ is
continuous guarantees that the number of components in
$f(\tau^\mathfrak{M})$ does not exceed the number of components in
$\tau^\mathfrak{M}$, which will be used in the case of logics with
connectedness predicates.

\begin{figure}[t]
\setlength{\unitlength}{.07cm}
\begin{center}
\begin{picture}(140,20)
\multiput(0,15)(20,0){7}{\circle{1.5}}
\multiput(10,0)(20,0){5}{\circle{1.5}} \put(120,0){\circle{1.5}}
\multiput(9,1.5)(20,0){2}{\vector(-2,3){8}}
\multiput(69,1.5)(20,0){2}{\vector(-2,3){8}}
\multiput(11,1.5)(20,0){3}{\vector(2,3){8}}
\put(91,1.5){\vector(2,3){8}} \put(48,1){\vector(-2,1){26}}
\put(52,1){\vector(2,1){26}}
\put(120,2){\vector(0,1){11}}
\put(137,14){\text{\small depth 0}} %
\put(137,-1){\text{\small depth 1}}
\end{picture}
\end{center}
\caption{Quasi-saw.}\label{f:qsaw}
\end{figure}
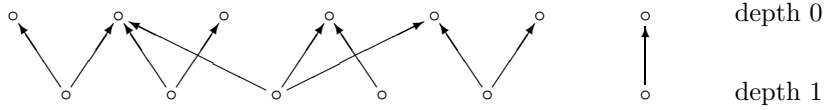

For the topological logics with $\cB$-terms only, say $\cBCmcc$, even
simpler Aleksandrov models are enough.
Call a quasi-order $(W,R)$ a \emph{quasi-saw} if $W=W_0\cup W_1$, for
some disjoint $W_0$ and $W_1$, and $R$ is the reflexive closure of a
subset of $W_1 \times W_0$. In this case we also say that the points
in $W_i$ are \emph{of depth} $i$ in $(W,R)$; see Fig.~\ref{f:qsaw}.
Aleksandrov models over quasi-saws will be
called \emph{quasi-saw models}.

\begin{lemma}\label{broom-lemma}
For every finite Aleksandrov model $\mathfrak{A} =
(T_A, \RegC(T_A), \cdot^\mathfrak{A})$, with $T_A$ induced by a quasi-order $(W,R_A)$, there is a
quasi-saw model $\mathfrak{B} = (T_B, \RegC(T_B),\cdot^\mathfrak{B})$ such that
$T_B$ is induced by $(W,R_B)$ with $R_B\subseteq R_A$ and, for every
$\cB$-term $\tau$, \textup{(i)} $\tau^\mathfrak{B} =
\tau^\mathfrak{A}$, and \textup{(ii)} $\tau$ has the same number of
components in $\mathfrak{A}$ and $\mathfrak{B}$.
\end{lemma}
\proof
Let $W_0$ be the set of maximal points in $(W, R_A)$---the set of
points from the final clusters in $(W, R_A)$, to be more
precise---i.e.,
\begin{equation*}
W_0 = \{ v\in W \mid vR_Au \text{ implies } uR_Av, \text{ for all } u\in W\}.
\end{equation*}
In every final cluster $C \subseteq W_0$ with $|C|\ge 2$ we select some
point and denote by $U$ the set of all such selected points. Then we set $V_0
= W_0 \setminus U$ and $V_1 = W \setminus V_0$, and
define $R_B$ to be the reflexive closure of $R_A \cap (V_1 \times
V_0)$. Clearly, $(W,R_B)$ is a quasi-saw, with $V_0$
and $V_1$ being the sets of points of depth 0 and  1, respectively.
For every variable $r$, let
$r^\mathfrak{B} = r^\mathfrak{A}$.
As the extension of a $\cB$-term $\tau$ in $\mathfrak A$ is regular closed
and $\mathfrak A$ is finite, it is straightforward to show:
\begin{align}\label{eq:rc:max-cluster}
& \text{ if } y\in\tau^\mathfrak{A} \text{ then there exists } z \in V_0
\text{ such that } yR_Az \text{ and } z\in\tau^\mathfrak{A} .
\end{align}
We now prove (i) and (ii) by induction on the construction of $\tau$.

\begin{enumerate}[(i)]

\item The basis of induction follows from the definition.

\textit{Case $\tau = -\tau_1$}. We have $x\in
(\tc{(\compl{\tau_1})})^\mathfrak{A}$ iff
\begin{align*}
&\text{\phantom{iff} [def.]} && \exists y\in W\ \bigl(xR_Ay \text{ and } y\notin \tau_1^\mathfrak{A}\bigr)\\
&\text{iff [\eqref{eq:rc:max-cluster}]} && \exists y\in V_0\ \bigl(xR_Ay \text{ and } y\notin \tau_1^\mathfrak{A}\bigr)\\
&\text{iff [IH]} && \exists y\in V_0 \ \bigl(xR_By \text{ and } y\notin \tau_1^\mathfrak{B}\bigr)\\
&\text{iff [def.]} && x\in(\tc{(\compl{\tau_1})})^\mathfrak{B}.
\end{align*}

\textit{Case $\tau = \tau_1 + \tau_2$}. We have
$(\tau_1 + \tau_2)^\mathfrak{A} =
\tau_1^\mathfrak{A} \cup \tau_2^\mathfrak{A} =
\tau_1^\mathfrak{B} \cup \tau_2^\mathfrak{B} =
(\tau_1 + \tau_2)^\mathfrak{B}$, with the middle equation following by IH.

\item As $R_B\subseteq R_A$, the number of connected
components of $\tau^\mathfrak{A}$ in $\mathfrak A$ cannot be greater
than the number of components of $\tau^\mathfrak{B}$ in $\mathfrak
B$. Conversely, suppose that $X$ is a component of $\tau^\mathfrak{A}$
in $\mathfrak{A}$. As $\tau^\mathfrak{A}$ is regular closed, it is the
closure under $R_A^-$ of the set of final clusters in $X$. It follows
immediately from the definition of $R_B$ that all non-final points in
$X$ have precisely the same $R_B$- and $R_A$-accessible final
points. This means that $X$ is connected in $(W,R_B)$ as well.\qed
\end{enumerate}

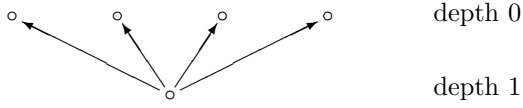
\begin{figure}[t]
\begin{center}
\setlength{\unitlength}{.07cm}
\begin{picture}(170,27)
\put(0,-7){%
\multiput(40,25)(20,0){4}{\circle{1.5}}%
\put(70,10){\circle{1.5}}%
\put(71,11.5){\vector(2,3){8}}%
\put(72,11){\vector(2,1){26}}%
\put(69,11.5){\vector(-2,3){8}}%
\put(68,11){\vector(-2,1){26}}%
\put(120,10){\small depth 1}%
\put(120,24){\small depth 0}%
}%
\end{picture}
\end{center}
\caption{$k$-fork for $k=4$.}\label{fig:nbroom}
\end{figure}

Recall now that every satisfiable $\cBCm$-formula is satisfiable in a finite Aleksandrov model, and so, by Lemma~\ref{broom-lemma}, in a quasi-saw model (over $\Regc$). The following lemma imposes restrictions on the branching factor and the number of points of depth 1 in such models.  Let us call a \emph{$k$-fork} any partial order of the form depicted in Fig.~\ref{fig:nbroom}.

\begin{lemma}\label{lma:boundedDegree}
If a $\cBCm$-formula $\varphi$ is satisfiable $($over $\Regc$$)$ then
it is satisfiable in a quasi-saw model over the disjoint union of
$n$-many $k_i$-forks, $1 \leq i \leq n$, where $n\le |\varphi|$ and
each $k_i$ does not exceed the largest value $k$ such that the
predicate $C^k$ occurs in $\varphi$.
\end{lemma}
\begin{proof}
Without loss of generality, we may assume that all the literals in $\varphi$ involving equality are of the form $\tau = \zero$ or $\tau \neq
\zero$. Suppose that $\mathfrak M\models\varphi$ and $\Lambda$ is the set of all literals of $\varphi$
that are true in $\mathfrak M$. By Lemmas~\ref{prop:KripkeSemantics} and~\ref{broom-lemma},
there is a quasi-saw model $\fB=(T_B, \RegC (T_B),\cdot^\fB)$ with $T_B$ induced by a partial order $(W_B,R_B)$ and  $\fB \models \Lambda$ (and so $\fB \models \varphi)$.

We construct a quasi-saw model $\fA=(T_A, \RegC (T_A),\cdot^\fA)$ induced by a disjoint union of forks $(W_A,R_A)$ and a map $f\colon T_A\to T_B$ as follows:
\begin{enumerate}[$\bullet$]
\item for each literal $(\tau \neq \zero)\in\Lambda$, we
select a point $x\in \tau^\fB$ of depth $0$, add a fresh $1$-fork $(\{u,v\},\{(u,v)\}^*)$ to $(W_A,R_A)$, where $R^*$ denotes the reflexive closure of $R$, and set $f(u) = x$, $f(v) = x$;

\item for each literal $C^k(\tau_1, \dots, \tau_k)\in\Lambda$, either (i) there is a point $x$ of depth $0$ in $\fB$ with $x \in \tau_1^\fB \cap
\dots \cap \tau_k^\fB$ or (ii) there exists a point $y$ of depth $1$ in $\fB$ such that $y \in \tau_1^\fB \cap
\dots \cap \tau_k^\fB$, in which case there are (not necessarily distinct) points $x_1,\dots,x_k$ of depth $0$ with $x_i \in \tau_i^\fB$; in the former case we add a fresh $1$-fork $(\{u,v\},\{(u,v)\}^*)$ to $(W_A, R_A)$ and set $f(u) = x$ and $f(v) = x$; in the latter case we add a fresh $k$-fork $(\{u,v_1,\dots,v_k\},\{(u,v_i) \mid 1 \leq i \leq k)\}^*)$ to $(W_A, R_A)$ and set $f(u) = y$ and $f(v_i) = x_i$, for $1 \leq i \leq k$.
\end{enumerate}
Define $\cdot^\fA$ by taking $v \in r^\fA$ iff $f(v) \in r^\fB$, for every $v$ of depth $0$, and $u \in r^\fA$ iff there is $v \in r^\fA$ of depth 0 with $uR_Av$, for every $u$ of depth 1. By definition, $r^\fA$ is regular closed in $\fA$, and it is easily checked that $\fA \models \Lambda$.
\end{proof}

As an immediate consequence of Lemma~\ref{lma:boundedDegree} we obtain the following:
\begin{corollary}\label{NB-complexity}
$\Sat(\cBCm{},\Regc)$, $\Sat(\cBC{},\Regc)$, $\Sat(\cB{},\Regc)$ and
$\Sat(\RCCE{},\Regc)$ are all \NP-complete.
\end{corollary}

The reader may wonder at this point whether lower complexities can be
achieved if we consider various sub-logics of the logics mentioned in this corollary. The answer is that they can. For example, maximal tractable (polynomial) fragments of $\RCCE$ were identified in \cite{Renz99,Renz&Nebel99}, and it was shown in \cite{aledThesis} that
the problem of determining the satisfiability of a conjunction of atomic
formulas of $\RCCE{}$ over the class $\Regc$ is \NLogSpace-complete.

We close this section with some remarks about logics interpreted over
connected spaces. The language $\cBC$
can distinguish between connected and disconnected spaces, because the
formula
\begin{equation}\label{eq:connectedSpaces}
\neg C(r, -r) \ \wedge \ (r \neq \zero) \  \wedge \ (r \neq \one)
\end{equation}
is satisfiable, but only in models over disconnected spaces.  By
contrast, the languages $\cB$ and $\RCCE$ cannot distinguish between
connected and disconnected spaces:
\begin{align*}
\Sat ( \RCCE, \Regc ) &= \Sat ( \RCCE, \ConR),\\
\Sat ( \cB, \Regc ) &= \Sat ( \cB, \ConR).
\end{align*}
Indeed, suppose that an $\RCCE$-formula $\varphi$ is satisfied in a
quasi-saw model $\fA$ with the underlying quasi-order $(W_A,R_A)$
constructed in the proof of Lemma~\ref{lma:boundedDegree}.  To make
this model connected, we can simply add to $W_A$ a new point $w$ of
depth 0 and extend $R_A$ by the arrows $uR_Aw$, for all $u$ of depth 1
in $\fA$. It is easy to see that, under the same valuation as in
$\fA$, $\varphi$ is satisfiable in the extended connected model. For
the less expressive language $\cB$, it suffices to add a new point of
depth 1 to $\fA$ and connect it to all points of depth $0$; details
are left to the reader.

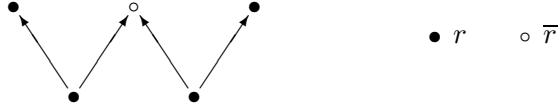
\begin{figure}[t]
\setlength{\unitlength}{.8mm}
\begin{center}
\begin{picture}(40,20)(0,-2)

\put(0,-2){%
\put(3,0){%
}%

\put(70,10){\circle*{1.7}}%
\put(85,10){\circle{1.5}}%
\put(73,9){$r$}%
\put(88,9){$\overline{r}$}

\multiput(0,15)(40,0){2}{\circle*{1.7}}
\multiput(20,15)(20,0){1}{\circle{1.5}}
\multiput(10,0)(20,0){2}{\circle*{1.7}}
\multiput(9,1.5)(20,0){2}{\vector(-2,3){8}}
\multiput(11,1.5)(20,0){2}{\vector(2,3){8}}
}%
\end{picture}
\end{center}
\caption{A quasi-saw model for $\neg c(r) \land
c(\one)$.}\label{55:two-fork}
\end{figure}

On the other hand, equipping even the weakest spatial logics such
as $\RCCE$ or $\cB$ with the connectedness predicates (or even just
interpreting them over connected spaces) invalidates various
model-theoretic properties employed above---most notably the
`tree-model' property, heavily used in the proof of
Lemma~\ref{lma:boundedDegree}. Consider, for example, the
$\cBc$-formula $\neg c(r) \land c(\one)$. Its smallest satisfying
quasi-saw model is illustrated in Fig.~\ref{55:two-fork}. Note that
this model cannot be transformed to a forest, because the underlying
frame must stay connected.
Indeed, as we shall see in the next section, to satisfy
$\cBc$-formulas, or $\cBC$-formulas in connected spaces, quasi-saw
models with \emph{exponentially} many points in the length of the
formulas may be required. It is these phenomena that are responsible
for the increased complexity of satisfiability which we shall
encounter below.


\section{Computational complexity}
\label{sec:complexity}

We are now in a position to prove tight
complexity results for spatial logics in the range between $\cBc$ and
$\conTc$.
%
%
The $\NExpTime$ upper bound for all the logics considered in this
paper was obtained in~\cite{PrattHartmann02}:

\begin{theorem}[\cite{PrattHartmann02}]\label{theo:conTcInNexptime}
$\Sat(\conTc,\All)$ is in \NExpTime{}.
\end{theorem}

The idea of the proof is based on the following observations. Let
$\varphi$ be any $\conTc$-formula. Evidently, $\varphi$ is satisfiable
if and only if there exists a set $\Phi$ of $\conTc$-literals, involving
all the atoms occurring in $\varphi$, such that: (i) $\Phi$ is
satisfiable; and (ii) $\Phi \cup \{\varphi\}$ is satisfiable in
propositional logic (where we treat all the atoms as propositional variables). Since propositional satisfiability can be checked
in NP, it suffices to restrict
attention to $\conTc$-formulas which are conjunctions of
literals---i.e.~those of the form:
\begin{align}
\label{conjunct0}
\bigl(\rho = \zero \bigr) \ \ \land \
\ \bigwedge_{i = 1}^m \bigl( \tau_i \ne \zero \bigr) \ \ \land \ \
\bigwedge_{i = 1}^n c^{\leq k_i}(\sigma_i) \ \ \land \ \
\bigwedge_{i = 1}^p c^{\geq k'_i}(\sigma_i').
\end{align}
The conjuncts of the form
$c^{\geq k'_i}(\sigma_i')$ can be eliminated using the following lemma~\cite{PrattHartmann02}:
\begin{lemma}[\cite{PrattHartmann02}]\label{lma:bigModels}
Let $\varphi$ be any $\conTc$-formula, and $\tau$ an $\SFU$-term.
Then, for every $n\ge 0$, there exists an $\conT$-formula $\psi$, with $|\psi|$ bounded by a
polynomial function of $n+ |\tau|$, such that $\varphi \wedge \psi$
is satisfiable if and only if $\varphi \wedge c^{\geq 2^n}(\tau)$ is
satisfiable.
\end{lemma}

By repeated applications of Lemma~\ref{lma:bigModels}, it is then a
straightforward matter to transform~\eqref{conjunct0}, in polynomial
time, into an equisatisfiable formula of the form
\begin{align}
\label{conjunct1}
\psi \quad=\quad
\bigl(\rho = \zero \bigr) \ \ \land \
\ \bigwedge_{i = 1}^m \bigl(\tau_i \ne \zero\bigr) \ \ \land \ \
\bigwedge_{i = 1}^n c^{\leq k_i}(\sigma_i).
\end{align}
Suppose that $\psi$ is true in some model $\fM$ over a space $T$,
and let $\Theta$ be the set of terms occurring in $\psi$.  Let
$\fA$ and $f$ be as guaranteed by
Lemma~\ref{prop:KripkeSemantics}. Then $\tau^\mathfrak{A} =
f(\tau^\mathfrak{M})$ for all $\tau \in \Theta$, and $f$ is
continuous, whence $\fA \models \psi$.  Thus, if
$\psi$ is satisfiable, then it is satisfied over a topological space
whose size is bounded by an exponential function of $|\psi|$, which gives the \NExpTime{} upper bound of Theorem~\ref{theo:conTcInNexptime}.

We turn next to the language $\conT$. By similar reasoning to the
above, we may without loss of generality confine attention to the
problem of determining the satisfiability of formulas of the form
\begin{align}\label{conjunct}
\psi \quad=\quad
\bigl(\rho = \zero \bigr) \ \ \land \
\ \bigwedge_{i = 1}^m \bigl(\tau_i \ne \zero\bigr) \ \ \land \ \
\bigwedge_{i = 1}^n
\bigl(c(\sigma_i)\land(\sigma_i\ne\zero)\bigr).
\end{align}

\begin{theorem}\label{theorem:S4uc:ExpTime}
$\Sat(\conT,\All)$ is in \ExpTime{}.
\end{theorem}
\begin{proof}
The proof is by reduction to the satisfiability problem for
propositional dynamic logic $\mathcal{PDL}$ with converse and
nominals, which is known to be \ExpTime-complete~\cite[Section
  7.3]{deGiacomo95}.  Let $\psi$ be as in (\ref{conjunct}). Take two
atomic programs $\alpha$ and $\beta$ and, for each $\sigma_i$, a
nominal $\ell_i$. For a term $\tau$, denote by $\tau^\ddagger$ the
$\mathcal{PDL}$-formula obtained by replacing in $\tau$, recursively,
each sub-term $\ti{\vartheta}$ with $[\alpha^{\ast}]\vartheta$. Thus
the transitive and reflexive accessibility
relation of the modal logic ${\bf S4}$ is
simulated by $\alpha^{\ast}$,
and the universal modality $\forall$ (see the end
of Section~\ref{sec:without}) is simulated by $[\gamma]$, where
$\gamma = (\beta \cup \beta^- \cup \alpha \cup
\alpha^-)^{\ast}$. Consider now the formula
\begin{equation*}
\psi'\quad=\quad[\gamma]\neg \rho^\ddagger \ \ \wedge \ \
\bigwedge_{i=1}^m \langle\gamma\rangle \tau_i^\ddagger \ \ \wedge \
\ \bigwedge_{i=1}^n \Bigl(\langle\gamma\rangle (\ell_i \wedge
\sigma_i^\ddagger) \ \wedge \ [\gamma](\sigma_i^\ddagger
\rightarrow \langle (\alpha \cup
\alpha^-;\sigma_i^\ddagger?)^\ast\rangle \ell_i)\Bigr).
\end{equation*}
It is easy to see that $\psi'$ is satisfiable
if and only if $\psi$ is satisfiable: the first conjunct of $\psi'$ states that
$\rho$ is empty, the second that all $\tau_i$ are non-empty, the third
states that each $\sigma_i$ holds at a point where $\ell_i$ holds and
that from each $\sigma_i$-point there is a path (along
$\alpha\cup\alpha^-$) to $\ell_i$ which lies entirely within
$\sigma_i$.
\end{proof}


If $\psi$ is parsimonious in its use of connectedness, we can do
somewhat better.  Denote by $\conT^1${} the set of $\conT${}-formulas
with \emph{at most one} occurrence of an atom of the form $c(\tau)$.

\begin{theorem}\label{thm:one-connected}
$\Sat(\conT^1,\All)$ is \PSpace{}-complete.
\end{theorem}
\begin{proof}
The lower bound follows from \cite{Ladner77}. We sketch a nondeterministic \PSpace{} algorithm
recognizing $\Sat(\conT^1,\All)$.
We may assume
without loss of generality that the input $\psi$ has the
form~\eqref{conjunct}, with $n \leq 1$.  If $n = 0$, i.e.~if $\psi$
does not contain a conjunct of the form $c(\sigma) \land (\sigma \ne
\zero)$, then a standard satisfiability checking algorithm for ${\bf S4}_u$
is applied. Assume, then, that $n= 1$ in~\eqref{conjunct}; and write
$\sigma$ for $\sigma_1$. Set
$\mathbf{B} = \{\ti{\compl{\rho}},\compl{\rho}\} \cup \{\tau, \compl{\tau}
\mid \tau\in\textit{term}(\psi)\}$, where $\textit{term}(\psi)$
is the set of all sub-terms of $\psi$. A subset $\tp$ of
$\mathbf{B}$ is called a \emph{type for} $\psi$ if
$\ti{\compl{\rho}} \in \tp$ and $\tau\in\tp$ iff
$\compl{\tau}\notin\tp$, for all $\compl{\tau}\in\mathbf{B}$.

Now, guess a type $\tp_{\sigma}$ containing $\sigma$ and start $(m+1)$
${\bf S4}$-tableau procedures (see, e.g.,
\cite{Fitting83,Gore99}) with inputs $\tau_1 \cap
\ti{\compl{\rho}}$, $\tau_2 \cap \ti{\compl{\rho}},\ldots,\tau_m
\cap \ti{\compl{\rho}}$, and $\bigcap \tp_{\sigma} \cap
\ti{\compl{\rho}}$ in the usual way, expanding nodes branch-by-branch, and
recovering the space once branches are checked. We may as well assume
that the nodes of these tableaux are types. Suppose $\tp$ is a type
occurring in one of them.  If $\sigma\in\tp$, it suffices to check
that $\tp$ can be connected by a $\sigma$-path of $\le 2^{|\psi|}$
points to $\tp_\sigma$. This can be done in \PSpace{} by the following
non-deterministic subroutine. We start with $\tp$ and count from 1 to
$2^{|\psi|}$. At each step we guess a new type $\tp'$ with
$\sigma,\ti{\compl{\rho}}\in \tp'$, and
check that
\begin{enumerate}[$\bullet$]
\item either (i) $\tau \in \tp'$ for all $\ti{\tau} \in \tp$, or (ii)
  $\tau \in \tp$, for all $\ti{\tau} \in \tp'$ (in the former case,
  $\tp'$ is accessible from $\tp$, in the latter, the other way
  around);
\item an ${\bf S4}$-tableau with root $\tp'$ can be constructed (which
  can be discarded after completion). Note that, although this tableau
  may contain types $\tp''$ with $\sigma \in \tp''$, these types can
  never threaten the connectedness of $\sigma$, since they are all
  accessible from the root $\tp'$ of the tableau, and so are connected
  to both $\tp$ and $\tp_\sigma$, by the transitivity of the
  accessibility relation.
\end{enumerate}
If both checks are successful and $\tp' = \tp_\sigma$, the subroutine
succeeds; if $\tp' \ne \tp_\sigma$ we set $\tp = \tp'$ and continue to
the next step (provided that the step number $< 2^{|\psi|}$, otherwise
the subroutine fails). Clearly, this subroutine succeeds if there is a
$\sigma$-path connecting $\tp$ and $\tp_\sigma$ and fails in every
computation otherwise; moreover, it requires only polynomial memory to
store the tableau for $\tp'$ and the step number.
%
%
\end{proof}

To reduce notational clutter
we denote, for any topological space $T$, the (singleton)
frame-class $\{(T, 2^T)\}$ simply by $T$, and the (singleton)
frame-class $\{(T, \RegC(T))\}$ simply by $\RegC(T)$. This notation is not
entirely uniform, but it should be obvious what is meant.

As shown in \cite{Shehtman99} (see also Theorem~\ref{theo:CalmostInsensitive} below), $\Sat(\SFU,\Con) = \Sat(\SFU,\R^n)$ for
any $n\geq 1$. Recalling now that the modal logic
${\bf S4}$ is \PSpace-hard, we immediately obtain the following:
\begin{corollary}\label{SFUoverR}
$\Sat(\SFU,\Con)$ and $\Sat(\SFU,\R^n)$ are all \PSpace-complete for any $n\geq 1$.
\end{corollary}

The proof of Theorem~\ref{thm:one-connected} can be generalized in various ways. For example, assume that
$\psi= \psi_{1} \wedge \psi_{2}$ is an $\conT$-formula in which
$\psi_{2}= \bigwedge_{1\leq i < j\leq n}(\sigma_{i}\cap \sigma_{j}=\zero)$, where $\{\sigma_{1},\ldots,\sigma_{n}\}$
is the collection of all $\sigma_{i}$ such that $c(\sigma_{i})$ occurs in $\psi_1$. A straightforward extension
of the algorithm in the proof of Theorem~\ref{thm:one-connected} shows that satisfiability of $\psi$
is still in \PSpace. Thus, if the
connectedness predicate is applied only to regions
that are known to be pairwise disjoint, then it does not matter how
many times this predicate occurs in the formula: satisfiability is
still in \PSpace.



Our next theorem gives matching lower bounds for
Theorem~\ref{thm:one-connected} and Corollary~\ref{SFUoverR}.
\begin{theorem}\label{pspace-lower}
$\Sat (\cBCc{}^1,\Regc)$ is \PSpace{}-hard. In fact, the problems
$\Sat (\cBC{},\ConR)$ and $\Sat (\cBC{},\RegC(\R^n))$ for all $n \geq 1$ are \PSpace{}-hard.
\end{theorem}
\begin{proof}
Let $L$ be a language in \PSpace. Then there is a
polynomial-space-bounded \emph{deterministic} Turing machine $M$
recognizing $L$. Without loss of generality, we may assume that, given
some input $\vec{a} \in L$ on the tape, $M$ starts in the \emph{initial
state}, reaches the \emph{accepting state} (with the resulting tape
being empty and the head positioned over the first cell) and then
moves to the \emph{halting state}, from which no transition is
possible. Moreover, throughout the computation the machine never goes
to the left of the first cell and to the right of the $s$'th cell,
where $s = p(|\vec{a}|)$ for some polynomial
$p(\cdot)$.

Let $Q$ and $\Sigma$ be the set of states and the alphabet of $M$,
respectively. The instructions of $M$ are of the form $(q,a)\to
(q',a',d)$, $d \in \{+1,0,-1\}$, with their standard
meaning.
A configuration of $M$ is a word
$\mathfrak{c}$ of the form
\begin{align}\label{config1}
a_1,\dots, a_{i-1}, (q,a_i), a_{i+1}, \dots, a_s,
\end{align}
where $a_1,\dots,a_s$ ($a_j\in \Sigma$) is the current contents of the tape, $q\in Q$ the current state, and $i$ the current position of
the head. If a configuration $\mathfrak c'$ is obtained from a
configuration $\mathfrak c$ by applying one instruction of $M$ then
we write $\mathfrak c \to \mathfrak c'$.

It will be convenient for us to represent $M$ as the following set $\mathcal{T}$ of 4-tuples, where $t,b$ are two fresh auxiliary symbols (see Fig.~\ref{fig:TM:tiles}):
\begin{enumerate}[$\bullet$]
\item $(a,t,a,t)$ and $(a,b,a,b)$, for every $a\in \Sigma$,
\item $(a',(q',b),(q',a'),b)$ and $(a',t,(q',a'),(q',t))$, for all $a'\in \Sigma$ and $q'\in Q$,
\item $((q,a),t,(q',a'),b)$, for every instruction $(q,a) \to (q',a',0)$
in $M$,
\item $((q,a),t,a',(q',b))$, for every instruction $(q,a) \to
(q',a',-1)$ in $M$,
\item $((q,a),(q',t),a',b)$, for every instruction $(q,a) \to
(q',a',+1)$ in $M$.
\end{enumerate}
\begin{figure}[ht]
\mbox{}\hfill\begin{picture}(360,140)(-40,-40)
\put(-30,55){%
\put(0,0){\line(0,1){40}}%
\put(0,0){\line(1,0){40}}%
\put(0,0){\line(1,1){40}}%
\put(0,40){\line(1,0){40}}%
\put(40,0){\line(0,1){40}}%
\put(40,0){\line(-1,1){40}}%
\small%
\put(18,30){$t$}%
\put(18,5){$t$}%
\put(5,18){$a$}%
\put(30,18){$a$}%
}%
\put(-30,5){%
\put(0,0){\line(0,1){40}}%
\put(0,0){\line(1,0){40}}%
\put(0,0){\line(1,1){40}}%
\put(0,40){\line(1,0){40}}%
\put(40,0){\line(0,1){40}}%
\put(40,0){\line(-1,1){40}}%
\small%
\put(18,30){$b$}%
\put(18,5){$b$}%
\put(5,18){$a$}%
\put(30,18){$a$}%
}%
\put(-25,-8){$a\in \Sigma$}
\put(65,30){%
\put(0,0){\line(0,1){40}}%
\put(0,0){\line(1,0){40}}%
\put(0,0){\line(1,1){40}}%
\put(0,40){\line(1,0){40}}%
\put(40,0){\line(0,1){40}}%
\put(40,0){\line(-1,1){40}}%
\small%
\put(18,30){$t$}%
\put(18,5){$b$}%
\put(2,18){$q,\!a$}%
\put(25,18){$q'\!\!,\!a'$}%
}%
\put(40,15){$(q,a) \to (q',a',0)$}
\put(160,55){%
\put(0,0){\line(0,1){40}}%
\put(0,0){\line(1,0){40}}%
\put(0,0){\line(1,1){40}}%
\put(0,40){\line(1,0){40}}%
\put(40,0){\line(0,1){40}}%
\put(40,0){\line(-1,1){40}}%
\small%
\put(18,30){$t$}%
\put(13,4){$q'\!\!,b$}%
\put(2,18){$q,\!a$}%
\put(30,18){$a'$}%
}%
\put(160,-10){%
\put(0,0){\line(0,1){40}}%
\put(0,0){\line(1,0){40}}%
\put(0,0){\line(1,1){40}}%
\put(0,40){\line(1,0){40}}%
\put(40,0){\line(0,1){40}}%
\put(40,0){\line(-1,1){40}}%
\small%
\put(13,30){$q'\!\!,b$}%
\put(18,5){$b$}%
\put(5,18){$a'$}%
\put(25,18){$q'\!\!,\! a'$}%
}%
\put(130,42){$(q,a) \to (q',a',-1)$}
\put(165,-24){$a'\in \Sigma$}%
\put(165,-34){$q'\in Q$}%
\put(270,55){%
\put(0,0){\line(0,1){40}}%
\put(0,0){\line(1,0){40}}%
\put(0,0){\line(1,1){40}}%
\put(0,40){\line(1,0){40}}%
\put(40,0){\line(0,1){40}}%
\put(40,0){\line(-1,1){40}}%
\small%
\put(18,30){$t$}%
\put(12,4){$q'\!\!,t$}%
\put(5,18){$a'$}%
\put(25,18){$q'\!\!,\! a'$}%
}%
\put(270,-20){%
\put(0,0){\line(0,1){40}}%
\put(0,0){\line(1,0){40}}%
\put(0,0){\line(1,1){40}}%
\put(0,40){\line(1,0){40}}%
\put(40,0){\line(0,1){40}}%
\put(40,0){\line(-1,1){40}}%
\small%
\put(13,30){$q'\!\!,t$}%
\put(18,5){$b$}%
\put(2,18){$q,\!a$}%
\put(30,18){$a'$}%
}%
\put(240,-34){$(q,a) \to (q',a',+1)$}%
\put(275,42){$a'\in \Sigma$}%
\put(275,32){$q'\in Q$}%
\end{picture}\hfill\mbox{}%
\caption{Tile types $\mathcal{T}$ for the Turing machine $M$.}\label{fig:TM:tiles}
\end{figure}
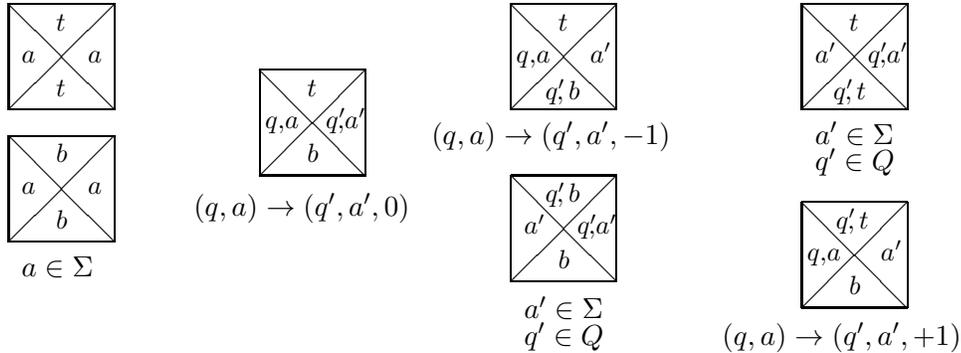
We call these 4-tuples \emph{tile types} and, for each $T \in \mathcal{T}$, denote its four
components by $\textit{left}(T)$, $\textit{top}(T)$,
$\textit{right}(T)$ and $\textit{bot}(T)$, respectively.
Configurations of $M$ will be encoded on the left- and
right-hand sides of the tile types in sequences
$T_{k_1},\dots,T_{k_s}$ such that
\begin{equation}\label{eq:tm:tiling}
\textit{top}(T_{k_s}) = t, \qquad \textit{top}(T_{k_i}) = \textit{bot}(T_{k_{i+1}}), \text{ for } 1 \leq i < s,\qquad\text{and}\qquad
\textit{bot}(T_{k_1}) = b.
\end{equation}
By the definition of $\mathcal{T}$, every such sequence $T_{k_1},\dots,T_{k_s}$ gives rise to two configurations $\mathfrak c = \textit{left}(T_{k_1}),\dots,\textit{left}(T_{k_s})$ and $\mathfrak c' =
\textit{right}(T_{k_1}),\dots,\textit{right}(T_{k_s})$ of $M$ with $\mathfrak c \to \mathfrak c'$.

Now we describe the computations of $M$ in terms of
\cBC{}-formulas. While constructing the formulas,
we will assume that $\mathfrak A$ is a connected quasi-saw model induced by
$(W,R)$ and $W_0$ is the set of points of
depth $0$ in $(W,R)$.

We need $s$ variables $T_k^1,\dots,T_k^s$, for
each $T_k\in \mathcal{T}$, and three additional variables $B^0$, $B^1$ and
$B^2$. Consider the following \cBC{}-formulas: 
\begin{align}
\label{TM1}
& B^0 + B^1 + B^2 = \one,\\
\label{TM2}
& B^0 \cdot B^1 = \zero,\quad B^1 \cdot B^2 = \zero, \quad B^2 \cdot B^0 = \zero.
\end{align}
If the conjunction of \eqref{TM1}--\eqref{TM2} holds in $\mathfrak{A}$
then every point $x\in W_0$ is precisely in one of
$(B^\ell)^\mathfrak{A}$, $0 \leq \ell \leq 2$.  We will use the
$B^\ell$ to introduce `direction' in our quasi-saw model in the
following sense.  If $x_1,x_2\in W_0$ and $zRx_i$, $i = 1,2$, then
there are only three possibilities:
\begin{enumerate}[$\bullet$]
\item $x_1$ and $x_2$ are regarded as `identical' whenever $x_1,x_2\in (B^\ell)^\fA$, for $0 \leq \ell \leq 2$,
\item $x_2$ is a `successor' of $x_1$  whenever $x_1\in (B^\ell)^\fA$ and
$x_2 \in (B^{\ell \oplus 1})^\fA$, for $0 \leq \ell \leq 2$,
\item $x_1$ is a `successor' of $x_2$ whenever $x_1 \in (B^{\ell \oplus 1})^\fA$ and $x_2\in (B^\ell)^\fA$, for $0 \leq \ell \leq 2$,
\end{enumerate}
where $\oplus$ denotes addition modulo 3.
(Here we remind the reader that $\tau_1 \cdot \tau_2 = \zero$ holds
in a quasi-saw model $\mathfrak A$ iff $\tau_1^{\mathfrak A}$ and
$\tau_2^{\mathfrak A}$ contain no common points of depth 0. This
means, in particular, that $(\tau_1 \cdot \tau_2)^\mathfrak{A} = \emptyset$ may hold
even though $\tau_1^\mathfrak{A} \cap \tau_2^\mathfrak{A} \ne
\emptyset$, i.e., $\mathfrak{A} \models C(\tau_1,\tau_2)$.)

Suppose also that the following formulas hold in $\mathfrak{A}$:
\begin{align}
\label{TM3}
& \textstyle\sum_{T_k\in \mathcal{T}} T_k^i = \one, &&1 \leq  i \leq s,\\
\label{TM4}
& T_{k_1}^i \cdot T_{k_2}^i = \zero, &&1 \leq  i \leq s, \ \ T_{k_1},T_{k_2}\in \mathcal{T}, \ \ k_1 \ne k_2,\\
\label{TM5} &T_{k_1}^i \cdot T_{k_2}^{i+1} = \zero, \quad
&&1 \leq  i < s, \ \ \textit{top}(T_{k_1}) \ne \textit{bot}(T_{k_2}),\ \ T_{k_1},T_{k_2}\in \mathcal{T},\\
\label{TM6}
&T_k^1 = \zero,&&\textit{bot}(T_k) \ne b,\  \ T_k\in \mathcal{T},\\
\label{TM7}
&T_k^s = \zero, &&\textit{top}(T_k) \ne t,\ \ T_k\in \mathcal{T}.%
\end{align}
Then, by~\eqref{TM3}--\eqref{TM4}, for every $x\in W_0$ there is a unique sequence of tile types $T_{k_1}, \dots, T_{k_s}$ with $x
\in (T^i_{k_i})^{\mathfrak{A}}$, for $1 \leq i \leq s$. In this
case we set $\textit{left}^i(x) =
\textit{left}(T_{k_i})$, $\textit{top}^i(x)
=\textit{top}(T_{k_i})$, etc. We also set $\textit{left}(x) = \textit{left}^1(x), \dots,
\textit{left}^s(x)$,
$\textit{right}(x)=\textit{right}^1(x),\dots,\textit{right}^s(x)$.
Then, by~\eqref{eq:tm:tiling} and~\eqref{TM5}--\eqref{TM7}, both $\textit{left}(x)$ and
$\textit{right}(x)$ are configurations of $M$ and $\textit{left}(x)
\to \textit{right}(x)$.

Consider now the following formulas, for $1 \leq i \leq s$, $0 \leq \ell \leq 2$, $T_{k_1},T_{k_2}\in \mathcal{T}$:
\begin{align}\label{TM8}
& \neg C(B^{\ell} \cdot T_{k_1}^i, \ B^{\ell\oplus 1} \cdot T_{k_2}^i),
\qquad\textit{right}(T_{k_1})\ne\textit{left}(T_{k_2}),
\\ \label{TM8b}
& \neg C(B^\ell \cdot T_{k_1}^i, \ B^\ell \cdot T_{k_2}^i),\qquad k_1 \ne k_2.
\end{align}
Suppose that all of \eqref{TM1}--\eqref{TM8b} are true in $\mathfrak
A$. It easy to see that, if $x,y \in W_0$ and there exists $z\in W$
with $zRx$ and $zRy$, then:
\begin{align}
& \text{$\textit{right}(x)= \textit{left}(y)$ whenever
   $x\in (B^\ell)^\mathfrak{A}$ and $y\in (B^{\ell\oplus 1})^\mathfrak{A}$,
   for $0 \leq \ell \leq 2$};
\label{eq:easy1}\\
& \text{$\textit{left}(x)= \textit{left}(y)$ and
       $\textit{right}(x)= \textit{right}(y)$ whenever
       $x,y \in (B^\ell)^\mathfrak{A}$, for $0 \leq \ell \leq 2$}.
\label{eq:easy2}
\end{align}
Finally, we require the following formulas:
\begin{align}
& T_{i_1}^1 \cdot \ \cdots \ \cdot T_{i_s}^s \ne \zero,\label{TM10}\\
& T_{j_1}^1 \cdot \ \cdots \ \cdot T_{j_s}^s \ne \zero,\label{TM11}
\end{align}
where $\textit{left}(T_{i_1}),\dots, \textit{left}(T_{i_s})$ is the
initial configuration (with $\vec{a}$ written on the tape) and
$\textit{right}(T_{j_1}),\dots, \textit{right}(T_{j_s})$ is the
accepting configuration (with empty tape and the head scanning the
first cell). Denote the conjunction of \eqref{TM1}--\eqref{TM11} by
$\Psi(M,\vec{a})$. Clearly, the length of this \cBC{}-formula  is
polynomial in the size of $M$ and $\vec{a}$.  We proceed to show:
({\em i}) if $\Psi(M,\vec{a})$ is satisfiable over $\ConR$,  then $M$
accepts $\vec{a}$; ({\em ii}) if $M$ accepts $\vec{a}$, then
$\Psi(M,\vec{a})$ is satisfiable over $\RegC(\R^n)$ for any $n \geq
1$. This proves the theorem.

Suppose $\Psi(M,\vec{a})$ is satisfiable over $\ConR$.  By
Lemmas~\ref{prop:KripkeSemantics} and~\ref{broom-lemma}, it is
satisfied in some model $\mathfrak{A}$ over a finite connected  quasi-saw
$(W,R)$. From~\eqref{TM10} and~\eqref{TM11}, choose points $u, u'$ of
depth 0 such that $\textit{left}(u)$ is the initial configuration and
$\textit{right}(u')$ the accepting configuration. Since $(W,R)$ is
connected, there exists a sequence $u_0, v_1, \ldots, u_{m-1}, v_m,
u_m$ with the points $u_i$ of depth 0 and the points $v_j$ of depth 1, such that:
$u_0 = u$, $u_m = u'$, and, for all $i$ ($1 \leq i \leq m$), $(v_i,
u_{i-1}) \in R$ and $(v_i, u_i) \in R$. We may assume without loss of
generality that the $u_i$ and $v_j$ are all distinct.
It should be noted that $\textit{left}(u_i)$ is not
the accepting configuration for any $i \leq m$. For brevity, we write
$\mathfrak c_i$ for $\textit{left}(u_i)$ and $\mathfrak{c}_{m+1}$ for $\textit{right}(u_m)$. We shall show that
$\mathfrak c_0, \ldots, \mathfrak c_m, \mathfrak{c}_{m+1}$ contains a sub-sequence that is
an accepting run of $M$. From~\eqref{eq:easy1} and ~\eqref{eq:easy2},
and the fact that $\textit{left}(x) \rightarrow \textit{right}(x)$ for
any point $x$, we see that, for all $i$ ($0 \leq i \leq m$), one of the
following conditions holds: ({\em i}) $\mathfrak c_i = \mathfrak
c_{i+1}$; ({\em ii})~$\mathfrak c_i \rightarrow \mathfrak c_{i+1}$; or
({\em iii}) $\mathfrak c_i \leftarrow \mathfrak c_{i+1}$. We shall presently
establish the following claim:
\begin{claim}\label{claim:TM}
If $0 \leq j \leq m$ and $\mathfrak c_j \leftarrow \mathfrak c_{j+1}$, then there
exists $k$ such that $j+ 1 < k \leq m$ and $\mathfrak c_k = \mathfrak c_j$.
\end{claim}
\noindent Taking this claim on trust for the moment, define the sub-sequence
$\mathfrak c_{j_0}, \mathfrak c_{j_1}, \ldots$ by setting $j_0$ to
be the largest $j \leq m$ such that $\mathfrak c_j = \mathfrak c_0$,
and, for $i \geq 0$, $j_{i+1}$ to be the largest $j \leq m$ such that
$\mathfrak c_j = \mathfrak c_{j_i +1}$, until we eventually reach
(say), $\mathfrak c_{j_K} = \mathfrak c_m$. It is then immediate
from the claim that $\mathfrak c_{j_i} \rightarrow \mathfrak
c_{j_i+1} = \mathfrak c_{j_{i+1}}$ for all $i$ ($0 \leq i < K$),
and we have the desired accepting run of $M$.

\begin{proofof}{Claim~\ref{claim:TM}}
The claim is proved by
(decreasing) induction on $j$. For $j = m$, the result is trivial,
since $\mathfrak c_{m+1}$ has no successor configurations. Assume, then,
that $0 \leq j < m$, and the claim holds for all larger values of
$j$ up to $m$. Let $k$ be the largest number ($j+1 \leq k \leq m +1$)
such that $\mathfrak c_{j+1} = \mathfrak c_k$. Thus, $k \leq m$ (since
$\mathfrak c_{j+1}$ is, by assumption, not the accepting configuration), and $\mathfrak
c_k \rightarrow \mathfrak c_{k+1}$ (since otherwise, using the
inductive hypothesis, we could find a larger value $k'$ with
$\mathfrak c_{j+1} = \mathfrak c_{k'}$).  Thus, $\mathfrak c_j
\leftarrow \mathfrak c_{j+1} = \mathfrak c_k \rightarrow \mathfrak
c_{k+1}$. But $M$ is deterministic, so $\mathfrak c_j = \mathfrak
c_{k+1}$, completing the induction, and proving the claim.
\end{proofof}

The proof of the converse direction is straightforward: for an accepting
computation $\mathfrak c_0 \to \dots \to \mathfrak c_{m+1}$ of $M$ on
$\vec{a}$, we construct a model $\mathfrak{A}$ over $\RegC(\R)$.
Define the closed intervals $I_0, \ldots, I_m$ in $\R$ by
\begin{equation*}
I_j = \begin{cases} (-\infty,0], & \text{ if } j =0, \\
[i-1, i], & \text{ if } 0 < j < m,\\
[m-1,+\infty), & \text{ if } j = m.
\end{cases}
\end{equation*}
Note that, given any valuation over $\RegC(\R)$ in which all the
variables $T_k^i$ are interpreted as unions of the intervals $I_0,
\ldots, I_m$, we may meaningfully write statements of the form
$\textit{left}(I_j)= \mathfrak{c}$ and $\textit{right}(I_j)=
\mathfrak{c}$, for $0 \leq j \leq m$, and $\mathfrak{c}$ a
configuration of $M$. Now define such a valuation
$\cdot^{\mathfrak{A}}$ in which, for all $j$ ($0 \leq j \leq m$), $I_j
\subseteq (B^\ell)^\mathfrak{A}$ if and only if $j \equiv \ell \pmod 3$; and, for
all $j$ ($0 \leq j < m$), $\textit{left}(I_j)= \mathfrak{c}_j$ and
$\textit{right}(I_j) = \mathfrak c_{j+1}$. It can be readily checked
that the resulting model satisfies $\Psi(M,\vec{a})$. For $n> 1$, we
may construct a model of $\Psi(M,\vec{a})$ over $\RegC(\R^n)$ by
cylindrification of $\fA$ in the obvious way.
\end{proof}

\begin{corollary}\label{cor:cBCconPSPACEcomplete}
$\Sat(\cBC,\ConR)$, $\Sat(\cBCc^1,\Regc)$ and $\Sat(\cBC,\RegC(\R^n))$ for any $n\geq 1$ are all \PSpace{}-complete.
\end{corollary}
\begin{proof}
Follows from Theorems~\ref{thm:one-connected}, \ref{pspace-lower} and Corollary~\ref{SFUoverR}.
\end{proof}

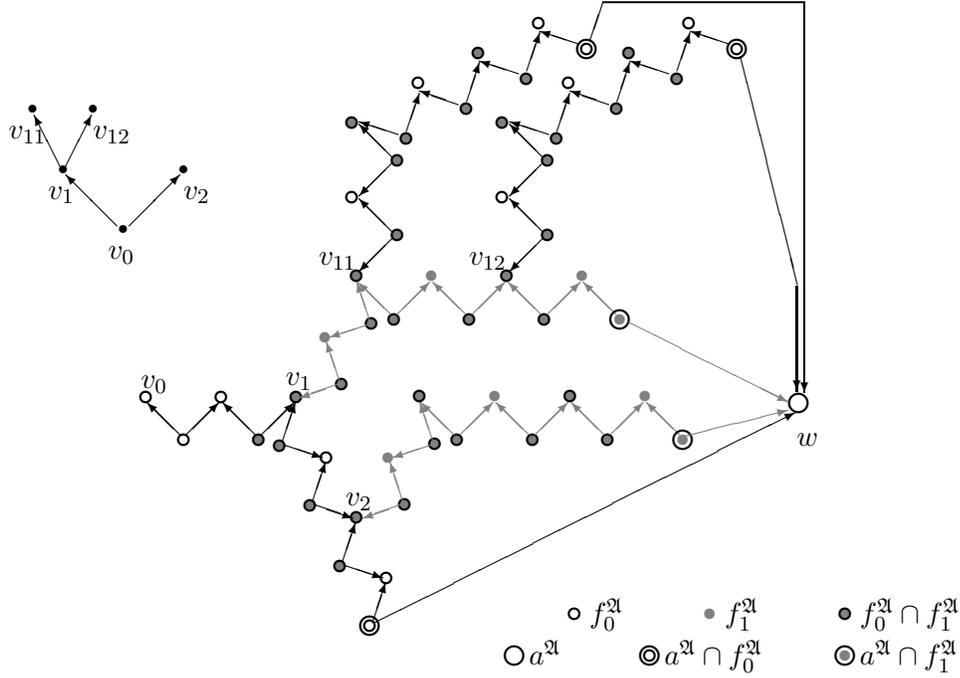
\begin{figure}[t]
\setlength{\unitlength}{.04cm}
\begin{center}
\begin{picture}(170,230)(15,-85)

\put(-50,70){%
\put(40,0){\circle*{2.5}}%
\put(38,1){\vector(-1,1){17}}%
\put(42,1){\vector(1,1){17}}%
\put(20,20){\circle*{2.5}}%
\put(60,20){\circle*{2.5}}%
\put(19,21){\vector(-1,2){8.5}}%
\put(21,21){\vector(1,2){8.5}}%
\put(10,40){\circle*{2.5}}%
\put(30,40){\circle*{2.5}}%
\put(35,-10){$v_0$}%
\put(15,10){$v_1$}%
\put(60,10){$v_2$}%
\put(2,30){$v_{11}$}%
\put(30,30){$v_{12}$}%
}%

{%
\put(-4,17){$v_0$}%
\put(44,18){$v_1$}%
\put(64,-22){$v_2$}%
\put(-2.5,14){\fzc} %
\put(22.5,14){\fzc} %
\put(35,0){\textcolor{gray}{\foc}}%
\multiput(10,0)(25,0){2}{\fzc}
\multiput(11,1.5)(25,0){2}{\vector(1,1){11}}
\multiput(9,1.5)(25,0){2}{\vector(-1,1){11}}
\put(35,-20){%
\multiput(12.5,34)(20,-40){2}{\textcolor{gray}{\foc}}%
\multiput(12.5,34)(20,-40){2}{\fzc}
\multiput(22.5,14)(20,-40){2}{\fzc}
\put(-15,5){%
\multiput(22,13)(10,-20){3}{\textcolor{gray}{\foc}}%
\multiput(22,13)(10,-20){4}{\fzc}
\multiput(23,14.5)(10,-20){4}{\vector(1,3){4.3}}
\multiput(23.5,13.2)(10,-20){3}{\vector(3,-1){12.9}}
\put(52,-47){\thicklines\circle{6}}%
\put(53.5,-46){\vector(2,1){140}}
\put(194.5,27){\thicklines\circle{6}}%
}}%
}%


\put(0,0){\textcolor{gray}{%
\put(55,58){\textcolor{black}{$v_{11}$}}%
\put(105,58){\textcolor{black}{$v_{12}$}}%
\put(25,0){%
\multiput(32,34)(10,20){1}{\foc}
\put(15,5){%
\multiput(22.5,13.5)(10,20){2}{\foc}
\multiput(22.5,13.5)(10,20){2}{\textcolor{black}{\fzc}}
\multiput(22,14.5)(10,20){2}{\vector(-1,3){4.3}}
\multiput(21.5,13.2)(10,20){2}{\vector(-3,-1){12.9}}
}}%
\put(70,40){%
\multiput(-2.5,14.5)(50,0){2}{\foc}
\multiput(-2.5,14.5)(50,0){2}{\textcolor{black}{\fzc}}
\multiput(22.5,14.5)(50,0){2}{\foc} %
\multiput(10,0)(25,0){4}{\foc}
\multiput(10,0)(25,0){3}{\textcolor{black}{\fzc}}
\multiput(11,1.5)(25,0){3}{\vector(1,1){11}}
\multiput(9,1.5)(25,0){4}{\vector(-1,1){11}}
\put(85,0){\textcolor{black}{\thicklines\circle{6}}}%
%
\put(88,-1){\vector(2,-1){53}}%
}}}%

\put(21,-40){\textcolor{gray}{%
\put(25,0){%
\multiput(32,34)(10,20){1}{\foc}
\put(15,5){%
\multiput(22.5,13.5)(10,20){2}{\foc}
\multiput(22.5,13.5)(10,20){2}{\textcolor{black}{\fzc}}
\multiput(22,14.5)(10,20){2}{\vector(-1,3){4.3}}
\multiput(21.5,13.2)(10,20){2}{\vector(-3,-1){12.9}}
}}%
\put(70,40){%
\multiput(-2.5,14.5)(50,0){2}{\foc}
\multiput(-2.5,14.5)(50,0){2}{\textcolor{black}{\fzc}}
\multiput(22.5,14.5)(50,0){2}{\foc} %
\multiput(10,0)(25,0){4}{\foc}
\multiput(10,0)(25,0){3}{\textcolor{black}{\fzc}}
\multiput(11,1.5)(25,0){3}{\vector(1,1){11}}
\multiput(9,1.5)(25,0){4}{\vector(-1,1){11}}
\put(85,0){\textcolor{black}{\thicklines\circle{6}}}%
\put(85,1){\vector(4,1){35}}%
}%
}}


\multiput(78,83)(50,0){2}{%
%
\put(-12,-2.5){\fzc} %
\put(-12,22.5){\textcolor{gray}{\foc}}
\put(-12,22.5){\fzc} %
\multiput(3,10)(0,-25){2}{\textcolor{gray}{\foc}}%
\multiput(3,10)(0,-25){2}{\fzc}
\multiput(1.5,9)(0,-25){2}{\vector(-1,-1){11}}
\multiput(1.5,11)(0,-25){2}{\vector(-1,1){11}}
\put(28,30){%
\put(2,15.5){\textcolor{gray}{\foc}}
\multiput(22,25.5)(-20,-10){3}{\fzc}
\put(25,-5){%
\multiput(-7,12)(-20,-10){3}{\textcolor{gray}{\foc}}%
\multiput(13,22)(-20,-10){4}{\fzc}
\multiput(10.5,23.5)(-20,-10){4}{\vector(-3,1){13}}
\multiput(-6.8,14.5)(-20,-10){3}{\vector(1,3){4.3}}
\put(13,22){\textcolor{black}{\thicklines\circle{6}}}%
%
}}%
}%

\put(98,93){%
\put(47.2,39.5){\line(1,3){4.3}}
\put(51.5,52.5){\line(1,0){67}} %
\put(118.5,52.5){\vector(0,-1){130}}
\put(97.2,34.5){\line(1,-4){19}} %
\put(116,-42){\vector(0,-1){35}}
}%

\put(214,-2){$w$}

\put(0,-8){
\put(140,-50){\fzc}%
\put(145,-53){$f_0^\mathfrak{A}$}%
\put(45,0){%
\put(140,-50){\textcolor{gray}{\foc}}%
\put(145,-53){$f_1^\mathfrak{A}$}%
}
\put(90,0){%
\put(140,-50){\textcolor{gray}{\foc}}%
\put(140,-50){\fzc}%
\put(145,-53){$f_0^\mathfrak{A} \cap f_1^\mathfrak{A}$}%
}
}%
\put(0,-22){%
\put(-20,0){
\put(140,-50){\textcolor{black}{\thicklines\circle{6}}}%
\put(145,-53){$a^\mathfrak{A}$}%
}%
\put(25,0){%
\put(140,-50){\fzc}%
\put(140,-50){\textcolor{black}{\thicklines\circle{6}}}%
\put(145,-53){$a^\mathfrak{A} \cap f_0^\mathfrak{A}$}%
}
\put(90,0){%
\put(140,-50){\textcolor{gray}{\foc}}%
\put(140,-50){\textcolor{black}{\thicklines\circle{6}}}%
\put(145,-53){$a^\mathfrak{A} \cap f_1^\mathfrak{A}$}%
}
}%
\end{picture}
\end{center}
\caption{First 4 steps of encoding the full binary tree using
7-saws.}\label{f:s-i-saw}
\end{figure}

Having established a lower bound for $\cBCc{}^1$, we now proceed to do
the same for the larger language $\cBCc$.  Observe that when
constructing a model for an $\conT^1$-formula with one positive
occurrence of $c(\tau)$, in the proof of
Theorem~\ref{thm:one-connected}, we could check the `connectability' of two
$\tau$-points by an (exponentially long) path using a
\PSpace{}-algorithm, because we did not need to keep in memory all the
points on the path. However, if two statements $c(\tau_1)$ and
$c(\tau_2)$ have to be satisfied, then, while connecting two
$\tau_1$-points using a path, one has to check whether the
$\tau_2$-points on that path can be connected by a path, which, in
turn, can contain another $\tau_1$-point, and so on.  The crucial idea
in the proof below is to simulate infinite binary
(\emph{non-transitive}) trees using quasi-saws. Roughly, the
construction is as follows.  We start by representing the root $v_{0}$
of the tree as a point also denoted by $v_{0}$ (see
Fig.~\ref{f:s-i-saw}), which is forced to be connected to an auxiliary
point $w$ by means of some $c(\tau_0)$. On the connecting path from
$v_0$ to $w$ we represent the two successors $v_1$ and $v_2$ of the
root, which are forced to be connected in turn to $w$ by some other
$c(\tau_1)$. On each of the two connecting paths, we again take two
points representing the successors of $v_1$ and $v_2$,
respectively. We treat these four points in the same way as $v_0$,
reusing $c(\tau_0)$, and proceed \emph{ad infinitum}, alternating
between $\tau_0$ and $\tau_1$ when forcing the paths which generate
the required successors. Of course, we also have to pass certain
information from a node to its two successors. Such
information can be propagated along connected regions. Note now that
all points are connected to $w$. To distinguish between the
information we have to pass from distinct nodes of even (respectively,
odd) level to their successors, we have to use \emph{two}
connectedness formulas of the form $c(f_i + a)$, $i=0,1$, in such a
way that the $f_i$ points form initial segments of the paths to $w$
and $a$ contains $w$. The $f_i$-segments are then used locally to pass
information from a node to its successors without conflict. We now
present the reduction in more detail.

\begin{theorem}\label{thm:BRCC-8:ExpTime}
$\Sat( \cBCc , \Regc)$ and $\Sat(
\cBCc , \ConR)$ are \ExpTime{}-hard.
\end{theorem}
\begin{proof}
The proof is by reduction of the following problem. Denote by
$\mathcal{D}_2^f$ the bimodal logic (with $\Box_1$ and $\Box_2$)
determined by  Kripke models based on the full infinite
binary tree $\mathfrak G= (V,R_1,R_2)$ with \emph{functional}
accessibility relations $R_1$ and $R_2$. Consider the \emph{global
consequence relation} $\models^f_2$ defined as follows: $\chi
\models^f_2 \psi$ iff $\mathfrak{K} \models \chi$ implies
$\mathfrak{K} \models \psi$, for every Kripke model $\mathfrak K$
based on $\mathfrak G$. This global consequence relation is \ExpTime-hard, see, e.g.,~\cite{Spaan}.
We construct a $\cBCc$-formula $\Phi(\chi,\psi)$, for any
$\mathcal{D}_2^f$-formulas $\chi$, $\psi$, such that (\textit{i})
$|\Phi(\chi,\psi)|$ is polynomial in $|\chi| + |\psi|$,
(\textit{ii}) if $\Phi(\chi,\psi)$ is satisfiable over $\Regc$ then $\chi \not\models^f_2 \psi$, and (\textit{iii}) if $\chi \not\models^f_2 \psi$ then $\Phi(\chi,\psi)$ is satisfiable over $\ConR$.
While constructing
$\Phi(\chi,\psi)$, we will assume that $\mathfrak A$ is a quasi-saw
model induced by $(W,R)$ and $W_0$ is the set of points of depth $0$
in $(W,R)$.

Let $\textit{sub}(\chi,\psi)$ be the closure under single negation of
the set of subformulas of $\chi$, $\psi$. For each $\varphi \in
\textit{sub}(\chi,\psi)$ we take a fresh variable $q_{\varphi}$, and
for each $\Box_{i}\varphi \in \mathit{sub}(\chi,\psi)$, a pair of
fresh variables $m_{\varphi}^{i,j}$,  $j=0,1$. 
We
also need fresh variables $a$ and $s_j^{i}$, for $j = 0,1$ and $0 \leq i \leq
6$. Let $d = s_{0}^{0}+s_{1}^{0}$.  Intuitively, $d$ simulates the
domain of the binary tree, where $s_{0}^{0}$ and $s_{1}^{0}$ stand for
nodes with even and, respectively, odd distance from the root. Suppose
that the following $\cBCc$-formulas hold in $\mathfrak{A}$
\begin{gather}
\label{eq:non-empty}
\bigl(a = s_0^6\bigr)  \ \land \ \bigl(a = s_1^6\bigr) \quad\land\quad \bigl(a \ne \zero\bigr)
\quad \land \quad c(f_0 + a) \quad \land \quad c(f_1 + a),\\
\label{eq:non-intersec:non-contact}
\bigwedge_{0 \leq k < k' \le 6} \bigl(s_j^k \cdot s_j^{k'} = \zero\bigr) \ \ \ \ \land \ \ \bigwedge_{\begin{subarray}{c}0 \leq k < k' \leq 6\\ |k - k'|>1\end{subarray}}  \hspace*{-1em}\neg C(s_j^k,s_j^{k'}),%
\end{gather}
where $f_j = s_j^0 + s_j^1 + s_j^2 + s_j^3 + s_j^4 + s_j^5$, for  $j=0,1$.
It follows that, for $j = 0,1$, if there is a point $x_0\in (s_j^0)^\mathfrak{A} \cap W_0$ then there is a (not necessarily unique) sequence of points $x_1,x_2,x_3,x_4,x_5$ from the same component of $f_j^\mathfrak{A}$ such that $x_i\in (s_j^i)^\mathfrak{A} \cap W_0$, $1 \leq i \leq 5$. Points $x_2$ and $x_4$ will be used to construct similar sequences for the two successors of the node represented by $x_0$:
if~\eqref{eq:non-empty}--\eqref{eq:non-intersec:non-contact} and
\begin{align}
\label{eq:init}
& s_0^{2i} ~\le~ s_1^0\qquad\text{and}\qquad s_1^{2i} ~\le~ s_0^0,\qquad \text{for } i=1,2,%
\end{align}
hold in $\mathfrak{A}$ and $x_0 \in (s_j^0)^\mathfrak{A} \cap W_0$, then one can recover from $\mathfrak{A}$ the infinite binary tree with the root at $x_0$. The formula
\begin{align}\label{1}
\bigl(q_{\neg\psi} \cdot s^0_0 \ne \zero\bigr) \quad \land\quad  \bigl(d ~\le~ q_\chi\bigr)
\end{align}
ensures then that there is $x_0 \in (s_0^0)^\mathfrak{A} \cap W_0$, the root of the tree, in which
$\neg \psi$ holds,
and $\chi$ holds everywhere in the tree, while the
formulas
\begin{align}\label{2}
d \cdot q_{\neg \varphi}  = d \cdot (-q_{\varphi}),\qquad\qquad
d \cdot q_{\varphi_1 \land \varphi_2}  =  d \cdot  (q_{\varphi_1} \cdot q_{\varphi_2}),
\end{align}
for all $\neg \varphi, \varphi_1\land\varphi_2 \in \textit{sub}(\chi,\psi)$, capture the meaning of the Boolean connectives from $\textit{sub}(\chi,\psi)$ relativized to $d$. The formulas, for all $\Box_{i}\varphi \in \textit{sub}(\chi,\psi)$ and $j= 0,1$,
\begin{align}
\label{5}%
& \neg C(f_j \cdot m_{\varphi}^{i,j},\ f_j \cdot (-m_{\varphi}^{i,j})),\\
\label{4}%
& s^0_j \cdot q_{\Box_i \varphi} = s^0_j \cdot m_{\varphi}^{i,j},\\
\label{4b}
& s_j^{2i} \cdot m_{\varphi}^{i,j} =  s_j^{2i} \cdot q_{\varphi}
%
\end{align}
are used to propagate information regarding $\Box_{i}\varphi$ along the components of $f_j$ using the markers $m_{\varphi}^{i,j}$. 
We define $\Phi(\chi,\psi)$ to be the conjunction of all the above
formulas.  Clearly, $|\Phi(\chi,\psi)|$ is polynomial in $|\chi| +
|\psi|$ and contains only two occurrences of the connectedness
predicate, $c$, in~\eqref{eq:non-empty}.

\bigskip

Suppose that $\Phi(\chi,\psi)$ is satisfied over
$\Regc$; we proceed to show that $\chi \not\models^f_2 \psi$.
By Lemmas~\ref{prop:KripkeSemantics} and~\ref{broom-lemma}, $\Phi(\chi,\psi)$ is satisfied in
a finite quasi-saw model $\mathfrak A$ induced by some $(W,R)$. Denote by $W_0$ the set of points of
depth $0$ in $(W,R)$. Our
aim is to construct, by induction, a Kripke model $\mathfrak K$ based
on the full infinite binary tree $\mathfrak G = (V,R_1,R_2)$ such that
$\mathfrak K \models \chi$ and $\mathfrak K,v_0 \not\models \psi$, for
the root $v_0$ of $\mathfrak G$. The points of $V$ will be
\emph{copies} of some points in $d^{\mathfrak A}\cap W_0$. If $v\in V$ is a copy of $x\in W$, then we write $x =
\varkappa(v)$.

\underline{\emph{Step} $0$.} Take some $x_0 \in
(q_{\neg\psi} \cdot s^0_0)^{\mathfrak A} \cap W_0$. It exists by~\eqref{1}. Take a fresh $v_0$ and let $\varkappa(v_0)=x_0$ and $\mathfrak G^0
=(\{v_0\},\emptyset,\emptyset)$ and set $v_0\models p$ for each
propositional variable $p \in \textit{sub}(\chi,\psi)$ such that
$x_0 \in q_p^{\mathfrak A}$.

\smallskip

\underline{\emph{Step} $n+1$.} Suppose that we have already
constructed $\mathfrak G^n =(V^n,R_1^n,R_2^n)$ and defined a
valuation in $\mathfrak G^n$ for the variables in
$\mathit{sub}(\chi,\psi)$. Let $v$ be a point of minimal co-depth
in $\mathfrak G^n$ which does not have $R_1$- and $R_2$-successors
yet, and let $\varkappa(v)=x \in (s_j^0)^{\mathfrak A}\cap W_0$, for some
$j\in \{0,1\}$.
We have $x \in f_j^{\mathfrak A}$, and so
by~\eqref{eq:non-empty}, there is a finite
path $\pi$ from $x$ to some $w \in (s_j^6)^{\mathfrak A}\cap W_0$
such that all the points on this path belong to $(f_j+a)^{\mathfrak A}$, and so to $\bigcup_{k =
0}^6 (s_j^k)^{\mathfrak A}$. Fix such a path
$\pi$. By the first conjunct
of~\eqref{eq:non-intersec:non-contact}, $x \ne w$ and in view of the second conjunct
of~\eqref{eq:non-intersec:non-contact}, of all
$(s_j^k)^{\mathfrak A}$ on the path $\pi$ the
set $(s_j^0)^{\mathfrak A}$ can only be in contact with (in fact,
externally connected to) $(s_j^1)^{\mathfrak A}$. Take the last
point $y_1 \in (s_j^1)^{\mathfrak A}\cap W_0$
on $\pi$. By a
similar argument, the next point of depth $0$ on $\pi$ can only be
from $(s_j^2)^{\mathfrak A}$. Let $x_1$ be the last point
in $(s_j^2)^{\mathfrak A}\cap W_0$ on $\pi$. In the same way we find the
last point $y_2 \in (s_j^3)^{\mathfrak A}$ of depth $0$ on $\pi$,
and then the last point $x_2 \in (s_j^4)^{\mathfrak A}\cap W_0$
on $\pi$. Let $v_1,v_2$ be fresh copies of $x_1,x_2$,
respectively, i.e., $\varkappa(v_i) = x_i$, $i = 1,2$. Now we set
$\mathfrak G^{n+1} =(V^n \cup \{v_1, v_2 \},R_1^n \cup \{ (v,
v_1) \},R_2^n \cup \{ (v,
v_2) \})$, and
$v_i\models p$ for each propositional variable $p \in
\mathit{sub}(\chi,\psi)$ such that $x_i \in q_p^{\mathfrak A}$, for
$i=1,2$. Note that, by~\eqref{eq:init}, $x_1,x_2\in (s^0_{j\oplus
1})^{\mathfrak A}$, where $\oplus$ denotes addition modulo 2, and we can move on to Step $n+2$.

\smallskip

\underline{\emph{Step} $\omega$.} Finally, we set $V=
\bigcup_{n<\omega} V^n$, $R_1= \bigcup_{n<\omega} R_1^n$,  $R_2=
\bigcup_{n<\omega} R_2^n$, $\mathfrak G =(V,R_1,R_2)$. Clearly,
$\mathfrak G$ is the full binary tree with functional $R_1$ and $R_2$.
The Kripke model $\mathfrak K$ is based on $\mathfrak G$ with the
valuation defined by the inductive procedure above.

It remains to show by induction that, for every $\varphi\in
\mathit{sub}(\chi,\psi)$ and every $v \in V$,
\begin{equation*}
\varkappa(v) \in q_{\varphi}^{\mathfrak A} \qquad \text{iff} \qquad \mathfrak K, v \models \varphi.
\end{equation*}
The basis of induction follows from the definition, and the case of
Boolean connectives from~\eqref{2}. Suppose now that $v$ is the point
considered at Step $n+1$ above. Let $i = 1$ or $2$ and $v_i\in V$ be
such that $(v,v_i)\in R_i$ (it is defined uniquely as $R_i$ is
functional). We have $\varkappa(v) \in (s_j^0)^{\mathfrak A}$, for
either $j = 0$ or $j=1$, and, by~\eqref{eq:init}, $\varkappa(v_i)\in
(s_{j\oplus 1}^0)^{\mathfrak A}$. Then $\varkappa(v) \in
q_{\Box_i\varphi}^{\mathfrak A}$ iff, by~\eqref{4}, $\varkappa(v) \in
(m_{\varphi}^{i,j})^{\mathfrak A}$ iff, by~\eqref{5}, $\varkappa(v_i)
\in (m_{\varphi}^{i,j})^{\mathfrak A}$ iff, by~\eqref{4b},
$\varkappa(v_i) \in q_{\varphi}^{\mathfrak A}$, iff, by IH, $\mathfrak
K, v_i \models \varphi$ iff, by functionality of $R_i$, $\mathfrak K,
v \models \Box_i\varphi$.

\bigskip

\begin{figure}[t]
\setlength{\unitlength}{0.8mm}
\centering{%
\begin{picture}(130,36)(0,-10)
\multiput(0,15)(20,0){7}{\circle{1.5}}
\multiput(10,0)(20,0){6}{\circle{1.5}}
\multiput(9,1.5)(20,0){6}{\vector(-2,3){8}}
\multiput(11,1.5)(20,0){6}{\vector(2,3){8}}
%
%
\put(0,-16){%
\put(-2,35){$y_0^v$}%
\put(18,35){$y_1^v$}%
\put(34,38){\begin{tabular}{c}$y_2^v$\\[-2pt]{\scriptsize ($= y_0^{v_1}$)}\end{tabular}}%
\put(58,35){$y_3^v$}%
\put(74,38){\begin{tabular}{c}$y_4^v$\\[-2pt]{\scriptsize ($= y_0^{v_2}$)}\end{tabular}}%
\put(98,35){$y_5^v$}%
\put(118,35){$w^v$ {\scriptsize $= w$}}%
}%
\put(-5,2){%
\put(13,-8){$z_0^v$}%
\put(33,-8){$z_1^v$}%
\put(53,-8){$z_2^v$}%
\put(73,-8){$z_3^v$}%
\put(93,-8){$z_4^v$}%
\put(113,-8){$z_5^v$}%
}%
\end{picture}
}%
\caption{A 7-saw for $v$.}\label{f:7-saw}
\end{figure}
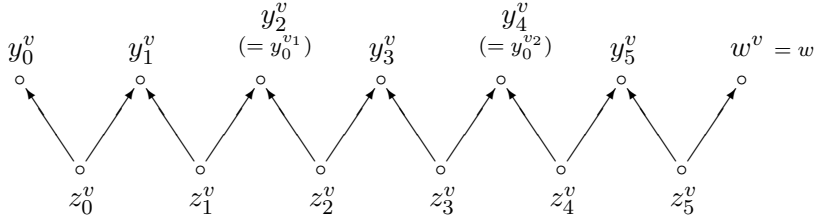

Suppose, finally, that $\chi \not\models^f_2 \psi$; we proceed to
show that $\Phi(\chi,\psi)$ is satisfied over $\ConR$, thus completing
the proof of the theorem.
Let $\mathfrak K$ be a model for $\mathcal{D}_2^f$ based on the full infinite binary
tree $\mathfrak G = (V,R_1,R_2)$ with root $v_0$ such that $\mathfrak
K \models \chi$ and $\mathfrak K,v_0 \not\models \psi$.  We
construct a connected quasi-saw model $\mathfrak A$ satisfying
$\Phi(\chi,\psi)$.
%
The model $\mathfrak A$ will be induced by the quasi-saw $(W,R)$ constructed by induction from (infinitely many copies of)
the $7$-saws shown in Fig.~\ref{f:7-saw}. For each node $v$ of the
infinite binary tree $\mathfrak{G}$, we take a fresh 7-saw $\mathfrak
S^v =(S^v,R^v)$, where $S^v = \{ y_i^v, z_i^v \mid 0 \le i \le
5 \}\cup\{w^v\}$, $z_i^v R^v y_i^v$, for $0 \le i \le
5$, $z_i^v R^v y_{i+1}^v$, for $0 \le i <
5$, and $z_5^v R^v w^v$, and identify the following points
\begin{equation*}
y_2^{v} = y_0^{v_1}, \qquad y_4^{v} = y_0^{v_2}, \qquad w^{v_1} = w^{v_2} = w^{v},
\end{equation*}
if $v_1$ and $v_2$ are the $R_1$- and $R_2$-successors of $v$. We
present our construction in a step-by-step manner (see
Fig.~\ref{f:s-i-saw}).

\smallskip

\underline{\emph{Step} $0$.} We set $W^0 = \{ y_0^{v_0}, w \}$, $R^0 = \emptyset$ and define a
valuation $\cdot^{{\mathfrak A}_0}$ on $W^0$ by taking
\begin{enumerate}[$\bullet$]
\item $(s_0^0)^{{\mathfrak A}_0} = \{ y_0^{v_0} \}$, \quad $(s_1^0)^{{\mathfrak A}_0} = \emptyset$,\hfill\break
 $(s_0^i)^{{\mathfrak A}_0} = (s_1^i)^{{\mathfrak A}_0} = \emptyset$, for $0 < i \leq 5$,\quad and\quad $a^{{\mathfrak A}_0}  = (s_0^6)^{{\mathfrak A}_0} = (s_1^6)^{{\mathfrak A}_0} = \{ w \}$;
%
%
\item $q_{\varphi}^{{\mathfrak A}_0} = \bigl\{ y_0^{v_0} \mid \mathfrak K,v_0 \models \varphi\bigr\}$, for $\varphi \in \mathit{sub}(\chi,\psi)$;
\item $(m^{i,0}_{\varphi})^{{\mathfrak A}_0} = \{ y_0^{v_0} \}$, if $\mathfrak K,v_0 \models \Box_i\varphi$, and $(m^{i,0}_{\varphi})^{{\mathfrak A}_0} = \emptyset$,
otherwise,\quad and
$(m^{i,1}_{\varphi})^{{\mathfrak A}_0} = \emptyset$, for $i=1,2$ and $\Box_i\varphi \in \mathit{sub}(\chi,\psi)$.
\end{enumerate}

\underline{\emph{Step} $n+1$.} Suppose now that $v$ is a node of
minimal co-depth in $\mathfrak G$ such that the constructed
quasi-saw $(W^n,R^n)$ contains $y_0^v$ but does not contain a
copy of $\mathfrak S^v$ for $v$ and let $v_1$ and $v_2$ be the $R_1$- and
$R_2$-successors of $v$. Then we take a fresh 7-saw $\mathfrak S^v$
and `hook' it to $(W^n,R^n)$ by identifying the first point of
$\mathfrak S^v$ with $y_0^v$ and its last point with $w$
(see~Fig.~\ref{f:s-i-saw}):
\begin{align*}
W^{n+1} & =  W^n \cup \{ y_1^v, y_0^{v_1}, y_3^v, y_0^{v_2}, y_5^v\}\cup \{z_i^v \mid 0 \leq i \leq 5 \},\\
\notag{} R^{n+1} & = R^n \cup \{ (z_{2i-1}^v,y_0^{v_i}), (z_{2i}^v,y_0^{v_i}) \mid i = 1,2\} \cup \{ (z_0^v,y_0^v), (z_5^v,w)  \}\\
& \hspace*{2.6em} \cup \{ (z_i^v,y_i^v), (z_{i-1}^v,y_i^v) \mid i = 1,3,5 \},
\end{align*}
and define the valuation $\cdot^{\mathfrak{A}_{n+1}}$ by taking
$p^{\mathfrak{A}_{n+1}} = p^{\mathfrak{A}_n} \cup
p^{\mathfrak{A}_v}$, where $\cdot^{\mathfrak{A}_v}$ is defined for
the new points as follows:
\begin{enumerate}[$\bullet$]
\item $(s_j^0)^{{\mathfrak A}_v} = \{ z_0^v \}$ \quad and \quad $(s_{j\oplus 1}^0)^{{\mathfrak A}_v} = \{ z_{2i-1}^v, y_0^{v_i}, z_{2i}^v \mid i = 1,2  \}$;
\item $(s_j^{2i})^{{\mathfrak A}_v} = \{ z_{2i-1}^v,y_0^{v_i},z_{2i}^v
\}$\quad and\quad $(s_{j\oplus 1}^{2i})^{{\mathfrak A}_v} = \{
 z_0^v \mid y_0^v \in (s^{2i}_{j \oplus
 1})^{\mathfrak{A}_n}\}$,\quad for $i = 1,2$;
\item $(s_j^i)^{{\mathfrak A}_v} = \{ z_{i-1}^v,y_i^v,z_i^v \}$\quad and\quad $(s_{j\oplus 1}^i)^{{\mathfrak A}_v} =
\emptyset$,\quad for $i = 1,3,5$;
\item $a^{\mathfrak{A}_v} = (s_j^6)^{\mathfrak{A}_v} = (s_{j\oplus 1}^6)^{\mathfrak{A}_v} = \{ z_5^v \}$,
%
%
\item $q_{\varphi}^{{\mathfrak A}_v} = \bigl\{ z_0^v \mid \mathfrak K,v \models \varphi\bigr\} \cup \bigl\{ z_{2i-1}^v, y_0^{v_i}, z_{2i}^v \mid \mathfrak K,v_i \models \varphi, i = 1, 2\bigr\}$, for $\varphi \in \mathit{sub}(\chi,\psi)$;
\item $(m^{i,j}_{\varphi})^{{\mathfrak A}_v} = \bigcup_{i = 0}^6 (s_j^i)^{{\mathfrak
A}_v}$,
if $\mathfrak K,v \models \Box_i\varphi$, and $(m^{i,j}_{\varphi})^{{\mathfrak A}_v} = \emptyset$, otherwise, \quad
and\\[6pt]
$(m^{i,j\oplus 1}_{\varphi})^{{\mathfrak A}_v} = \bigl\{z_0^v \mid
y_0^v \in (m^{i,j\oplus 1}_{\varphi})^{{\mathfrak A}_n} \bigr\} \cup
\bigl\{ z_{2i-1}^v,y_0^{v_i},z_{2i}^v \mid \mathfrak K,v_i \models
\Box_i\varphi\bigr\}$,\\[6pt] for $\Box_i\varphi \in
\mathit{sub}(\chi,\psi)$, $i=1,2$;
\end{enumerate}
where $j = 0$ if $v$ is of \emph{even} co-depth in $\mathfrak G$ and
$j = 1$ otherwise, and $\oplus$ denotes addition modulo 2.

\smallskip

\underline{\emph{Step} $\omega$.} Finally, we set $W=
\bigcup_{n<\omega} W^n$, $R= \bigcup_{n<\omega} R^n$ and $\cdot^\mathfrak{A} = \bigcup_{n < \omega}
\cdot^{\mathfrak{A}_n}$.

Note that $y^v_1$ and $y^v_3$ are required to make the set of points
in $f_j^\mathfrak{A}$ representing a node $v$ of $\mathfrak{G}$
disconnected from the subset of $f_j^\mathfrak{A}$ representing
another node $v'$ of $\mathfrak{G}$ and thus satisfy~\eqref{5};
$y^v_5$ are required to satisfy the last conjunct
of~\eqref{eq:non-intersec:non-contact}. We leave the remaining
details to the reader.
\end{proof}

\begin{corollary}
$\Sat( \cBCc, \Regc)$  and $\Sat( \cBCc, \ConR)$ are \ExpTime{}-complete.
\label{cor:BRCC-8:ExpTimecomplete}
\end{corollary}
\begin{proof}
Follows from Theorems~\ref{theorem:S4uc:ExpTime} and~\ref{thm:BRCC-8:ExpTime}.
\end{proof}

The argument of Theorem~\ref{thm:BRCC-8:ExpTime} goes through
unproblematically when we restrict attention to the spaces $\R^n$ for
$n \geq 3$, thus showing that $\Sat( \cBCc , \RegC(\R^n))$ is
\ExpTime{}-hard for these values of $n$.  If $n=2$, however, this
approach fails.  The difficulty is that the construction of the model
$\fA$ of $\Phi(\chi,\psi)$ from the Kripke structure $\mathfrak K$
will result in sets $(s_j^k)^\fA$ that have infinitely many components
lying in a bounded region of the plane. These infinitely many
components will give rise to accumulation points, meaning that we can
no longer guarantee the truth of the formulas $\neg C(s_j^k,
s_j^{k'})$ of~\eqref{eq:non-intersec:non-contact} and~\eqref{5}.
Fortunately, the proof can be rescued, either by means of the finite
model property for $\models_2^f$, or, alternatively, via the
following explicit construction based on polynomial-space alternating
Turing machines. We remind the reader, in this connection, that
$\textsc{APSpace} = \textsc{ExpTime}$ (see
e.g.,~\cite{Kozen06}).
\begin{theorem}\label{thm:BRCC-8:ExpTime:bis}
The problems $\Sat( \cBCc , \RegC(\R^n))$ for all $n \geq 2$
are \ExpTime{}-hard.
\end{theorem}
\begin{proof}
Let $M$ be an alternating Turing machine that uses $\leq p(n)$ cells
on each input of length $n$, for some polynomial $p(\cdot)$. Let $q_Y$
and $q_N$ be the accepting and rejecting states of $M$, respectively
(they have no transitions from them). Without loss of generality we
assume that every branch of the computation tree of $M$ is of
\emph{finite length} and its final state is either $q_Y$ or $q_N$
(indeed, for if it is not the case we can augment $M$ with a subroutine
that at each step of $M$ increases the current step number written on
an auxiliary tape and makes $M$ go into the rejecting state $q_N$
whenever the step number reaches the total number of configurations of
$M$; as $M$ uses a polynomial working tape, we need only a polynomial
number of cells on that auxiliary tape to represent the maximum number
of configurations of $M$). All states of $M$ (except $q_Y$ and $q_N$)
are divided into existential and universal states. We assume without
loss of generality that there are exactly two transitions from each
state, i.e., for $q\in Q\setminus\{q_Y,q_N\}$, $a\in\Sigma$ and $j =
1,2$, we have $(q,a)\to_j (q',a',d)$, where $d \in \{-1,0,+1\}$.  We
assume that $M$ never goes beyond special markers at the beginning and
the end of the tape. We denote $p(n)$ by $s$, and
employ propositional variables with the following intuitive meanings:
\begin{enumerate}[$\bullet$]
\item $H_{q,i}$, for $1 \leq i \leq s$ and $q\in Q$, to say that the head
is scanning
the $i$th cell and the state is $q$,
\item $S_{a,i}$, for $1 \leq i \leq s$ and $a\in \Sigma$, to say that the $i$th cell contains $a$,
\item $A$ to say that the computation is accepting.
\end{enumerate}
Let $\chi_M$ be the conjunction of the following formulas, for all $i$, $1 \leq i \leq s$,
\begin{align*}
& H_{q,i} \land S_{a,i} \to \Box_j (H_{q', i+d} \land S_{a',i}), \qquad&&\text{for } (q,a)\to_j (q',a',d), \ j = 1,2, \ d\in \{-1,0,+1\},\\
& H_{q,i} \land S_{a,k} \to \Box_j S_{a,k}, \qquad&&\text{for each } k \ne i \text{ and }  j = 1,2,\\
& H_{q_Y,i} \to A,\\
& H_{q,i} \land (\Diamond_1 A \land \Diamond_2 A) \to A,&&\text{for each universal state }q\in Q,\\
& H_{q,i} \land (\Diamond_1 A \lor \Diamond_2 A) \to A,&&\text{for each existential state }q\in Q.
\end{align*}
Let $\vec{a} = a^1,\dots,a^n$ be an input. For convenience, let $a^i$ be blank, for each $n < i \leq s$. Write
\begin{equation*}
\psi_{\vec{a}} \ \ = \ \ (H_{q_0,1} \land S_{a^1,1} \land \dots \land S_{a^s,s}) \to A,
\end{equation*}
where $q_0$ is the initial state of $M$. Clearly, $M$ accepts
$\vec{a}$ iff $\chi_M\models_2^f\psi_{\vec{a}}$. Moreover,
since we assume
Turing machines to terminate each branch of a
computation in $q_Y$ or $q_N$ after a finite number of steps,
it follows that $M$
accepts $\vec{a}$ if and only if $A$ is satisfied at the root of every finite
binary tree such that
\begin{enumerate}[(b)]
\item[(a)] $\chi_M$ is satisfied in every point of the tree,
\item[(b)] every point satisfying $H_{q,i}$,
for $q\in Q\setminus\{q_Y,q_N\}$ and $1 \leq i \leq s$, has a pair of
successors (in particular: is not a leaf node),
\item[(c)] $H_{q_0,1} \land S_{a^1,1} \land \dots\land S_{a^s,s}$ is satisfied at the root.
\end{enumerate}
Denote by $\Phi'(\chi_M,\psi_{\vec{a}})$ the conjunction of formulas~\eqref{eq:non-empty}--\eqref{eq:non-intersec:non-contact} and \eqref{1}--\eqref{4b}, constructed for $\chi_M$ and $\psi_{\vec{a}}$ in the proof of Theorem~\ref{thm:BRCC-8:ExpTime}, together with
the following replacement of~\eqref{eq:init}, for $1 \leq k \leq s$ and $q'\in Q\setminus\{q_Y,q_N\}$:
\begin{equation}
\label{eq:init:prime}
q_{H_{q',k}} \cdot s_0^{2i} ~\le~ s_1^0\qquad\text{and}\qquad q_{H_{q',k}} \cdot s_1^{2i} ~\le~ s_0^0,\qquad \text{for } i=1,2.%
\end{equation}
%

If $\Phi'(\chi_M,\psi_{\vec{a}})$ is satisfied in a model over $\RegC(\R^n)$, for $n \geq 2$, then it follows from the proof of Theorem~\ref{thm:BRCC-8:ExpTime} that $\chi_M\not\models_2^f\psi_{\vec{a}}$, and thus $M$ does not accept $\vec{a}$. Conversely, if $M$ does not accept $\vec{a}$, then there is a finite binary tree satisfying (a)--(c); we then construct a model of $\Phi'(\chi_M,\psi_{\vec{a}})$ over $\RegC(\R^2)$ by representing nodes of the finite binary tree together with the `sink' node $w$ (see Fig.~\ref{f:s-i-saw}) by rectangles of decreasing dimensions.
\end{proof}

Having established a lower bound for $\cBCc{}$, we now proceed to do
the same for the larger language $\cBCcc$.

\begin{theorem}\label{nexptime}
$\Sat(\cBCcc,\Regc)$ is \NExpTime{}-hard.
In fact,
$\Sat(
\cBCcc , \ConR)$ and $\Sat( \cBCcc , \RegC(\R^n))$ for all $n \geq 2$
are \NExpTime{}-hard.
\end{theorem}
\begin{proof}
The proof is by reduction of the \NExpTime-complete $2^d \times 2^d$
\emph{tiling problem} \cite{EmdeBoas97}: Given $d < \omega$, a finite set $\mathcal{T}$
of \emph{tile types}---i.e., 4-tuples of colours $T =
(\textit{left}(T),\textit{top}(T),\linebreak\textit{right}(T),
\textit{bot}(T))$---and a $T_0\in \mathcal{T}$, decide whether
$\mathcal{T}$ can tile the $2^d\times 2^d$ grid in such a way that
$T_0$ is placed onto $(0,0)$. In other words, the problem is to decide
whether there is a function $f$ from $\{(n,m) \mid n,m<2^d\}$ to
$\mathcal{T}$ such that $\textit{top}(f(n,m) ) =
\textit{bot}(f(n,m+1))$, for all $n, m+1<2^d$,
$\textit{right}(f(n,m)) = \textit{left}(f(n+1,m))$, for all $n+1,m<2^d$, and $f(0,0)=T_0$.
We construct a \cBCcc{}-formula $\Theta(\mathcal{T},T_0,d)$ such that
(\textit{i}) $|\Theta(\mathcal{T},T_0,d)|$ is polynomial in $|\mathcal{T}|$ and
$d$, (\textit{ii}) if $\Theta(\mathcal{T},T_0,d)$ is satisfiable over $\Regc$ then
$\mathcal{T}$ tiles the $2^d\times 2^d$ grid, with $T_0$ being
placed onto $(0,0)$, and (\textit{iii}) if
$\mathcal{T}$ tiles the $2^d\times 2^d$ grid, with $T_0$ being
placed onto $(0,0)$, then $\Theta(\mathcal{T},T_0,d)$ is satisfiable over $\RegC(\R^n)$, $n \geq 2$. While constructing the formula,
we will assume that $\mathfrak A$ is a quasi-saw model induced by
$(W,R)$ and $W_0$ is the set of points of
depth $0$ in $(W,R)$.

We partition all points
of $W_0$ with the help of a pair of variable
triples $H^0,H^1,H^2$ and $V^0,V^1,V^2$. Suppose that
the formulas, for $0 \leq \ell \leq 2$,
\begin{align}\label{eq:tiling:B}
& H^0 + H^1 + H^2 = \one,\\
\label{eq:tiling:Ba}
& H^0 \cdot H^1 = \zero, \qquad H^1 \cdot H^2 = \zero,\qquad H^2 \cdot H^0 = \zero,
\end{align}
and their $V$-counterparts hold in $\mathfrak{A}$.
Then every point in $W_0$ is in exactly
one of the $(H^\ell)^{\mathfrak A}$, $0 \leq \ell \leq 2$, and exactly one of the
$(V^\ell)^{\mathfrak A}$, $0 \leq \ell \leq 2$ (these variables play the same role as the $B^\ell$ in the proof of Theorem~\ref{pspace-lower}).

To encode coordinates of the tiles in binary, we take a pair of
variables $X_j$ and $Y_j$, for each $j$, $d \geq j \geq 1$. For $n
< 2^d$, let $n_X$ be the $\cB$-term $X'_{d} \cdot X'_{d-1} \cdot \
\cdots \ \cdot X'_1$, where $X'_j = X_j$ if the $j$th bit in the
binary representation of $n$ is $1$, and $X'_j = -X_j$ otherwise.
For a point $u\in W_0$, we denote by $X(u)$ the binary $d$-bit
number $n$, called the \emph{$X$-value} of $u$, such that $u\in
n_X^\mathfrak{A}$ (note that the $X$-value is defined uniquely for the points in $W_0$); the $j$th bit of $X(u)$ is denoted by $X_j(u)$.
The term $m_Y$, the \emph{$Y$-value} $Y(u)$ of $u$ and its $j$th bit
$Y_j(u)$ are defined analogously. For a point $u\in W_0$ we
write $\coor{u}$ for $(X(u),Y(u))$. We will use the variables $X_i$
and $Y_j$ to generate the $2^d \times 2^d$ grid, which consists of
pairs $(n_X,m_Y)$, for $n,m < 2^d$. Consider the following
formulas, for $0 \leq \ell  \leq 2$,
\begin{align}\label{eq:tiling:sameB}
& \neg C(X_k \cdot H^\ell, \ (-X_k) \cdot H^\ell), && d \geq k \geq 1,\\ %
\label{eq:tiling:counter:upper:1}
& \neg C(X_j \cdot (-X_k) \cdot H^\ell, \ (-X_j) \cdot H^{\ell \oplus 1}), && d \geq j > k \geq 1,\\
\label{eq:tiling:counter:upper:0}
& \neg C((-X_j) \cdot (-X_k) \cdot H^\ell, \ X_j \cdot H^{\ell \oplus 1}), && d \geq j > k \geq 1, \\
\label{eq:tiling:counter:1}
& \neg C((-X_k) \cdot X_{k-1} \cdot \ \cdots \ \cdot X_1 \cdot H^\ell, \ \ (-X_k) \cdot H^{\ell \oplus 1}), && d \geq k >  1,\\
\label{eq:tiling:counter:lower}
& \neg C((-X_k) \cdot X_{k-1} \cdot \ \cdots \ \cdot X_1 \cdot H^\ell, \ \ X_i \cdot H^{\ell \oplus 1}),  && d \geq k > i \geq 1,\\
\label{eq:tiling:counter:end:2d}
& \neg C(X_d \cdot \ \cdots \ \cdot X_1 \cdot H^\ell,\ H^{\ell\oplus 1}),
\end{align}
where $\oplus$ denotes addition modulo 3. Suppose that $\mathfrak{A}$ satisfies~\eqref{eq:tiling:B}--\eqref{eq:tiling:counter:end:2d}.
If $u,v\in W_0$ and $zRu$ and $zRv$, for some $z\in
W$, then (cf.~\eqref{eq:easy1} and~\eqref{eq:easy2})
\begin{enumerate}[$\bullet$]
\item $X(v) = X(u)$ whenever $u,v\in (H^\ell)^\mathfrak{A}$, for $0 \leq \ell \leq 2$,
\item $X(v) = X(u) + 1 < 2^d$ whenever $u\in
(H^\ell)^\mathfrak{A}$ and $v\in (H^{\ell\oplus
1})^\mathfrak{A}$, for $0 \leq \ell \leq 2$,
\item $X(u) = X(v) + 1 < 2^d$ whenever $v\in
(H^\ell)^\mathfrak{A}$ and $u\in (H^{\ell\oplus
1})^\mathfrak{A}$, for $0 \leq \ell \leq 2$.
\end{enumerate}
Indeed, if $u,v\in (H^\ell)^\mathfrak{A}$ then, by~\eqref{eq:tiling:sameB}, $X(u) = X(v)$. If $u\in (H^\ell)^\mathfrak{A}$ and $v\in (H^{\ell\oplus 1})^\mathfrak{A}$ then, by~\eqref{eq:tiling:counter:end:2d}, $X(u) < 2^d - 1$. Let $k$ be the minimal number such that $u\in (-X_k)^\mathfrak{A}$, i.e., $u\in ((-X_k) \cdot X_{k-1} \cdot \ \cdots \ \cdot X_1 \cdot H^\ell)^\mathfrak{A}$. Then, by~\eqref{eq:tiling:counter:1}, $v\in X_k^\mathfrak{A}$ and, by~\eqref{eq:tiling:counter:lower}, $v\in (-X_i)^\mathfrak{A}$, for all $i$, $k > i \geq 1$. By~\eqref{eq:tiling:counter:upper:1} and~\eqref{eq:tiling:counter:upper:0}, $v\in (-X_j)^\mathfrak{A}$ iff $u\in (X_j)^\mathfrak{A}$, for all $d \geq j > k$. It follows that $X(v) = X(u) + 1$. The case of  $v\in (H^\ell)^\mathfrak{A}$ and $u\in (H^{\ell\oplus 1})^\mathfrak{A}$ is similar.

Suppose now that~\eqref{eq:tiling:B}--\eqref{eq:tiling:counter:end:2d} with their the $Y$-counterparts ($X_i$ and $H^\ell$ replaced by $Y_i$ and $V^\ell$, respectively)
hold in $\mathfrak{A}$. It
follows that for every $u,v\in W_0$ with $zRu$ and $zRv$, for some $z\in W$, we have
\begin{equation}\label{eq:tiling:9-neighb}
|X(u) - X(v)|\leq 1 \qquad\text{and}\qquad |Y(u) - Y(v)|\leq 1,
\end{equation}
which means that every point of depth 0 may be surrounded
by all of its 8-neighbours (in the sense that they have a common predecessor of depth 1). We are, however, interested only in the 4-neighbours since only the 4-neighbours restrict the tile that can be assigned to the point. Formally, given a pair $(n,m)$ in the $2^d\times 2^d$ grid, denote by $\neighb{n,m}$ the set that
consists of $(n,m)$ and its (at most four) neighbours in the grid, i.e., $(n-1,m)$, $(n+1,m)$, $(n,m-1)$, $(n,m+1)$; see Fig.~\ref{fig:neighbours}.
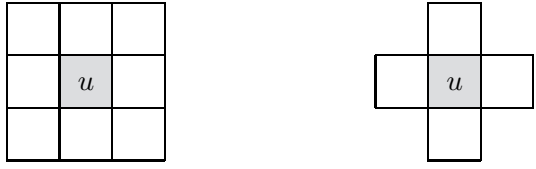
\begin{figure}[t]
\setlength{\unitlength}{0.7mm}
\centering{%
\begin{picture}(100,30)
\put(10,10){\textcolor{Gray!30}{\rule{7mm}{7mm}}}
\multiput(0,0)(10,0){4}{\line(0,1){30}}%
\multiput(0,0)(0,10){4}{\line(1,0){30}}%
\put(13.5,13.5){$u$}
\put(70,0){%
\put(10,10){\textcolor{Gray!30}{\rule{7mm}{7mm}}}
\multiput(0,10)(30,0){2}{\line(0,1){10}}%
\multiput(10,0)(10,0){2}{\line(0,1){30}}%
\multiput(0,10)(0,10){2}{\line(1,0){30}}%
\multiput(10,0)(0,30){2}{\line(1,0){10}}%
\put(13.5,13.5){$u$}
}%
\end{picture}
}%
\caption{8-neighbours vs.\ 4-neighbours.}\label{fig:neighbours}
\end{figure}
Let $G$ be a fresh variable, which will represent the points of the grid. Suppose now that in addition the
formulas
\begin{equation}\label{eq:tiling:cross}
\neg C(G \cdot X_1 \cdot Y_1, \ G \cdot (-X_1) \cdot (-Y_1)),\qquad \neg C(G \cdot (-X_1) \cdot Y_1, \ G \cdot X_1 \cdot (-Y_1))
\end{equation}
hold in $\mathfrak{A}$. Then, by~\eqref{eq:tiling:cross},
if $u,v\in G^\mathfrak{A}\cap W_0$ and $zRu$ and $zRv$, for some $z\in
W$, then either $X_1(u) = X_1(v)$ or $Y_1(u) = Y_1(v)$, and thus, by~\eqref{eq:tiling:9-neighb}, $\coor{u} \in \neighb{\coor{v}}$ and $\coor{v} \in \neighb{\coor{u}}$.

Suppose now that the following formulas are true in $\mathfrak
A$ as well:
\begin{multline}\label{eq:tiling:perimeter}
G \cdot 0_X \cdot 0_Y \ne \zero, \qquad
G \cdot (2^d-1)_X \cdot (2^d -1)_Y \ne \zero,\\
 c(G \cdot (0_X + (2^d -1)_Y)), \qquad c(G \cdot ((2^d -1)_X + 0_Y)).
\end{multline}
These constraints guarantee that in the connected set $(G \cdot (0_X + (2^d
-1)_Y))^{\mathfrak A}$ there are points $u_{(0,n)}$ and
$u_{(n,2^d-1)}$, for $n <2^d$, such that $\coor{u_{(0,n)}} =
(0,n)$ and $\coor{u_{(n,2^d-1)}} = (n,2^d-1)$. Similarly for the
connected set $(G \cdot (2^d -1)_X + 0_Y)^{\mathfrak A}$. This gives us the
border of the $2^d \times 2^d$ grid we are after (see Fig.~\ref{f:tiling:two} (\textit{a})).
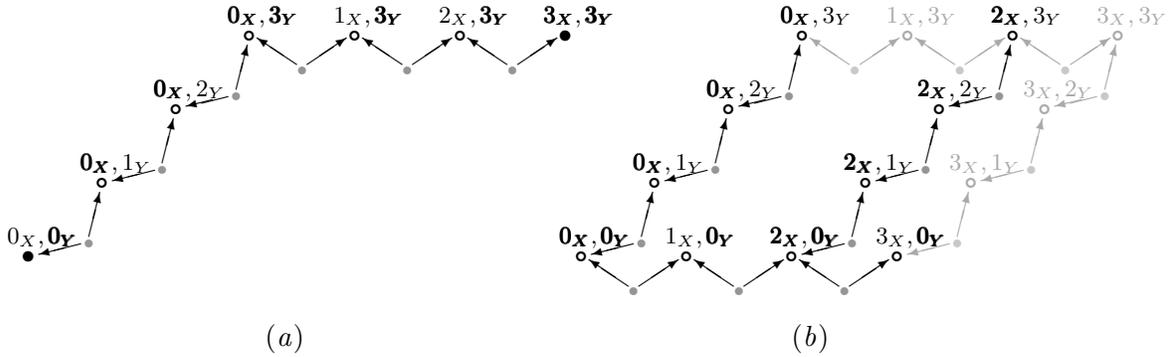
\begin{figure}[b]
\setlength{\unitlength}{0.7mm}
\centering{%
\begin{picture}(250,70)(-10,0)
\put(-5,10){
%
\multiput(42,42)(14,14){1}{%
\multiput(10,0.5)(20,0){3}{\textcolor{Gray}{\circle*{1.5}}}
\multiput(9,1.5)(20,0){3}{\vector(-3,2){7.5}}
\multiput(11,1.5)(20,0){3}{\vector(3,2){7.5}}
}%
\multiput(0,0)(14,14){3}{%
\multiput(0,0)(20,0){1}{%
\put(11.5,9.5){\textcolor{Gray}{\circle*{1.5}}}%
\put(11.5,10.8){\vector(1,4){2.1}}%
\put(10,9.5){\vector(-4,-1){8}}%
}%
}%
\thicklines
\put(0,7){\circle*{2}}%
\put(102,49){\circle*{2}}%
\multiput(14,21)(14,14){3}{\circle{1.7}}%
\multiput(62,49)(20,0){2}{\circle{1.7}}%
%
\footnotesize
\put(-4,9.5){%
\put(0,0){$0_X,\pmb{0_Y}$}%
}%
\put(10,23.5){%
\put(0,0){$\pmb{0_X},1_Y$}%
}%
\put(24,37.5){%
\put(0,0){$\pmb{0_X},2_Y$}%
}%
\put(38,51.5){%
\put(0,0){$\pmb{0_X},\pmb{3_Y}$}%
\put(20,0){$1_X,\pmb{3_Y}$}%
\put(40,0){$2_X,\pmb{3_Y}$}%
\put(60,0){$\pmb{3_X},\pmb{3_Y}$}%
}%
}%
%
\put(100,10){%
\multiput(0,0)(14,14){1}{%
\multiput(10,0.5)(20,0){3}{\textcolor{Gray}{\circle*{1.5}}}
\multiput(9,1.5)(20,0){3}{\vector(-3,2){7.5}}
\multiput(11,1.5)(20,0){3}{\vector(3,2){7.5}}
}%
\multiput(42,42)(14,14){1}{\textcolor{Gray!80}{%
\multiput(10,0.5)(20,0){3}{\textcolor{Gray!50}{\circle*{1.5}}}
\multiput(9,1.5)(20,0){3}{\vector(-3,2){7.5}}
\multiput(11,1.5)(20,0){3}{\vector(3,2){7.5}}}
}%
\multiput(0,0)(14,14){3}{%
\multiput(0,0)(40,0){2}{%
\put(11.5,9.5){\textcolor{Gray}{\circle*{1.5}}}%
\put(11.5,10.8){\vector(1,4){2.1}}%
\put(10,9.5){\vector(-4,-1){8}}%
}%
}%
\multiput(0,0)(14,14){3}{%
\multiput(60,0)(40,0){1}{\textcolor{Gray!80}{%
\put(11.5,9.5){\textcolor{Gray!50}{\circle*{1.5}}}%
\put(11.5,10.8){\vector(1,4){2.1}}%
\put(10,9.5){\vector(-4,-1){8}}}%
}%
}%
\thicklines
\multiput(0,7)(20,0){4}{\circle{1.7}}%
\multiput(0,0)(14,14){3}{%
\multiput(14,21)(40,0){2}{\circle{1.7}}
}%
\multiput(62,49)(40,0){2}{\textcolor{Gray!80}{\circle{1.7}}}%
\multiput(74,21)(14,14){2}{\textcolor{Gray!80}{\circle{1.7}}}%
\footnotesize
\put(-4,9.5){%
\put(0,0){$\pmb{0_X},\pmb{0_Y}$}%
\put(20,0){$1_X,\pmb{0_Y}$}%
\put(40,0){$\pmb{2_X},\pmb{0_Y}$}%
\put(60,0){$3_X,\pmb{0_Y}$}%
}%
\put(10,23.5){%
\put(0,0){$\pmb{0_X},1_Y$}%
\put(40,0){$\pmb{2_X},1_Y$}%
\put(60,0){\textcolor{Gray!80}{$3_X,1_Y$}}%
}%
\put(24,37.5){%
\put(0,0){$\pmb{0_X},2_Y$}%
\put(40,0){$\pmb{2_X},2_Y$}%
\put(60,0){\textcolor{Gray!80}{$3_X,2_Y$}}%
}%
\put(38,51.5){%
\put(0,0){$\pmb{0_X},3_Y$}%
\put(20,0){\textcolor{Gray!80}{$1_X,3_Y$}}%
\put(40,0){$\pmb{2_X},3_Y$}%
\put(60,0){\textcolor{Gray!80}{$3_X,3_Y$}}%
}%
}%
\put(40,0){(\textit{a})}%
\put(140,0){(\textit{b})}%
\end{picture}
}%
\caption{Satisfying (\textit{a}) $c(G\cdot(0_X + (2^d -1)_Y))$ and (\textit{b}) $c(G \cdot ((-X_1) + 0_Y))$, for $d= 2$, in a quasi-saw model.}\label{f:tiling:two}
\end{figure}
And the
constraints
\begin{equation}
c(G \cdot ((-X_1) + 0_Y)),\qquad
c(G \cdot (X_1 + 0_Y)),\qquad c(G \cdot (0_X + (-Y_1))),\qquad c(G \cdot (0_X + Y_1))
\end{equation}
ensure that we can find inner points of the grid (see Fig.~\ref{f:tiling:two} (\textit{b})).

It is to be noted,
however, that in general $u \ne v$ even if $\coor{u}=\coor{v}$. In
other words, the constructed points do not necessarily form a proper
$2^d \times 2^d$ grid. Let
\begin{equation*}
\mathbf{b} = \bigl(X_1 \cdot (-Y_1)\bigr) + \bigl((-X_1) \cdot
Y_1\bigr)\qquad\text{and}\qquad\mathbf{w} = \bigl((-X_1) \cdot
(-Y_1)\bigr) + \bigl(X_1 \cdot Y_1\bigr).
\end{equation*}
Points in $\mathbf{b}^{\mathfrak A}$ and $\mathbf{w}^{\mathfrak A}$
can be thought of as \emph{black} and \emph{white} squares of a
chess board. Observe that if $u,v\in
\mathbf{b}^{\mathfrak A}\cap W_0$ and $\coor{u} \ne \coor{v}$ then
$u$ and $v$ cannot belong to the same component of
$\mathbf{b}^{\mathfrak A}$. Thus, there are at least $2^{d-1}$
components in both $\mathbf{b}^{\mathfrak A}$ and
$\mathbf{w}^{\mathfrak A}$. Our next pair of constraints
\begin{equation}
c^{\leq 2^{d-1}}(\mathbf{b})\qquad\text{and}\qquad c^{\leq 2^{d-1}}(\mathbf{w})
\end{equation}
says that $\mathbf{b}^{\mathfrak A}$ and $\mathbf{w}^{\mathfrak A}$
have precisely $2^{d-1}$ components. In particular,
if $u,v\in W_0$ belong to the same component of
$\mathbf{b}^{\mathfrak A}$ then $\coor{u} = \coor{v}$. This gives
a proper $2^d \times 2^d$ grid on which we encode the tiling
conditions. The formulas
\begin{align}
&\textstyle\sum_{T_k\in\mathcal{T}} T_k = G,\\
& T_{k_1} \cdot T_{k_2} = \zero, \quad \text{ for  } T_{k_1},T_{k_2}\in\mathcal{T}, k_1\ne k_2,
\end{align}
say that every point in $G^\mathfrak{A}\cap W_0$ is covered by precisely one tile and
\begin{equation}
\neg C(H^{\ell} \cdot V^{\ell'} \cdot T_{k_1}, \ H^{\ell} \cdot
V^{\ell'} \cdot T_{k_2}),\qquad  \text{ for }  T_{k_1},T_{k_2}\in\mathcal{T}, k_1\ne k_2,
\end{equation}
for $0 \leq \ell,\ell'  \leq 2$, says
that all points in the same component of $(H^\ell \cdot
V^{\ell'})^{\mathfrak A}$ are covered by the same tile. That the
colours of adjacent tiles match is ensured by
\begin{align}
&\neg C(H^\ell \cdot T_{k_1}, \ H^{\ell \oplus 1} \cdot  T_{k_2}), && %
\text{ for } T_{k_1},T_{k_2}\in\mathcal{T} \text{ with } \textit{right}(T_{k_1}) \ne
\textit{left}(T_{k_2}),\\
&\neg C(V^\ell \cdot T_{k_1}, \ V^{\ell \oplus 1} \cdot  T_{k_2}), && %
\text{ for } T_{k_1},T_{k_2}\in\mathcal{T} \text{ with } \textit{top}(T_{k_1}) \ne
\textit{bot}(T_{k_2}),
\end{align}
for $0 \leq \ell \leq 2$. Finally, we have to say that $(0,0)$ is covered with $T_0$:
\begin{equation}
0_X\cdot 0_Y \leq T_0.
\end{equation}
One can check that the conjunction $\Theta(\mathcal{T},T_0,d)$ of
these \cBCcc{}-formulas is as required. For instance, a quasi-saw
model $\mathfrak{A}$ satisfying $\Theta(\mathcal{T},T_0,d)$ (if a
tiling exists) is shown in Fig.~\ref{f:tiling-model}~(\textit{a}): all points
belong to $G^\mathfrak{A}$, solid white points to
$\mathbf{w}^\mathfrak{A}$, solid black to $\mathbf{b}^\mathfrak{A}$
and grey to both of the sets. Note that each point of
depth 0 has at most 4 predecessors (of depth 1) and the same number of
neighbours (points of depth 0 that share a common point of depth 1).

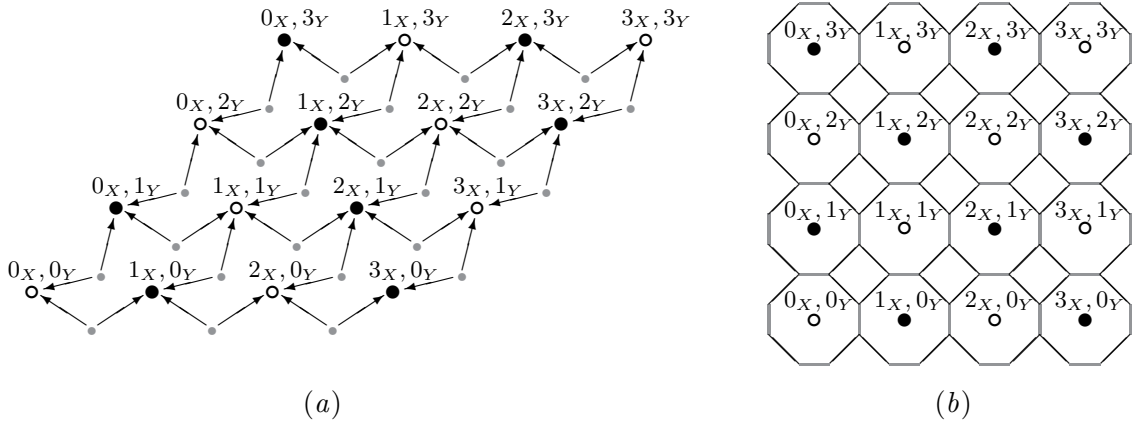
\begin{figure}[t]
\setlength{\unitlength}{0.8mm}
\centering{%
\begin{picture}(250,73)(-10,-3)
\put(-5,10){%
\multiput(0,0)(14,14){4}{%
\multiput(10,0.5)(20,0){3}{\textcolor{Gray}{\circle*{1.5}}}
\multiput(9,1.5)(20,0){3}{\vector(-3,2){7.5}}
\multiput(11,1.5)(20,0){3}{\vector(3,2){7.5}}
}%
\multiput(0,0)(14,14){3}{%
\multiput(0,0)(20,0){4}{%
\put(11.5,9.5){\textcolor{Gray}{\circle*{1.5}}}%
\put(11.5,10.8){\vector(1,4){2.1}}%
\put(10,9.5){\vector(-4,-1){8}}%
}%
}%
\thicklines
\multiput(0,0)(28,28){2}{%
\multiput(0,7)(40,0){2}{\circle{1.7}}
\multiput(20,7)(40,0){2}{\circle*{2}}
\multiput(34,21)(40,0){2}{\circle{1.7}}
\multiput(14,21)(40,0){2}{\circle*{2}}
}%
\footnotesize
\put(-4,9.5){%
\put(0,0){$0_X,0_Y$}%
\put(20,0){$1_X,0_Y$}%
\put(40,0){$2_X,0_Y$}%
\put(60,0){$3_X,0_Y$}%
}%
\put(10,23.5){%
\put(0,0){$0_X,1_Y$}%
\put(20,0){$1_X,1_Y$}%
\put(40,0){$2_X,1_Y$}%
\put(60,0){$3_X,1_Y$}%
}%
\put(24,37.5){%
\put(0,0){$0_X,2_Y$}%
\put(20,0){$1_X,2_Y$}%
\put(40,0){$2_X,2_Y$}%
\put(60,0){$3_X,2_Y$}%
}%
\put(38,51.5){%
\put(0,0){$0_X,3_Y$}%
\put(20,0){$1_X,3_Y$}%
\put(40,0){$2_X,3_Y$}%
\put(60,0){$3_X,3_Y$}%
}%
}%
\put(125,5){%
\thicklines
\multiput(0,0)(0,15){4}{%
\multiput(-7.5,5)(15,0){5}{\textcolor{Gray}{\thicklines\line(0,1){5}}}
\multiput(-7.5,10)(15,0){4}{\textcolor{black}{\thinlines\line(1,1){5}}}
\multiput(-2.5,15)(15,0){4}{\textcolor{Gray}{\thicklines\line(1,0){5}}}
\multiput(2.5,15)(15,0){4}{\textcolor{black}{\thinlines\line(1,-1){5}}}
\multiput(7.5,5)(15,0){4}{\textcolor{black}{\thinlines\line(-1,-1){5}}}
\multiput(-2.5,0)(15,0){4}{\textcolor{black}{\thinlines\line(-1,1){5}}}
}%
\multiput(-2.5,0)(15,0){4}{\textcolor{Gray}{\thicklines\line(1,0){5}}}
\multiput(0,0)(0,30){2}{%
\multiput(0,7.5)(30,0){2}{\circle{1.7}}
\multiput(15,7.5)(30,0){2}{\circle*{2}}
\multiput(15,22.7)(30,0){2}{\circle{1.7}}
\multiput(0,22.5)(30,0){2}{\circle*{2}}
}%
\footnotesize\put(-1,0){%
\put(-4,9.5){%
\put(0,0){$0_X,0_Y$}%
\put(15,0){$1_X,0_Y$}%
\put(30,0){$2_X,0_Y$}%
\put(45,0){$3_X,0_Y$}%
}%
\put(-4,24.5){%
\put(0,0){$0_X,1_Y$}%
\put(15,0){$1_X,1_Y$}%
\put(30,0){$2_X,1_Y$}%
\put(45,0){$3_X,1_Y$}%
}%
\put(-4,39.5){%
\put(0,0){$0_X,2_Y$}%
\put(15,0){$1_X,2_Y$}%
\put(30,0){$2_X,2_Y$}%
\put(45,0){$3_X,2_Y$}%
}%
\put(-4,54.5){%
\put(0,0){$0_X,3_Y$}%
\put(15,0){$1_X,3_Y$}%
\put(30,0){$2_X,3_Y$}%
\put(45,0){$3_X,3_Y$}%
}%
}%
}%
\put(40,-3){(\textit{a})}%
\put(145,-3){(\textit{b})}%
\end{picture}
}%
\caption{Satisfying $\Theta(\mathcal{T},T_0,d)$, for $d= 2$, in (\textit{a}) a quasi-saw model and (\textit{b})  $\R^2$.}\label{f:tiling-model}
\end{figure}

The second statement of the theorem follows from the
fact that the model constructed above can be embedded in $\RegC(\R^n)$
for all $n \geq 2$: indeed, in $\R^2$, we may represent
tiles as octagons (see Fig.~\ref{f:tiling-model}(\textit{b})), taking
$G$ to be the union of all those tiles. Note that the
tiles' corners are cut to ensure that every tile has
at most 4 neighbours and thus $\mathbf{b}^{\mathfrak A}$ and
$\mathbf{w}^{\mathfrak A}$ have precisely $2^{d-1}$ components.
\end{proof}

As a consequence of Theorems~\ref{theo:conTcInNexptime} and~\ref{nexptime} we obtain:
\begin{corollary}\label{cor:nexptimeComplete}
$\Sat(\cBCcc,\Regc)$ and $\Sat(\cBCcc,\ConR)$ are \NExpTime{}-complete.
\end{corollary}

We now consider lower complexity bounds for the languages $\cBc$ and
$\cBcc$, which will be obtained by reduction of satisfiability in
$\cBCc$ and $\cBCcc$,
respectively. The basic idea is that two connected closed
sets are in contact if and only if their union is connected and they are non-empty; in other words, the
formula
\begin{equation*}
c(\tau_1) \land c(\tau_2) \to \bigl(C(\tau_1,\tau_2)\
\leftrightarrow \ c(\tau_1 + \tau_2) \land (\tau_1 \ne\zero) \land (\tau_2\ne\zero)\bigr)
\end{equation*}
is a
$\cBCc$-validity. Thus, if a $\cBCcc$-formula contains a subformula
$C(\tau_1,\tau_2)$, where $\tau_1$ and $\tau_2$ denote
\emph{connected} non-empty regions, then we may replace that subformula by
$c(\tau_1 + \tau_2)$, thus eliminating an occurrence of $C$.
The problem we face, of course, is that this `reduction' of $C$ to $c$
cannot be directly applied to our formulas in which $\tau_1$ and
$\tau_2$ are not necessarily connected and non-empty. The next three lemmas show how
to overcome this problem.

We write $\varphi[\psi]$ to indicate that
$\varphi$ contains a \emph{positive} occurrence of
$\psi$; then $\varphi[\psi/\chi]$ denotes the
result of replacing all positive occurrences of $\psi$ in $\varphi$ by $\chi$.
Recall that topological spaces are allowed to be empty.
There is exactly one frame based on the
empty topological space, and exactly one model over that frame.
(It is trivial to check whether a formula $\varphi$ is satisfied in this model.)
Accordingly, we define:
\begin{equation*}
\epsilon_\varphi =
\begin{cases}
\zero = \one, & \text{if $\varphi$ is satisfied over the empty space},\\
\bot, & \text{otherwise.}
\end{cases}
\end{equation*}
\begin{lemma}\label{lma:poss2}
Let $\varphi[C(\tau_1,\tau_2)]$ be a
\cBCcc-formula, and $t, t_1, t_2$ fresh variables. Then $\varphi$ is equisatisfiable \textup{(}in $\Regc$ and $\ConR$\textup{)} with the formula
\begin{equation*}
\varphi^*  =  \epsilon_\varphi \vee
\Bigl(
\varphi[C(\tau_1,\tau_2)/(t = \zero)] \ \ \wedge\ \
   \Bigl((t = \zero)  \rightarrow
  c(t_1 + t_2) \ \land \bigwedge_{i = 1,2} \bigl((t_i \ne\zero) \land (t_i \leq \tau_i) \land c(t_i)\bigr) \Bigr) \Bigr) .
\end{equation*}
\end{lemma}
\begin{proof}
If $\varphi$ is satisfiable over the empty space, there is nothing to
prove; so assume otherwise.  Evidently, $\models \bigwedge_{i = 1,2}
\bigl((t_i \ne\zero) \land (t_i \leq \tau_i) \land c(t_i) \bigr)
\wedge c(t_1 + t_2) \rightarrow C(\tau_1,\tau_2)$, whence $\models
\varphi^* \rightarrow \varphi$. Conversely, let $\fA \models \varphi$,
for an Aleksandrov model $\mathfrak{A} =
(T,\RegC(T),\cdot^\mathfrak{A})$; we then construct $\fA^* =
(T,\RegC(T),\cdot^{\mathfrak{A}^*})$, where $r^{\mathfrak{A}^*} =
r^\mathfrak{A}$, for all variables occurring in $\varphi$. If $\fA
\not \models C(\tau_1,\tau_2)$, then setting $t^{\fA^*} = T$ secures
$\fA^* \models \varphi^*$. If, on the other hand, $\fA \models
C(\tau_1,\tau_2)$, then there is $z \in \tau_1^\mathfrak{A} \cap
\tau_2^\mathfrak{A}$.  As the space is Aleksandrov, there is a
\emph{minimal} open subset $V$ of $\fA$ containing $z$.  Set
$t^{\fA^*} = \emptyset$ and $t_i^{\fA^*} = \tau_i^\fA \cap V^-$ ($1
\leq i \leq 2$).  It is routine to check that both of the
$t_i^{\fA^*}$ are regular closed and connected. Moreover, their sum is
connected, since they share the point $z$. This secures $\fA^* \models
\varphi^*$.
\end{proof}%

We note that the above lemma does not work in any $\R^n$, as
intersecting sets do not necessarily share an open subset.

Suppose $T$ is a topological space, and $S$ a regular closed subset of
$T$. Then $S$ is itself a topological space (with the subspace
topology), which has its own regular closed algebra. The following
lemma is tedious to show in full generality, but simple for finite
Aleksandrov spaces (which, by Lemma~\ref{prop:KripkeSemantics}, is
sufficiently general for our purposes). We state it here without
proof.
\begin{lemma}\label{lma:regularSubspaceLemma}
If $S \in \RegC(T)$, then $\RegC(S) = \{ S \cdot X \mid X \in \RegC(T)
\}$. Furthermore, denoting the Boolean operations in $\RegC(S)$ by
$+_S$, $\cdot_S$ and $-_S$, etc., we have, for any $X, Y
\in \RegC(S)$\textup{:} \textup{(}i\textup{)} $X +_S Y= X + Y$,
\textup{(}ii\textup{)} $X \cdot_S Y = X \cdot Y$,
\textup{(}iii\textup{)} $-_S X= S \cdot (-X)$, \textup{(}iv\textup{)}
$\one_S = S$ and $\zero_S = \zero$; \textup{(}v\textup{)} $X$
has the same number of components in $S$ and $T$.
\end{lemma}%
For a formula $\varphi$ and a
variable $s$, define $\varphi_{\mid s}$ to be the result of replacing
every maximal term $\tau$ occurring in $\varphi$ by the
term $s\cdot \tau$. For any model $\fM = (T, \RegC(T), \cdot^\fM)$,
define $\fM_{\mid s}$ to be the model over the topological
space $s^\fM$ (with the subspace topology) obtained by setting
$r^{\fM_{\mid s}} = (s \cdot r)^\fM$ for all variables $r$.
\begin{lemma}
\label{lma:subspaceEquisatisfiability}
For any \cBCcc-formula  $\varphi$, $\fM \models \varphi_{\mid s}$ \ iff \
$\fM_{\mid s} \models \varphi$.
\end{lemma}
\begin{proof}
Using Lemma~\ref{lma:regularSubspaceLemma} (\emph{i})--(\emph{iv}),
we can show by structural induction on terms that $(s\cdot \tau)^\fM =
\tau^{\fM_{\mid s}}$, for any $\cB$-term $\tau$.  The result then
follows by Lemma~\ref{lma:regularSubspaceLemma} (\emph{v}).
\end{proof}

\begin{lemma}\label{lma:poss3}
Let $\varphi[\neg C(\tau_1,\tau_2)]$ be a \cBCcc-formula, and $s, t, t_1,
t_2$ fresh variables. Then $\varphi$ is
equisatisfiable
in $\Regc$ 
with the formula
\begin{multline*}
\varphi^*  = \epsilon_\varphi \vee
\Bigl(\ (s \ne \zero) \ \land \ \varphi[\neg C(\tau_1,\tau_2)/(t = \zero)]_{\mid s} \
\land\\
   \bigl((t\cdot s = \zero)  \ \rightarrow  \
\neg c(t_1 + t_2) \ \land \bigwedge_{i = 1,2}
         \bigl(c(t_i) \land (\tau_i \cdot s \leq t_i)\bigr)\bigr)\Bigr),
\end{multline*}
and $\varphi$ is equisatisfiable in $\ConR$ with $\varphi^* \land c(s)$.
\end{lemma}
\begin{proof}
Again, we may assume that $\varphi$ is not satisfiable over the empty space.
Evidently, $\bigwedge_{i = 1,2} \left(c(t_i) \land (\tau_i \cdot s
\leq t_i) \right) \wedge \neg c(t_1 + t_2) \rightarrow \neg C(\tau_1
\cdot s, \tau_2 \cdot s)$ is a \cBCcc-validity. So any model $\fA$ of
$\varphi^*$ is a model of $\varphi_{\mid s}$,
whence, by Lemma~\ref{lma:subspaceEquisatisfiability}, $\fA_{\mid s}
\models \varphi$. Note that $\mathfrak{A}_{\mid s}$ is connected whenever $\mathfrak{A}\models c(s)$.
Conversely, suppose $\fA
\models \varphi[\neg C(\tau_1,\tau_2)]$, for a quasi-saw model $\fA$
induced by $(W,R)$. Let $W_i$ ($i = 0,1$) be the set of points of
depth $i$ in $(W,R)$. Without loss of generality, we may assume that
every point in $W_0$ has an $R$-predecessor in $W_1$.  If $\fA \models
C(\tau_1, \tau_2)$, let $\fA^*$ be exactly like $\fA$ except that
$s^{\fA^*}$ and $t^{\fA^*}$ are both the whole space. Then $\fA^*
\models \varphi^*$. On the other hand, if $\fA \not \models
C(\tau_1, \tau_2)$
then, for $i = 1,2$,
we add an extra point $u_i$ to $W$ to connect up
the points in $\tau_i^\fA$ (if there are any). Formally, let $W^* = W \cup \{u_1, u_2\}$,
where $u_1, u_2 \not \in W$, and let $R^*$ be the reflexive closure of
the union of $R$ and $\{(z,u_i) \mid z \in \tau_i^\fA \cap W_1 \}$,
for $i = 1,2$.  Note that
$(W^*, R^*)$ is connected if $(W, R)$ is connected.
Clearly, $W$ is a regular closed subset of the
topological space $(W^*, R^*)$.  Now define the model $\fA^*$ over
$(W^*, R^*)$ by setting $s^{\fA^*} = W$, $t^{\fA^*} = \emptyset$, $t_i
^{\fA^*} = \tau_i^\fA \cup \{u_i \}$ ($i = 1,2$), and $r^{\fA^*} =
r^\fA$ for all other variables $r$. Thus, $\fA = \fA^*_{\mid s}$,
whence, by Lemma~\ref{lma:subspaceEquisatisfiability}, $\fA^* \models
\varphi_{\mid s}$, and so $\fA^*
\models \varphi[\neg C(\tau_1,\tau_2)/t \neq \zero]_{\mid s}$. By
construction, $\fA^* \models \bigwedge_{i = 1,2} \left(c(t_i) \land
(\tau_i \cdot s \leq t_i)\right) \wedge \neg c(t_1 + t_2)$. Thus,
$\fA^*\models\varphi^*$.
\end{proof}

As a consequence of Lemma~\ref{lma:poss2} and~\ref{lma:poss3}, we obtain
\begin{theorem}\label{thm:Bc:ExpTime}
$\Sat(\cBc,\Regc)$ and $\Sat(\cBc,\ConR)$ are both
\ExpTime{}-complete.\\ $\Sat(\cBcc,\Regc)$ and $\Sat(\cBcc,\ConR)$ are both
\NExpTime{}-complete.
\end{theorem}
\begin{proof}
Given a $\cBCc$-formula, by repeated applications of Lemmas~\ref{lma:poss2} and~\ref{lma:poss3}, we can compute an
equisatisfiable $\cBc$-formula in polynomial time.  Similarly, given
a $\cBCcc$-formula, we can compute an equisatisfiable
$\cBcc$-formula in polynomial time.  It then follows from
Theorem~\ref{thm:BRCC-8:ExpTime} that $\Sat(\cBc,\Regc)$ is
\ExpTime{}-hard, and from Theorem~\ref{nexptime} that
$\Sat(\cBcc,\Regc)$ is \NExpTime{}-hard.
Noting that Lemmas~\ref{lma:poss2} and~\ref{lma:poss3}, and
Theorems~\ref{thm:BRCC-8:ExpTime} and~\ref{nexptime} all
hold when we restrict attention to connected spaces, we obtain the
remaining statements of the theorem.
\end{proof}

We mention here that although Lemma~\ref{lma:poss3} does not hold for
$\R^n$, a simple modification of the proof of
Theorem~\ref{thm:Bc:ExpTime} can be used to show the following:
\begin{theorem}\label{thm:Bc:ExpTime:R}
$\Sat(\cBc,\RegC(\R^n))$ is
\ExpTime{}-hard and
$\Sat(\cBcc,\RegC(\R^n))$ is
\NExpTime{}-hard, for any $n \geq 3$.
\end{theorem}
\begin{proof}
Consider the formula $\Phi'(\chi_M,\psi_{\vec{a}})$ constructed in the
proof of Theorem~\ref{thm:BRCC-8:ExpTime:bis}. It contains only
negative occurrences of the predicate $C$. So, we can iteratively apply the
transformation of Lemma~\ref{lma:poss3} to obtain a $\cBc$-formula
$\Phi^*$. Denote by $s$ the product of all regions $s$ that have been
used to relativize the formula. It follows from the proof of
Theorem~\ref{thm:BRCC-8:ExpTime:bis} that if $M$ does not accept
$\vec{a}$ then $\Phi'(\chi_M,\psi_{\vec{a}})$ is satisfiable in a
quasi-saw model, for which we construct a new quasi-saw model
$\mathfrak{A}^*$ over $(W^*,R^*)$ as described in the proof of
Lemma~\ref{lma:poss3}. It follows from the proof of
Lemma~\ref{lma:poss3} that $\mathfrak{A}^*\models\Phi^*$. We then
embed the graph of the quasi-saw model $\mathfrak{A}^*$ into $\R^n$ by
associating with each point of depth 0 an $n$-dimensional ball with
attached cylinders, which touch only if there is a respective point of
depth 1. As $\Phi^*$ is a $\cBc$-formula and
$\mathfrak{A}^*\models\Phi^*$, $\Phi^*$ is satisfied in the
constructed model over $\RegC(\R^n)$. Conversely, if $\Phi^*$ is
satisfiable over $\RegC(\R^n)$ then, by
Lemmas~\ref{prop:KripkeSemantics} and~\ref{broom-lemma}, it is
satisfied in a quasi-saw model $\mathfrak{A}$. As $\bigwedge_{i = 1,2}
\left(c(t_i) \land (\tau_i \cdot s \leq t_i) \right) \wedge \neg c(t_1
+ t_2) \rightarrow \neg C(\tau_1 \cdot s, \tau_2 \cdot s)$ is a
validity, $\fA$ is a model of $\Phi'(\chi_M,\psi_{\vec{a}})_{\mid s}$,
whence, by Lemma~\ref{lma:subspaceEquisatisfiability}, $\fA_{\mid s}
\models \Phi'(\chi_M,\psi_{\vec{a}})$ and thus, $M$ does not accept
$\vec{a}$.
%

For the second statement of the Theorem, we proceed in a similar
fashion, using Lemma~\ref{lma:poss3} to remove all (negative)
occurrences of $C$ in the formula $\Theta(\mathcal{T},T_0,d)$
constructed in the proof of Theorem~\ref{nexptime}.
\end{proof}


\section{Topological logics over Euclidean spaces}
\label{sec:Euclidean}

So far, we have been mainly concerned with the computational complexity of various
topological logics interpreted over the very general classes of frames
$\All$, $\Con$, $\Regc$ and $\ConR$. In this section, we discuss
what happens when we restrict consideration to the {\em specific}
topological spaces $\R$, $\R^2$ or $\R^3$.

For languages without connectedness predicates, there is little work
to do. For the languages $\cB$ and $\RCCE$, we have, for all $n \geq
1$:
\begin{align*}
\Sat(\cB, \RegC(\R^n))   &= \Sat(\cB, \Regc),\\
\Sat(\RCCE, \RegC(\R^n)) &= \Sat(\RCCE, \Regc).
\end{align*}
These equations can be established by embedding any model over $\Regc$
into one over the domain $\RegC(\R^n)$ in such a way that the
satisfaction of formulas in the relevant language is unaffected.  A
suitable embedding for $\RCCE$ is described in~\cite{Renz98}; the case
of $\cB$ is even more straightforward.

The corresponding equations fail for the languages $\cBC$, $\cBCm$ and
$\SFU$.  For example, we saw that the
$\cBC$-formula~\eqref{eq:connectedSpaces} is satisfiable over
frames in $\Regc$, but only when the underlying space is disconnected.
Thus, $\Sat(\cBC, \RegC(\R^n)) \neq \Sat(\cBC, \Regc)$, whence
$\Sat(\SFU, \R^n) \neq \Sat(\SFU, \All)$, for all $n \geq 1$. However,
it turns out that~\eqref{eq:connectedSpaces} is, so to speak, the only
fly in the ointment:
\begin{theorem}[\cite{Shehtman99}]\label{theo:CalmostInsensitive}
Let $T$ be any connected, dense-in-itself, separable metric space.
Then $\Sat(\SFU, T) = \Sat(\SFU, \Con)$.
\end{theorem}
\noindent
Hence, $\Sat(\SFU, \cK) = \Sat(\SFU, \Con)$ for any class of frames
$\cK$ included in $\Con$ and containing $\R^n$ for any $n$, and
similarly, $\Sat(\cBC, \cK) = \Sat(\cBC, \ConR)$ for any class of
frames $\cK$ included in $\ConR$ and containing $\RegC(\R^n)$ for any
$n$.

The predicate $c$, however, changes the above picture dramatically.
\begin{theorem}
The problems $\Sat(\RCCEc, \RegC(\R))$, $\Sat(\RCCEc, \RegC(\R^2))$ and \linebreak
$\Sat(\RCCEc, \RegC(\R^3))$ are all different.
\label{obs:rccInsensitive}
\end{theorem}
\begin{proof}
Let $r_{i}$ ($1 \leq i \leq 5$) and $r_{i,j}$ ($1 \leq i < j \leq 5$)
be variables.  As we observed in Section~\ref{sec:intro}, the formula
\begin{equation*}
\varphi_1 ~=~
\bigwedge_{1 \leq i \leq 3} c(r_i) \wedge
\bigwedge_{1 \leq i < j \leq 3} \mathsf{EC}(r_i,r_j)
\end{equation*}
is not satisfiable over $\RegC(\R)$, since the non-empty, regular
closed, connected subsets of $\R$ are exactly the closed, non-punctual
intervals. On the other hand, $\varphi_1$ is visibly satisfiable over
$\RegC(\R^n)$ for all $n \geq 2$.

\bigskip

\noindent
Now consider the formula
\begin{multline*}
\varphi_2 ~=~ \bigwedge \Bigl\{\mathsf{DC}(r_{i,j},r_{k,l}) \ \Big| \
1 \leq i < j \leq 5, 1 \leq k < l \leq 5, \{i,j\} \cap \{k,l\} = \emptyset \Bigr\} \wedge{}\\
\bigwedge \Bigl\{\mathsf{TPP}(r_i,r_{j,k}) \ \Big| \
1 \leq i \leq 5, 1 \leq j < k \leq 5, i \in \{j,k\} \Bigr\} \wedge
\bigwedge_{1 \leq i < j \leq 5} c(r_{i,j}).
\end{multline*}
Think of the regions (assigned to) $r_1, \ldots, r_5$ as `vertices'
and the regions (assigned to) $r_{i,j}$ ($1 \leq i < j \leq 5$) as
`edges', with $r_{i,j}$ connecting $r_i$ and $r_j$. The literals of
$\varphi_2$ having the form $\mathsf{TPP}(r_{i},r_{j,k})$ ensure that
`vertices' (are non-empty and) are contained in any `edges' which
involve them. The literals of $\varphi_2$ having the form
$\mathsf{DC}(r_{i,j},r_{k,l})$ ensure that `edges' lacking a common
`vertex' are not in contact. It is easy to construct a model for this
formula over $\RegC(\R^n)$ for all $n > 2$. To show that $\varphi_2$
is not satisfiable over $\RegC(\R^2)$, suppose otherwise; we show,
contrary to fact, that the graph $K_5$ has a plane embedding.

To avoid notational clutter, take $r_{i}$ ($1 \leq i \leq 5$) and
$r_{i,j}$ ($1 \leq i < j \leq 5$) to denote regular closed sets of
$\R^2$ satisfying $\varphi_2$. For any point $p$ in any of the
$r_{i,j}$, let $\varepsilon'_p$ be the minimum Euclidean distance to
any point $q$ in any $r_{k,l}$ such that $\{i,j\} \cap \{k,l\} \neq
\emptyset$; let $\varepsilon_p = \max(1, \varepsilon'_p)$;
and let $D_p$ be the closed disc centred on $p$ of radius
$\varepsilon_p/3$. For all $i, j$ ($1 \leq i < j \leq 5$), let
$r'_{i,j} = \bigcup \{ D_p \mid p \in r_{i,j} \}$. The following are
simple to verify: (i) $r'_{i,j}$ is regular closed; (ii) the open set
$\ti{(r'_{i,j})}$ is connected (hence path-connected) and contains
both $r_i$ and $r_j$; (iii) for all $k$, $l$ ($1 \leq k < l \leq 5$)
such that $\{i,j\} \cap \{k,l\} \neq \emptyset$, $r'_{i,j} \cap
r'_{k,l} = \emptyset$. For all $i$ ($1 \leq i \leq 5$), choose a point
$v_i \in r_i$ and, for all $i$, $j$, ($1 \leq i < j \leq 5$), join
$v_i$ and $v_j$ with an arc $\alpha_{i,j}$ lying in $r'_{i,j}$. It is
straightforward to draw these arcs in such a way that, for each $i$,
the arcs with $v_i$ as an endpoint meet only at $v_i$. However, if
$\{i,j\} \cap \{k,l\} = \emptyset$, the arcs $\alpha_{i,j}$ and
$\alpha_{k,l}$ do not intersect.
\end{proof}
Corresponding remarks apply to $\cB$:
\begin{theorem}
The problems $\Sat(\cBc, \RegC(\R))$, $\Sat(\cBc, \RegC(\R^2))$
and $\Sat(\cBc, \RegC(\R^3))$ are all different.
\label{obs:Binsensitive}
\end{theorem}
\begin{proof}
Almost identical to Theorem~\ref{obs:rccInsensitive}, noting that, for
$r$, $s$ connected, $\mathsf{DC}(r, s)$ if and only if $r+s$ is not
connected.
\end{proof}

\begin{figure}[t]
\begin{center}
\begin{picture}(0,0)%
\includegraphics{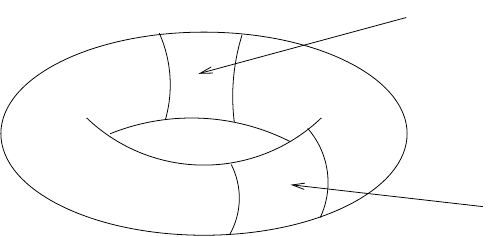}%
\end{picture}%
\setlength{\unitlength}{1973sp}%
\begin{picture}(4754,2260)(2318,-2694)
\put(6382,-626){$r_1$}%
\put(7072,-2456){$r_2$}%
\end{picture}
\end{center}
\caption{Two non-intersecting, connected, closed sets $r_1$ and $r_2$ on a
torus: note that $\compl{r}_1$ and $\compl{r}_2$
are connected, but $\compl{r}_1 \cap \compl{r}_2$ is not.}
\label{55:fig:torus}
\end{figure}
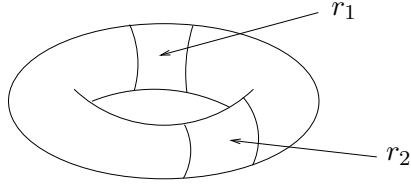

It is not known whether $\Sat(\cBc, \RegC(\R^3))$ and
$\Sat(\cBc,\ConR)$ are different. However, $\Sat(\SFU,\R^n) \neq
\Sat(\SFU,\Con)$ for all $n \geq 1$.  To see this, we rely on the
following fact:
\begin{theorem}[\cite{Newman64}, p.~137]
\label{55:theo:Newman137}
If $r_1$ and $r_2$ are non-intersecting closed sets in $\R^n$, and
points $x$ and $y$ are connected in $\compl{r}_1$ and also in
$\compl{r}_2$, then $x$ and $y$ are connected in $\compl{r}_1 \cap
\compl{r}_2$.
\end{theorem}
The formula
\begin{align}\label{55:eq:torus}
(r_1 \cap r_2 = \zero) \ \land \
   \bigwedge_{i = 1,2}\bigl((\tc{r}_i \subseteq r_i) \ \land \
c(\compl{r}_i)\bigr) \ \ \land \ \ \neg c(\compl{r}_1 \cap \compl{r}_2)
\end{align}
states that $r_1$ and $r_2$ are non-intersecting, closed, connected
regions having connected complements, such that the intersection of
their complements is not connected.  This formula is not satisfiable
over any $\R^n$, by Theorem~\ref{55:theo:Newman137}. However, it is
satisfiable over $T$ for many natural, connected topological spaces
$T$. For example, let $T$ be a torus, and let $r_1$ and $r_2$ be
interpreted as rings in $T$, arranged as in Fig.~\ref{55:fig:torus};
it is then obvious that $r_1$ and $r_2$
satisfy~\eqref{55:eq:torus}.

Thus, for all of our base languages $\cL$, the language $\cLc$ (and
therefore also $\cLcc$) is more sensitive than $\cL$ to the spatial
domain over which it is interpreted.  Since we know the complexity
of the satisfiability problems for $\cLc$ and $\cLcc$ for very general
classes of spatial domains, the question naturally arises as to the
complexity of these problems for spatial domains based on
low-dimensional Euclidean spaces.

For $n = 1$, we have a reasonably complete picture.
\begin{theorem}\label{thm:R:NP}
$\Sat(\cBc,\RegC(\R))$ is in \NP{}.
\end{theorem}
\begin{proof}
\emph{Future-past temporal
logic formulas} (FP-formulas, for short) are constructed from
propositional variables $p_i$, $i<\omega$, using the Boolean
connectives $\wedge$, $\neg$, $\top$, and temporal operators
$\Diamond_F$ (`some time in the future') and
$\Diamond_P$ (`some time in the past'). A model $(\R,V)$ for FP
consists of the real line $\R$ and a valuation $V$ mapping each
propositional variable $p_i$ to a subset $V(p_i)$ of $\R$. The
truth-relation $\models$ between pointed models $(\R,V,x)$ and
FP-formulas $\phi$ is defined as follows (the clauses for the
Boolean connectives are standard):
\begin{enumerate}[$\bullet$]
\item $(\R,V,x) \models p_i$ iff $x\in V(p_{i})$;
\item $(\R,V,x) \models \Diamond_F \phi$ iff there is $y>x$
    such that $(\R,V,y) \models \phi$, and symmetrically for $\Diamond_P \phi$.
\end{enumerate}
An FP-formula $\phi$ is satisfiable if there exists $(\R,V,x)$ with
$(\R,V,x)\models \phi$. Enumerating the region variables as $r_1,
r_2, \ldots$, we define a translation $\cdot^\ast$ from $\cB$-formulas
to FP-formulas as follows:
\begin{align*}
r^*_i & = p_i, &  (-\tau)^* & = \neg \tau^*, \\
(\tau_1 \cdot \tau_2)^* & = \tau^*_1 \wedge \tau^*_2, &
(\tau_1 + \tau_2)^* & = \tau^*_1 \vee \tau^*_2,\\
\zero^* & = \bot, &  \one^* & = \top,\\
(\tau_1 = \tau_2)^* & = \neg \Diamond_F \Diamond_P \neg
(\tau^*_1 \leftrightarrow \tau^*_2), &
(c(\tau))^* & = \neg \Diamond_F \Diamond_P(\tau^* \wedge \Diamond_F
(\neg \tau^* \wedge \Diamond_F \tau^*)),\\
(\neg \varphi)^* &= \neg \varphi^*, &
(\varphi_1 \wedge  \varphi_2)^* & = \varphi^*_1 \wedge  \varphi^*_2,\\
(\varphi_1 \vee  \varphi_2)^* & = \varphi^*_1 \vee  \varphi^*_2,
\end{align*}
where $\tau$, $\tau_1$, $\tau_2$ range over $\cB$-terms, and
$\varphi$, $\varphi_1$, $\varphi_2$ over $\cB$-formulas.  This
translation can clearly be computed in polynomial time. It is routine
to verify that a $\cBc$-formula $\varphi$ is satisfiable over
$\RegC(\R)$ if and only if $\varphi^*$ is satisfiable over the
temporal flow $(\R, <)$. But the satisfiability problem for FP-formulas over
$(\R, <)$ is known to be in \NP~(see, e.g.,~\cite{KK:gkwz},
Theorem~2.4).
\end{proof}

The exact complexities of the problems $\Sat(\cBcc,\RegC(\R))$,
$\Sat(\RCCEc,\RegC(\R))$ and \linebreak
$\Sat(\RCCEcc,\RegC(\R))$ are not known, though an upper
bound of $\PSpace$ is provided by our next theorem.
\begin{theorem}\label{thm:R:PSpace}
$\Sat(\conTc,\R)$ is in \PSpace{}.
\end{theorem}
\begin{proof}
The proof is by reduction to the propositional temporal logic with Since and Until over the
real line, for which satisfiability is known to be
\PSpace{}-complete~\cite{Reynolds-real}. Recall that \emph{linear temporal
logic formulas} (LTL-formulas, for short) are constructed from
propositional variables $p_i$, $i<\omega$, using the Boolean
connectives $\wedge$, $\neg$, $\top$, and binary temporal operators
$\since$ (`since') and $\until$ (`until'). A model $(\R,V)$ for LTL
consists of the real line $\R$ and a valuation $V$ mapping each
propositional variable $p_i$ to a subset $V(p_i)$ of $\R$. The
truth-relation $\models$ between pointed models $(\R,V,x)$ and
LTL-formulas $\phi$ is defined as follows (the clauses for the
Boolean connectives are standard):
\begin{enumerate}[$\bullet$]
\item $(\R,V,x) \models p_i$ iff $x\in V(p_{i})$;
\item $(\R,V,x) \models \phi \since \psi$ iff there exists $y<x$ such that $(\R,V,y) \models \psi$
and for all $z$ with $y<z< x$, $(\R,V,z) \models \phi$;
\item $(\R,V,x) \models \phi \until \psi$ iff there exists $y>x$ such that $(\R,V,y) \models \psi$
and for all $z$ with $x< z < y$, $(\R,V,z)
\models \phi$.
\end{enumerate}
An LTL-formula $\phi$ is satisfiable if there exists $(\R,V,x)$
with $(\R,V,x)\models \phi$.

Suppose
now that we are given an $\conTc$-formula $\varphi$. Without loss of generality, we may assume that
no nested interior and closure operators occur in $\varphi$ (by introducing additional variables, we can always transform $\varphi$ into an equisatisfiable formula of length $2|\varphi|$ without nested topological operators).
Let $K$ be the maximum number $k$ such that $c^{\leq k}(\tau)$ occurs in $\varphi$.

Our
\PSpace{} algorithm for checking the satisfiability of such a $\varphi$
proceeds as follows. To simplify the exposition, we
introduce a new unary predicate $\infty(\cdot)$, with the following
semantics (for models $\fM$ over $\R$):
\begin{equation*}
\fM \models \infty(\tau) \mbox{ iff }
\text{some bounded interval includes infinitely many components of $\tau^\fM$.}
\end{equation*}
(Note that $\infty$ is not part of the language $\conTc$.)
Now,
we first guess a conjunction $\chi$ of literals
of the forms $\tau_1 = \tau_2$, $\tau_1 \ne \tau_2$, $c^{= k}(\tau)$
and $c^{\geq K+1}(\tau)$ such that (i) for each subformula of
$\varphi$ of the form $\tau_1=\tau_2$, $\chi$ contains either
$\tau_1=\tau_2$ or $\tau_1\ne\tau_2$, and (ii) for each subformula of
the form $c^{\leq k}(\tau)$, it contains one of $c^{=0}(\tau)$, \dots,
$c^{=K}(\tau)$, $c^{\geq K+1}(\tau)$, where $c^{= k}(\tau)$
abbreviates $c^{\leq k}(\tau) \land c^{\geq k}(\tau)$. The literals in
$\chi$ give in an obvious way the truth-values to the atoms in
$\varphi$, and we check whether $\varphi$, as a propositional formula,
is true under this valuation. If this is indeed the case, it only
remains to verify whether $\chi$ is satisfiable over $\mathbb R$. This
can be done by expressing $\chi$ as an equisatisfiable LTL-formula
$\chi^*$.

To this end, we non-deterministically replace each literal of the form
$c^{\geq K + 1}(\tau)$ in $\chi$ with {\em either} $\infty(\tau)$ {\em
  or} the conjunction
\begin{equation*}
(\tau = r_1 \cup r_2) \  \land \  (\tc{r_1} \cap \tc{r_2} \cap \tau = \zero) \  \land \
     c^{=K}(r_1) \  \land \   (r_2 \neq \zero),
\end{equation*}
where $r_1$ and $r_2$ are fresh variables. It is not difficult to see
that $\chi$ is satisfiable if and only if some conjunction
$\hat{\chi}$ obtained in this way is also satisfiable; moreover, the
length of $\hat{\chi}$ is linear in the length of $\chi$.  Note that
$\hat{\chi}$ is not necessarily an $\conTc$-formula (because it may
contain occurrences of $\infty$); however, all terms in $\hat{\chi}$
are $\SFU$-terms.

Now, the meaning of \SFU-terms over $\mathbb R$ can be expressed with the
help of LTL-formulas: for an \SFU-term $\tau$ we construct the
LTL-formula $\tau^\ast$ according to the following definition:
\begin{align*}
r_i^{\ast} & = p_i, & \zero^\ast & = \bot, &  \one^\ast & =  \top,\\
\compl{\tau}^{\ast} & = \neg \tau^{\ast}, &
(\tau_{1} \cap \tau_{2})^\ast & = \tau_1^\ast\land\tau_2^\ast, & (\tau_{1} \cup \tau_{2})^\ast & = \tau_1^\ast\lor\tau_2^\ast, \\
(\ti{\tau})^\ast & = (\tau^{\ast} \since \top) \land \tau^\ast \land (\tau^{\ast} \until \top)\hspace*{-8em}
\end{align*}
and $\tc{\tau}$ is treated as an abbreviation for $\compl{\ti{\compl{\tau}}}$. As $\tau$ does not have nested interior and closure operators, the length of $\tau^\ast$ is linear in the length of $\tau$. Let $\mathfrak{M}$ be a topological model over $\R$. Set $V_\mathfrak{M}(p_i) = p_i^\mathfrak{M}$. It can be shown by induction on the structure of an \SFU-term $\tau$ that $(\R,V_\mathfrak{M},x)\models \tau^\ast$ if and only if $x\in\tau^\mathfrak{M}$.

We now map each literal $\psi$ of $\hat{\chi}$ into an LTL-formula
$\psi^*$ with (intuitively) the same meaning as $\psi$.  First, we
translate literals of the form $(\tau_1 = \tau_2)$ or
$(\tau_1\ne\tau_2)$ into LTL-formulas as follows:
\begin{align*}
(\tau_1 = \tau_2)^{\ast} & \ \ = \ \ \Box_F\Box_P  (\tau_1^{\ast} \leftrightarrow \tau_2^\ast),\\
(\tau_1 \ne \tau_2)^{\ast} & \ \ = \ \ \Diamond_F\Diamond_P \neg(\tau_1^{\ast} \leftrightarrow \tau_2^\ast),
\end{align*}
where $\Diamond_{F}\psi = \top \until \psi$, $\Box_F = \neg\Diamond_F\neg\psi$,
$\Diamond_{P}\psi = \top \since \psi$, $\Box_P\psi= \neg\Diamond_P\neg\psi$.

Next, we translate literals of the form $\infty(\tau)$
into LTL-formulas as follows:
\begin{equation}\label{eq:accumulation}
(\infty(\tau))^* = \Diamond_P \Diamond_F
\left(
(\neg (\tau^* \until \tau^*) \wedge \neg (\neg \tau^* \until \neg \tau^*)) \vee
(\neg (\tau^* \since \tau^*) \wedge \neg (\neg \tau^* \since \neg \tau^*))
\right).
\end{equation}
We claim that the formula on the right-hand side
of~\eqref{eq:accumulation} is true (at all points in $\R$) if and only
if there exists a bounded interval of $\R$ which includes infinitely
many components of $\tau$. For if $x \in \R$ is such that $(\R,V,x)
\models \neg (\tau^* \until \tau^*) \wedge \neg (\neg \tau^* \until
\neg \tau^*)$, under some valuation $V$, then every interval to the right of $x$ (i.e., whose
left-hand endpoint is $x$) contains infinitely many components of
$\tau$; and similarly for the second disjunct.  Conversely, if there
exists a bounded interval of $\R$ which includes infinitely many
components of $\tau$, then these components must have some
accumulation point $x$.  But then either every interval to the right
of $x$ contains infinitely many components of $\tau$, or every
interval to the left of $x$ does (or both); in the former case, $x$
satisfies $\neg (\tau^* \until \tau^*) \wedge \neg (\neg \tau^* \until
\neg \tau^*)$, in the latter case, the second disjunct
in~\eqref{eq:accumulation} is satisfied.

It is much harder to translate atoms of the form $c^{=k}(\tau)$.
To do this, we will require LTL-formulas of the form:
\begin{equation*}
\beta_\psi(\eta,\xi) \ \ = \ \ (\psi\land\eta) \until \bigl(\bigl((\psi\land\eta) \land ((\neg \psi \land \xi)\until\top)\bigr)\ \lor\ (\neg\psi\land \xi)\bigr).
\end{equation*}
It is readily seen that if $\beta_\psi(\eta,\xi)$ is true at a point $x$ then there is $y > x$ such that either (i) $\psi\land\eta$ is true everywhere in $(x,y)$\footnote{As usual, $(x,y]=\{z\in \R\mid x < z \leq y\}$ and $(x,y)=\{z \in \R \mid x < z < y\}$.} and $\neg\psi\land\xi$ is true at $y$, or (ii) $\psi\land\eta$ is true everywhere in $(x,y]$ and $\neg\psi\land\xi$ is true in $(y,z]$, for some $z > y$. In other words, if $\beta_\psi(\eta,\xi) \land \psi$ is true at $x$ then $\eta$ is true at all points from $(x,\infty)$ that belong to the same connected component of $\psi$ as $x$, while $\xi$ is true immediately to the right of that connected component.

To count the connected components of (extensions of) terms $\tau$,
we construct LTL-formulas $\theta_k^\tau$, where $k$ is a natural
number not exceeding $K$. Take $\lfloor\log_2 K\rfloor + 1$ fresh variables $v^\tau_n,
\dots,v^\tau_1$ to represent a connected component number in binary.
The formula $\theta_k^\tau$ contains the following conjuncts:
\begin{align*}
& \tau^* \land v^\tau_i \ \to \ \bigl(\neg\tau^*\until\top \ \lor \ \beta_{\tau^*}(v^\tau_i,\top) \ \lor \ \Box_F(\tau^*\land v^\tau_i)\bigr),\quad \text{ for } n \geq i \geq 1,\\
& \tau^* \land \neg v^\tau_i \ \to \ \bigl(\neg\tau^*\until\top \ \lor \ \beta_{\tau^*}(\neg v^\tau_i,\top) \ \lor \ \Box_F(\tau^*\land\neg v^\tau_i)\bigr),\quad \text{ for } n \geq i \geq 1,\\ %
& \tau^*\land v^\tau_j \land \neg v^\tau_h \ \to \ \textsf{\small next-int}_{\tau^*}(v^\tau_j)  \ \lor \ \Box_F\neg\tau^*,\quad \text{ for } n \geq j > h \geq 1,\\
& \tau^*\land \neg v^\tau_j \land \neg v^\tau_h  \ \to \ \textsf{\small next-int}_{\tau^*}(\neg v^\tau_j)  \ \lor \ \Box_F\neg\tau^*,\quad \text{ for } n \geq j > h \geq 1,\\
& \tau^*\land \neg v^\tau_h \land v^\tau_{h-1} \land \dots \land v^\tau_1 \ \to \ \textsf{\small next-int}_{\tau^*}(v^\tau_h)  \ \lor \ \Box_F\neg\tau^*, \quad \text{ for } n \geq h \geq 1,\\
& \tau^*\land \neg v^\tau_h \land v^\tau_{h-1} \land \dots \land v^\tau_1 \ \to \ \textsf{\small next-int}_{\tau^*}(\neg v^\tau_i)  \ \lor \ \Box_F\neg\tau^*,\quad \text{ for } n \geq h > i \geq 1,
\end{align*}
where
\begin{equation*}
\textsf{\small next-int}_{\tau^*}(\eta) \ \ = \ \ \beta_{\neg\tau^*}(\top,\eta) \ \lor \ 
\bigl(\tau^* \until \beta_{\neg\tau^*}(\top,\eta)\bigr).
\end{equation*}
It can be seen that if $\textsf{\small next-int}_{\tau^*}(\eta)\land \tau^*$ is
true at $x$ then $\eta$ is true at some $y > x$ that belongs to the next connected component of $\tau$ to the right of $x$.
The first two formulas
ensure that inside the current connected component of $\tau$, the bits
of the counter remain constant. The remaining conjuncts ensure proper counting; cf.~\eqref{eq:tiling:counter:upper:1}--\eqref{eq:tiling:counter:lower}. So, we set
\begin{multline*}
(c(\tau)^{= k})^{\ast} \ \  = \ \
\Diamond_F\Diamond_P\bigl(\textbf{0}_\tau\land\tau^* \land ((\tau^* \since \neg \tau^*) \lor \Box_P \neg \tau^* \lor \Box_P \tau^*) \ \ \land \ \  \theta_k^\tau \land \Box_F\theta_k^\tau \\
 \land \ \Diamond_F(\textbf{k}_\tau\land\tau^* \land ((\tau^* \until \neg \tau^*) \lor \Box_F\neg\tau^* \lor \Box_F\tau^*) \bigr),
\end{multline*}
where $\textbf{k}_\tau$ is the binary representation of $k$ using the
counter variables $v_n^\tau,\dots,v_1^\tau$ (e.g., $\textbf{0}_\tau = \neg v_n^\tau \land \dots \land \neg v_1^\tau$). Clearly, the length of $(c(\tau)^{= k})^{\ast}$ is polynomial in the length of $\tau$ and $\log_2 K$.

Finally, we construct the LTL-formula $\chi^*$ by replacing each
conjunct $\psi$ in $\hat{\chi}$ with $\psi^*$. Clearly, the length of
$\chi^*$ is polynomial in the length of $\varphi$. We leave it to the
reader to verify that $\chi^*$ is satisfiable if and only if
$\hat{\chi}$ is satisfiable over $\R$. Since the
satisfiability problem for LTL over $\R$ is
in \PSpace~\cite{Reynolds-real}, this completes the proof.
\end{proof}

\begin{corollary}
The problems $\Sat(\cBCc,\RegC(\R))$, $\Sat(\cBCcc,\RegC(\R))$,
$\Sat(\conT,\R)$ and \linebreak
$\Sat(\conTc,\R)$ are all
\PSpace{}-complete\textup{;} the problems $\Sat(\cBcc,\RegC(\R))$,
$\Sat(\RCCEc,\RegC(\R))$ and $\Sat(\RCCEcc,\RegC(\R))$ are all \textup{(}\NP{}-hard
and\textup{)} in \PSpace{}.
\label{cor:hisLastBow}
\end{corollary}
\begin{proof}
Theorems~\ref{pspace-lower} and~\ref{thm:R:PSpace}.
\end{proof}

This concludes our discussion of the complexity of satisfiability for
topological logics interpreted over $\R$.


Over $\R^2$, topological logics become harder to analyze. The
encodings used to obtain lower complexity bounds in
Section~\ref{sec:complexity} apply unproblematically to Euclidean
spaces of dimension at least 2.  In particular,
Theorem~\ref{thm:BRCC-8:ExpTime} states that $\Sat(\cBCc,
\RegC(\R^n))$ is \ExpTime{}-hard, for all $n \geq 2$, whence
$\Sat(\conT,\R^n)$ is \ExpTime{}-hard, for all $n \geq 2$.  Similarly,
Theorem~\ref{nexptime} states that $\Sat(\cBCcc,\RegC(\R^n))$ is
\NExpTime{}-hard, for all $n \geq2$, whence $\Sat(\conTc,\R^n)$ is
\NExpTime{}-hard, for all $n \geq 2$.

Upper bounds for $\Sat(\cBCc, \RegC(\R^n))$, $\Sat(\conT,\R^n)$,
$\Sat(\cBCcc, \RegC(\R^n))$ or $\Sat(\conTc,\R^n)$, where $n \geq 2$,
are not known. However, for the smaller language $\RCCE$, upper bounds
are known from the literature in the case where the spatial domain is
limited to certain well-behaved regions in $\R^2$. One such domain is
the set $\mathbf{D}$ of closed disc-homeomorphs in $\R^2$, with which
we began this paper.  We mention the following remarkable fact, the
proof of which is too involved to repeat here:
\begin{theorem}[\cite{sss03}]\label{theo:SSS}
The problem $\Sat(\RCCE,(\R^2,\mathbf{D}))$ is \NP{}-complete.
\end{theorem}


\begin{table}[t]\label{table:theBigBoard}
\begin{tabular}{|c|c|c|c|c|c|}\hline
& $\Regc$ & $\ConR$ & $\RegC(\R^n)$ & $\RegC(\R^2)$ & $\RegC(\R)$ \\[-4pt] %
& & & \small $n \geq 3$ &  &  \\\hline\hline %
\RCCE{} & \multicolumn{5}{c}{\CPLX[Cor.]{\NP{}}{NB-complexity}}\vline \\\hline  %
\RCCEc{} & \multicolumn{3}{c}{\CPLX[Cor.]{\NP{}}{NB-complexity}}\vline  & $\geq$\NP{} & \NPPSpace{} \\\hline  %
\!\RCCEcc{}\! & \multicolumn{3}{c}{\CPLX[Cor.]{\NP{}}{NB-complexity}}\vline  & $\geq$\NP{} & \NPPSpace{} \\\hline\hline  %
\cB{}  & \multicolumn{5}{c}{\CPLX[Cor.]{\NP}{NB-complexity}}\vline \\\hline  %
\cBc{}  & \CPLX{\Exp{}}{thm:Bc:ExpTime} & \CPLX{\Exp{}}{thm:Bc:ExpTime} & \CPLX{$\geq$\Exp{}}{thm:Bc:ExpTime} & $\geq$\NP{} & \CPLX{\NP}{thm:R:NP} \\\hline  %
\cBcc{} & \CPLX{\NExp{}}{thm:Bc:ExpTime} & \CPLX{\NExp{}}{thm:Bc:ExpTime} & \CPLX{$\geq$\NExp{}}{thm:Bc:ExpTime} & $\geq$\NP{} & \NPPSpace{} \\\hline\hline  %
\cBC{} & \CPLX[Cor.]{\NP}{NB-complexity} & \multicolumn{4}{c}{\CPLX[Cor.]{\PSpace{}}{cor:cBCconPSPACEcomplete}}\vline \\\hline  %
\cBCc{} & \CPLX[Cor.]{\Exp}{cor:BRCC-8:ExpTimecomplete} & \CPLX[Cor.]{\Exp}{cor:BRCC-8:ExpTimecomplete} &  \CPLX{$\geq$\Exp}{thm:BRCC-8:ExpTime} &  \CPLX{$\geq$\Exp}{thm:BRCC-8:ExpTime} & \PSpace{} \\\hline  %
\cBCcc{} & \CPLX[Cor.]{\NExp}{cor:nexptimeComplete} & \CPLX[Cor.]{\NExp}{cor:nexptimeComplete} & \CPLX{$\geq$\NExp}{nexptime} & \CPLX{$\geq$\NExp}{nexptime} & \PSpace{} \\\hline\hline  %
\cBCm{} & \CPLX[Cor.]{\NP}{NB-complexity} & \multicolumn{4}{c}{\PSpace{}}\vline \\\hline  %
\rule[0pt]{0pt}{12pt}\cBCmc{}& \Exp & \Exp & $\geq$\Exp & $\geq$\Exp & \PSpace{} \\\hline  %
\rule[0pt]{0pt}{12pt}\cBCmcc{} & \NExp & \NExp{} & $\geq$\NExp{} & $\geq$\NExp{} & \PSpace{} \\\hline\multicolumn{2}{c}{}\\[-2pt]\hline  %
\rule[0pt]{0pt}{12pt}& $\All$ & $\Con$ & $\R^n$, {\small $n \geq 3$}& $\R^2$ & $\R$ \\\hline\hline %
\SFU{} & \PSpace{}~{\scriptsize\cite{Ladner77,Nutt99,Arecesetal00}} & \multicolumn{4}{c}{\CPLX[Cor.]{\PSpace}{SFUoverR}}\vline \\\hline  %
\conT{} & \CPLX{\Exp}{theorem:S4uc:ExpTime} & \CPLX{\Exp}{theorem:S4uc:ExpTime} & $\geq$\Exp & $\geq$\Exp & \PSpace \\\hline  %
\conTc{} & \CPLX{\NExp}{theo:conTcInNexptime} & \CPLX{\NExp}{theo:conTcInNexptime} & $\geq$\NExp & $\geq$\NExp & \CPLX{\PSpace}{thm:R:PSpace} \\\hline  %
\end{tabular}
\caption{Satisfiability complexity for the topological logics
  considered in this paper.}
\end{table}

Finally, we remark that, if $n \geq 3$, no upper complexity bound is
currently known for the problem $\Sat(\cBc,\RegC(\R^n))$, or,
therefore, for any more expressive spatial logic.

\section{Conclusion}
\label{sec:conclusion}

In this paper, we have investigated the effect of augmenting various
topological logics in the qualitative spatial reasoning literature
with predicates able to express the property of connectedness.  We
considered three principal {\em base languages}: $\cB$, the language of
the variety of Boolean algebras; $\cBC$, the extension of the
well-known language $\RCCE$ with region-combining operations; and
$\SFU$, the extension of Lewis' system $\SF$ with a universal
operator, under the topological interpretation of McKinsey and
Tarski. And we considered two kinds of connectedness predicate:
$c(r)$, for `region $r$ is connected'; and $c^{\le k}(r)$, for `region
$r$ has at most $k$ connected components.'  For each base language
$\cL$, we defined the languages $\cLc$ (by adding the predicate $c$)
and $\cLcc$ (by adding the predicates $c^{\le k}(r)$ for $k\geq 1$);
and we considered the complexity of the satisfiability problems for
$\cL$, $\cLc$ and $\cLcc$ over various natural (classes of) spatial
domains, both very general---as in the case of $\Regc$, $\ConR$,
$\All$ and $\Con$---and also very specific---as in the case of
$\RegC(\R^n)$ and $\R^n$ for various $n$.

We showed that, whereas the base languages display a surprising
indifference to the frames over which they are interpreted,
the corresponding languages with connectedness predicates are highly
sensitive in this regard. We also showed that the addition of
connectedness predicates increases the complexity of satisfiability
over general classes of frames---typically from $\NP$ or
$\PSpace$ (for the base language $\cL$) to $\ExpTime$ (for the
corresponding language $\cLc$) and $\NExpTime$ (for the language
$\cLcc$). We observed that this increase in complexity is `stable':
over the most general classes of frames, the extensions of such
different formalisms as $\cB$ and $\SFU$ with connectedness predicates
are of the same complexity. We further observed that by restricting
these languages to formulas with just one connectedness constraint of
the form $c(r)$, we obtain logics that are still in \PSpace, while two
such constraints lead to \ExpTime-hardness. Finally, we turned our
attention to the complexity of the satisfiability problems for these
languages when interpreted over Euclidean spaces, summarizing what is
currently known and stating several open problems. The results
obtained are summarized in Table~\ref{table:theBigBoard}.

\section*{Acknowledgements}

The work on this paper was
partially supported by the U.K.\ EPSRC research grants
EP/E034942/1 and EP/E035248/1.

\bibliographystyle{plain}

\end{document}